\documentclass[iop]{emulateapj}

\usepackage{epstopdf}
\usepackage{amsmath}

\shortauthors{Constantin et al.}  

\shorttitle{The Power Sources of LLAGNs}

\slugcomment{Accepted in The Astrophysical Journal}

\begin{document}

\title{Dissecting the Power Sources of Low-Luminosity Emission-Line Galaxy
Nuclei via Comparison of \emph{HST}-STIS and Ground-Based Spectra\altaffilmark{1}}

\altaffiltext{1}{Based on observations made with the NASA/ESA {\it
Hubble Space Telescope}, obtained from the data archive at the Space
Telescope Science Institute. STScI is operated by the Association of
Universities for Research in Astronomy, Inc., under NASA contract NAS
5-26555.}

\author{Anca Constantin\altaffilmark{2}, Joseph
  C. Shields\altaffilmark{3},  Luis C. Ho\altaffilmark{4,5}, Aaron J.
  Barth\altaffilmark{6}, Alexei V. Filippenko\altaffilmark{7}, and Christopher A. Castillo\altaffilmark{2}}
  
\altaffiltext{2}{Department of Physics and Astronomy, James Madison 
  University, Harrisonburg, VA 22807}
  
\altaffiltext{3}{Department of Physics and Astronomy, Ohio University, 
 Athens, OH 45701}

\altaffiltext{4}{Kavli Institute for Astronomy and Astrophysics, Peking University, Beijing 100871, China}

\altaffiltext{5}{Department of Astronomy, Peking University, Beijing 100871, China}

\altaffiltext{6}{Department of Physics and Astronomy, University of California,
Irvine, CA 92697-4575}

\altaffiltext{7}{Department of Astronomy, University of California, Berkeley, 
CA 94720-3411}

\begin{abstract}
Using a sample of $\sim100$ nearby line-emitting galaxy nuclei, we have built the currently definitive atlas of spectroscopic measurements of H$\alpha$ and neighboring emission lines at subarcsecond scales.  We employ these data in a quantitative comparison of the nebular emission in \emph{Hubble Space Telescope (HST)} and ground-based apertures, which offer an order-of-magnitude difference in contrast, and provide new statistical constraints on the degree to which Transition Objects and low-ionization nuclear emission-line regions (LINERs) are powered by an accreting black hole at $\la10$ pc. 
We show that while the small-aperture observations clearly resolve the nebular emission, the aperture dependence in the line ratios is generally weak, and this can be explained by gradients in the density of the line-emitting gas: the higher densities in the more nuclear regions potentially flatten the excitation gradients, suppressing the forbidden emission.   The Transition Objects show a threefold increase in the incidence of broad H$\alpha$ emission in the high-resolution data, as well as the strongest density gradients, supporting the composite model for these systems as accreting sources surrounded by star-forming activity.    
The  narrow-line LINERs appear to be the weaker counterparts of the Type 1 LINERs, where the low accretion rates cause the disappearance of the broad-line component.   
The enhanced sensitivity of the \emph{HST} observations reveals a 30\% increase in the incidence of accretion-powered systems at $z \approx 0$.   A comparison of the strength of the broad-line emission detected at different epochs implies potential broad-line variability on a decade-long timescale, with at least a  factor of three in amplitude.  
\end{abstract}

\noindent
\keywords{galaxies: active --- galaxies: nuclei --- galaxies: emission
lines}

\section{Introduction}

Galaxy nuclei that exhibit emission-line activity are very common at
low redshifts. For a significant fraction (at least 20\%) of these
sources, the emission is believed to come from nonstellar
processes that in many cases can be associated with accretion of
matter onto massive black holes \citep[e.g.,][]{ho97a}, similar to
classical active galactic nuclei (AGNs) such as Seyferts and
quasars.  The physical nature of the large majority (about 2/3)
of the nearby emission-line nebulae, however, remains controversial.  In
the emission-line diagnostic diagrams that are used to classify
galaxy nuclei \citep[e.g.,][]{bpt, vei87, kewley06}, these ambiguous objects exhibit
high-ionization lines that are weaker than those of the accretion-type
sources (Seyferts), and low-ionization forbidden lines that are
stronger than those usually present in H~{\sc ii} regions (powered 
by hot, massive, young stars).  These properties are
characteristic of  low-ionization nuclear emission-line regions (LINERs;
Heckman 1980), including those lacking broad lines (LINER 2, by
analogy with Seyfert 2), and the LINER/H~{\sc ii} ``Transition''
sources \citep{ho93}.   For the sake of simplicity, in most of the text and figures herein we refer to Seyferts as S, to LINERs as L, and to Transition Objects as T.

Several excitation scenarios have been proposed for these low-luminosity active nuclei with an ambiguous nature, in particular LINERs.
The fact that the majority of them exhibit X-ray cores (70\% vs. 30\% in line-free galaxies), that they present radio cores with a detection rate that is similar to that of Seyferts,  that they show significant ultraviolet (UV) variability, and that 25\% of them exhibit broad-line emission \citep[e.g.,][and references therein]{ho08}, strongly argues that these sources are genuine AGNs.  Large statistical studies of the nearby galaxy nuclei also suggest a potential  H~{\sc ii} $\rightarrow$ T/S $\rightarrow$ L $\rightarrow$ passive galaxy evolutionary
sequence in the process of black hole growth within galaxies \citep{sch07, con08,
  con09, sch10}, and thus the possibility that black hole accretion (and hence AGN ionization) is important, if not dominant, in the ambiguous L and T nuclei.   In this picture, Transition objects and Seyferts are intermediate phases between the initial onset of accretion, usually swamped by star-forming gas and associated dust seen optically as an H~{\sc ii} system, and the final phase of accretion observed as LINERs.   

This proposed sequence does not imply, however, that every H~{\sc ii} galaxy at the present epoch must necessarily become a LINER;  it remains possible that some systems go through only the H~{\sc ii} or L phases, which might not be part of the larger progression (e.g., because systems of different activity types appear to have different distributions of host-galaxy morphologies).  
Observations of nearby galactic nuclei suggest other links between these different types of systems as well. For example, a luminous Seyfert 1 will eventually fade to a fainter Type 2 AGN as the accretion rate falls with time \citep[e.g.,][]{elitzur14}, as has been proposed to favor a (not necessarily temporal) sequence in which T nuclei are in between the L nuclei and passive galaxies \citep{ho09}.    This idea is not necessarily at odds with the proposed H~{\sc ii} $\rightarrow$ T/S $\rightarrow$ L $\rightarrow$ passive galaxy evolutionary sequence; while both T and L2 nuclei remain intermediate in their accretion power between bona-fide AGNs and galaxies lacking nonthermal emission, one type of nucleus could still be the manifestation of growing accretion strength while the other represents the waning of it.   Ages and properties of the associated stellar populations on scales of $\sim 100$ pc distinguish between them, with the T systems  being generally younger than the LINERs, reinforcing the idea of a T $\rightarrow$ S $\rightarrow$ L sequence.  Nevertheless, this information is sparse for the small nuclear regions.

On the other hand, works such as \citet[][hereafter EHF]{eracleous10} reveal that there is a serious energy-budget deficit in quite a large majority of the LINER systems in particular: the ionizing photon rate is on par with the Balmer H$\alpha$ photon rate for only very few L nuclei; that is, the ionizing photon rate does not meet the minimum requirement for photon balance (even in the best-case scenario in which all of the ionizing photons are absorbed by the nebula).   
Mechanisms alternative to accretion onto a black hole include
photoionization by hot, massive, young stars \citep{fil92, shi92, bar00}, hot evolved stars \citep[e.g.,][]{binette94}, 
clusters of planetary nebula nuclei \citep{tan00}, or shocks
\citep{dop95}, but they proved only moderately successful in explaining the
optical spectra of these objects.  These models, however, often require
exceptional physical conditions or finely tuned parameters, which are
not consistent with the statistical trends present in large samples of
such nuclei (e.g., the Palomar survey of nearby galaxies; Filippenko
\& Sargent 1985; Ho, Filippenko, \& Sargent 2003; and references
therein).  In particular, scenarios that invoke
excitation by hot, young stars are at odds with evidence
that the nearby galaxy nuclei contain predominantly {\it old} stellar
populations \citep{ho03, gon04, sar05}.

Particularly ambiguous in their interpretation remain the T-type
sources whose emission-line ratios are on the borderline between those of
Seyferts/LINERs and H~{\sc ii} regions.  A seemingly natural explanation for
these nuclei is the so-called composite picture, in which a central
accretion-type source (a weak AGN, maybe a LINER) is surrounded by
star-forming regions \citep[e.g.,][]{sf90}; such a structure would generally remain 
unresolved in typical ground-based apertures.  This scenario is supported by
the similarity in the statistical traits of L and T
nuclei, in both nuclear and host properties.  \citet{ho03} 
indicate that star formation may be more enhanced in the Transition
objects, and that potential projection effects might lead to greater
contamination of their nuclear emission.  \citet{ver97} and
\citet{gon99} additionally argue that the composite nature is revealed in the
emission-line profiles.  They found through a Gaussian decomposition
that the emission separates into narrow ($\sim 200$--270 km s$^{-1}$)
and relatively broad ($\sim 400$--700 km s$^{-1}$) components, with 
H~{\sc ii} and AGN (Seyfert 2)-like flux ratios, respectively.

The ambiguity in finding and defining the dominant ionization mechanism in these systems is clearly related to the fact that the majority of these studies employed ground-based spectroscopy, which, under typical seeing conditions, cannot resolve better than the central few hundred parsecs, even for nearby galaxies ($\la 40$ Mpc).    Moreover, studies like those of \citet{yan06}, \citet{yan12}, \citet{lemaux10}, and \citet{bongiorno10} reveal increasing evidence for LINER-like emission in a large number of galaxies that (1) have been probed with even larger apertures, such as the Sloan Digital Sky Survey \citep[SDSS;][]{york00} observations that employ a $3''$ diameter fiber, or (2) are far more distant, such as in SDSS and in high-redshift surveys like DEEP2 \citep{davies03}, zCOSMOS \citep{lilly07}, BOSS \citep{eisenstein11}, and distant-cluster studies \citep[e.g.,][]{lubin09}, thus covering a much larger  ($>$ kpc scale) ``nuclear" region.    There seems to be an increased consensus that, especially when the LINER-like emission is extended, the most likely ionization source originates in hot, evolved, post-asymptotic giant branch (post-AGB) stars \citep{stasinka08, sarzi10, cidfernandes11, capetti11, yan12}, analogous to the conclusion drawn by EHF based on energy-budget deficits, and consistent with earlier ideas proposed by \citet{binette94}.

Higher-resolution studies, such as with the {\it Hubble Space Telescope} (\emph{HST}), remain equally ambiguous in determining the nature of L nuclei.
Some LINERs exhibit a narrow-line region (NLR) that is strongly concentrated in the center \citep{walsh08}, while in some others it can extend to hundreds of parsecs \citep{shields07, masegosa11}.   Thus, even in the very nuclear regions of L nuclei, the post-AGB scenario could, in principle, still be dominant.  However,
the relatively small-number statistics of these studies preclude robust conclusions on the truly nuclear nature of these galaxies. Similarly, the T nuclei remain marginally studied at high spatial resolution.

With the work presented here, we aim to help solve this problem through a careful aperture comparison of the truly nuclear nebular emission for $\sim 100$ objects.   We have built the currently definitive spectral atlas of {\it HST} Space Telescope Imaging Spectrograph (STIS) data on H$\alpha$ and neighboring features that uniquely probe the emission-line properties at subarcsecond scales ($\sim 0\farcs 2$), which we statistically compare with spectral properties obtained with order-of-magnitude larger apertures from the ground ($\sim 2''$).   
In this way, the high angular resolution isolates the spectrum of the central few tens of
parsecs, while the degree of central concentration is constrained by
comparison with the large-aperture data.  If the composite picture
for the T sources is correct, then we expect to see a
difference in the nebular line emission in the two apertures, with a
more AGN-like behavior in the small-aperture spectra.  Moreover, when
an active nucleus is present, the ionizing radiation emerges from a
central source and falls off in density as $r^{-2}$, and if the gas
density gradients are small, the degree of ionization of the emitting
gas is expected to diminish with radius.  Thus, radial variations in
nebular line ratios should reflect this behavior because such gradients 
are not expected for distributed sources of ionization (e.g., shocks, hot
stars, turbulent mixing layers).  

Another feature that a truly nuclear aperture comparison can study in 
detail is the presence of broad emission lines, which are generally 
considered to be a clear signature of an accretion-type nucleus.   
A lack of broad emission in the ambiguous L and T objects 
can arise because the broad-line region (BLR) 
is simply absent, or because it is below the detection
threshold, which is set mostly by the degree of contamination by
starlight in the host galaxy.  Ground-based observations (e.g., the
Palomar survey) employ apertures that span typically many hundreds of
parsecs in size, and thus can intercept significant continuum and NLR
emission from circumnuclear regions.  Thus, a significant
improvement in the sensitivity to broad components should come from
observing the nuclear emission through smaller slit sizes.  Spectra
acquired through apertures that are an order of magnitude smaller
include significantly less surrounding starlight, thereby
providing greater contrast that should allow for a more
sensitive measure of the nuclear emission.

Our sample presents  a 4-fold increase in the statistics employed in the pilot study
by \citet[][; hereafter S07]{shields07} called the Survey of Nearby Nuclei with the
Space Telescope Imaging Spectrograph (SUNNS).  
The findings of this initial aperture comparison are surprising in both the expected gradients in the line ratios and the expected increase in sensitivity and thus detection of the BLR in these low-luminosity galaxy nuclei:  (1) the emission-line spectra often show little variation with aperture size, suggesting that the ambiguous  emission-line sources do not necessarily harbor an accretion component, and (2) the STIS spectra do not reveal any new signature for broad emission for the sources that appear narrow lined from the ground.   
Hence, the limited objects used in SUNNS do not readily confirm the expectations for the composite model of T nuclei, which do not seem to reveal an accretion-powered component in the small-aperture spectra.  Also, with the BLR detected and measured in only six sources, alternative explanations for the lack of new broad H$\alpha$ detections in the more sensitive \emph{HST} observations (e.g., those involving variability)  cannot be tested.   
Our much larger sample allows us to significantly more thoroughly examine these ideas and possible implications.   

Throughout this paper, unless otherwise indicated, we adopt the flat, $\Lambda$-dominated cosmology with $\Omega_m = 0.3$, and H$_0 = 72$ km s$^{-1}$ Mpc$^{-1}$ \citep{spergel03}.

\section{The nuclei sample and data processing} \label{data}

With the aim of expanding the S07 study to significantly higher number statistics, we searched the \emph{HST} archive for all galaxies known to have emission-line nuclei for which STIS spectroscopy covering the Balmer H$\alpha$ and the adjacent strong nebular emission lines was obtained and is publicly available.   We have identified such observations with the STIS G750M grating for 74 sources having high-quality ground-based nuclear spectroscopy and associated measurements of nebular emission from the Palomar Spectroscopic Survey of nearby galaxies \citep{fil85,ho97a}, in addition to the 23 sources presented by SUNNS.    
For 16 additional sources with STIS data, we have been able to obtain with the Multiple Mirror Telescope (MMT) new ground-based spectra of quality and resolution matching the Palomar data.  Measurements of emission lines for some of the MMT targeted galaxies exist in the literature, but they are not suitable for the analysis we propose; the published results are incomplete, inhomogeneous, and often have not taken the stellar continuum into account properly, as we are able to do now.  We were thus able to put together an \emph{HST} vs. ground-based aperture comparison of the nuclear nebular emission for a total of 113 objects. 

We present in this section details about both the new MMT spectroscopic program and the archival STIS observations, as well as the data reduction and the corresponding emission-line measurements.  Table~\ref{tbl-1} presents some basic properties and the STIS observational details of both the Palomar galaxies and the objects with new MMT spectra (see \S~\ref{mmt_data}). The SUNNS objects are discussed by S07 and their properties are not listed here, but they are included in the analysis.

\subsection{The MMT Observations} \label{mmt_data}

We have obtained high signal-to-noise ratio (S/N) spectrophotometry of 16 galaxy nuclei with the 6.5-m MMT, on the nights of 2008 Feb. 13, Nov. 5, and Nov. 6.  We employed the Blue Channel spectrograph with the 500 grooves mm$^{-1}$ grating, used in first order with
the $1''$ slit to cover 3800--7000~\AA\ with 3.6~\AA\ resolution; the slit was at the parallactic angle \citep{fil82}.  This setup allows us to measure the
red nebular lines ([O~{\sc i}] $\lambda$6300, H$\alpha$, [N~{\sc ii}]
$\lambda$6583, and [S~{\sc ii}] $\lambda\lambda$6716, 6731) that are of primary
interest for comparison with the {\it HST} measurements, as well as the blue part of the spectrum covering H$\beta$ and [O~{\sc iii}] $\lambda\lambda$4959, 5007,
which allow for a complete spectroscopic classification. The
resolution is similar to that employed in the Palomar spectroscopic
survey, which is sufficient to model and obtain a relatively clean separation of the continuum starlight
and nebular emission (i.e., for typical velocities in the centers of
galaxies, the stellar and nebular features are at least partially
resolved).  The seeing was $\sim 1\farcs0$--$1\farcs2$ during all three nights. 
The total exposure times varied between 25 and 40 min.  Each exposure was broken up into two subexposures to allow for the removal of cosmic rays (CRs).

\begin{deluxetable*}{lrlllclll}
\tablecolumns{9} \tablewidth{0pt} \tablecaption{The Nuclei Sample, \emph{HST}-STIS Observations
\label{tbl-1}} \tablehead{ 
\colhead{Object} & 
\colhead{Prop.} & 
\colhead{$D_L$\tablenotemark{a}} &
\colhead{Spectral} & 
\colhead{Aperture} &
\colhead{Central $\lambda$} & 
\colhead{Plate} &
\colhead{Palomar} &
\colhead{Notes\tablenotemark{e}}\\
\colhead{Name} & 
\colhead{ID} & 
\colhead{(Mpc)} &
\colhead{Type\tablenotemark{b}} & 
\colhead{Size} &
\colhead{(\AA)} & 
\colhead{Scale\tablenotemark{c}} &
\colhead{/MMT\tablenotemark{d}} &
\colhead{}} 

\startdata
  IC 356      &12187&18.90&  T2      &0$\farcs$25 $\times$ 0$\farcs$1 & 6581 &0.05  &P& (3)\\  
  NGC 193  & 8236 & 56.40 &L/L   &0$\farcs$25 $\times$ 0$\farcs$2 & 6768 &0.10 &M& (1)\\  
 NGC 315   & 8236 &59.60&  L1.9   &0$\farcs$25 $\times$ 0$\farcs$1 & 6768 &0.05  &P& (1)\\  
 NGC 383   & 8236 & 65.62 &L/L1.9   &0$\farcs$25 $\times$ 0$\farcs$1 & 6768 &0.05 &M& (1)\\
 NGC 541   & 8236 & 63.24 &L:T/T   &0$\farcs$25 $\times$ 0$\farcs$2 &6768  &0.10 &M& \\
 NGC 1052 & 7403 &19.48&  L1.9   &0$\farcs$25 $\times$ 0$\farcs$2 & 6581 &0.05  &P& (1)\\
 \enddata 
\tablecomments{For ApJ Editors: The full version of this table is presented as Table~\ref{tbl-1whole} at the end of the manuscript; {\it this is because emulateapj does not properly set multi-page tables inside the main text}.} 
\tablenotetext{a}{The mean distance of all redshift-independent values, from NASA/IPAC Extragalactic Database (NED).}
\tablenotetext{b}{Spectral class for the Palomar objects is from \citet{ho97a}; spectral class for the MMT objects is based on our measurements of emission-line fluxes in the MMT spectra (see Table~\ref{tbl-mmt-mmt}) and spectral classification criteria of  \citet{ho97a}/Kewley et al. (2006). With the exception of NGC 383, all of the MMT objects are of Type 2 (narrow emission lines only) based on their  MMT spectra.  Uncertain classifications are followed by a colon.}  
\tablenotetext{c}{In arcsec pixel$^{-1}$; the spatial pixel size on STIS is intrinsically
0$.\farcs$05, but some spectra were obtained with a
2-pixel binning readout mode along the spatial direction, producing a spatial scale of 0$\farcs$1 in the readout.}
\tablenotetext{d}{The origin of the ground-based data; P for Palomar, M for MMT.}
\tablenotetext{e}{On the \emph{HST} spectral measurements: (1) Broad H$\alpha$ is required to fit the H$\alpha$ 
+ [\ion{N}{2}] emission complex.  (2) Broad wings are apparent in the
H$\alpha$ + [\ion{N}{2}] blend; however, models
of two Gaussians for the [\ion{S}{2}] lines do not require a broad H$\alpha$.  (3) Poor-quality {\it HST}-STIS
spectra, no measurements.  (4) Not suitable for modeling or deblending the emission lines using [\ion{S}{2}] as an empirical template (see Sections ~\ref{mmt_data} and ~\ref{lines}).}

\end{deluxetable*}

Standard reductions including bias subtraction, flat-field correction, wavelength calibration, and flux calibration were performed  using IRAF.\footnote{IRAF is distributed by the National Optical Astronomical Observatory, which is operated by the Association of Universities for Research in Astronomy, Inc., under cooperative agreement with the National Science Foundation (NSF).}  For flux calibration, the Kitt Peak IRS spectroscopic standard stars G191B2B and BD+17$^\circ$4708 were observed in February, while BD+17$^\circ$4708, HD 19445, and HD 84937 were chosen during the November run; they were observed at the beginning, middle, and end of each night.  Spectra of He-Ne-Ar comparison lamps were obtained before and after each observation to calibrate the wavelength scale. 
The one-dimensional (1-D) spectra were extracted with a $1\arcsec \times 4\arcsec$ effective aperture for consistency with the Palomar survey.   We corrected for the telluric oxygen absorption lines near 6280 and 6860~\AA\ through division by normalized, intrinsically featureless spectra of the standard stars; differences in airmass were scaled appropriately.   This reduction procedure also corrected for continuum atmospheric extinction.

\begin{figure*}
\epsscale{1.175}
\plotone{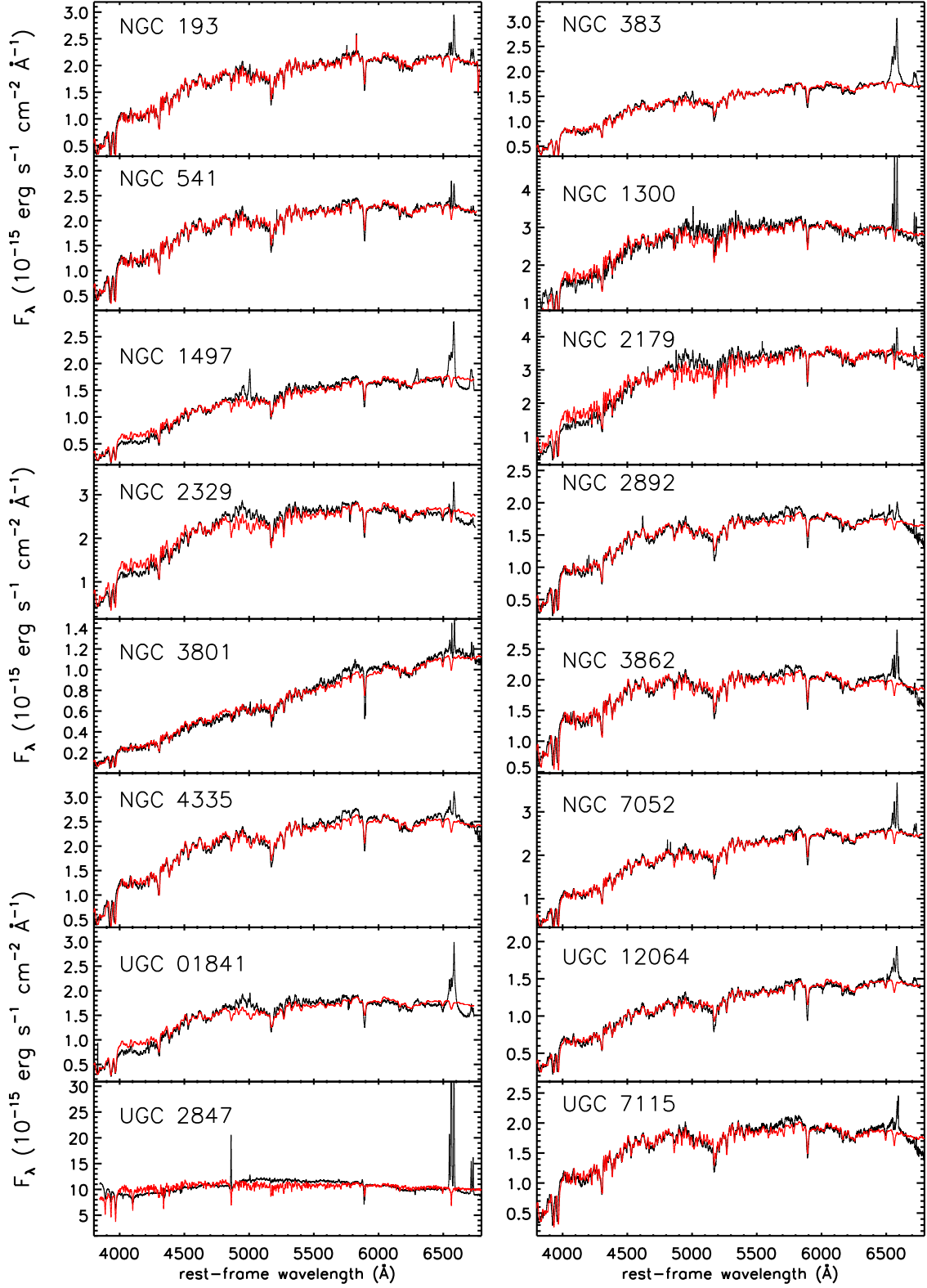}
\caption{Rest-frame spectra of the galaxy nuclei acquired with the MMT ({\it black}), with fits of the host stellar light overplotted ({\it red}).
\label{allmmt}}
\end{figure*}

The emission-line fluxes in the MMT spectra were measured after removal of the underlying stellar continuum, which is significant in the ground-based aperture data.
Host-galaxy starlight was modeled with six synthesized galaxy spectral templates, which were built from the library of stellar populations of \citet{bru03} using the  method described by \citet{tremonti04}.  Each template was broadened by convolution with a Gaussian to match the stellar velocity dispersion of the host galaxy. 
The final extracted nuclear MMT spectra corrected for telluric absorption are shown in Figure~\ref{allmmt}, along with their best-fitting multiple-stellar-population models.   Many nuclei display obvious nebular emission; however, especially in the blue part of the spectrum (e.g., [\ion{O}{3}] $\lambda\lambda$4959, 5007; H$\beta$), the emission lines are mostly buried in the stellar continuum.

We note that while the spectral fits are generally satisfactory for the continuum shape, the reddest part of the spectrum that hosts the [\ion{S}{2}] emission doublet fails to match the theoretical model for about half of the MMT sample (see Fig.~\ref{allmmt}).   In particular, the synthetic stellar continuum cuts through this doublet with the effect of artificially diminishing the total [\ion{S}{2}] fluxes as well as increasing the [\ion{S}{2}] $\lambda$6716/$\lambda$6731 ratios measured in the continuum-subtracted spectra.    Because there is no significant stellar absorption feature that affects this emission doublet, we decided to use the measurements of the [\ion{S}{2}] features in the original flux-calibrated spectra without subtraction of the underlying continuum.    For completeness, for the aperture comparison tests involving this particular emission doublet, we have compared the results considering measurements with and without the continuum subtraction, as well as with results obtained by simply removing these objects from the comparison, and the conclusions remained consistent within the errors.

The continuum-subtracted emission-line spectra have been modeled and measured with {\tt specfit}
\citep{kri94}, a method that employs line and continuum spectral fitting via an interactive ${\chi}^2$ minimization.
The formal uncertainties in the line fluxes include errors in the continuum subtraction and flux calibration, and they have been propagated to the line ratios.  The continuum was
approximated by a linear fit, while the emission features were modeled
by Gaussian profiles.  The fit was performed simultaneously for all
lines, assuming common velocity fields and widths for all features of the NLR.
For deblending the H$\alpha + $[\ion{N}{2}] emission region, we followed the technique of \citet{ho97a, ho97b} and assigned to both the [\ion{N}{2}] lines and the narrow H$\alpha$ emission the line profile that was predetermined by fitting the [\ion{S}{2}] lines.
When the [\ion{S}{2}] features presented
visibly strong broad wings, the narrow components were modeled with
multiple Gaussians, in order to determine the necessity of an
additional broad component for the Balmer H$\alpha$.  This procedure
is in principle able to account for potential asymmetries, velocity
blending, or multiple velocity peaks in the narrow lines, as well as for 
wings in their bases (see also Section~\ref{broad}).  Line doublets were
forced to share the same velocity widths.  The flux ratio for the
[\ion{N}{2}] doublet was constrained to the value determined by the
branching ratio.  As in \citet{ho97a}, the measurements of the line
flux ratios involving H$\alpha$ employed the Gaussian components
describing only the narrow-line emission.  The fluxes of the main narrow emission features measured in the MMT spectra are listed in Table \ref{tbl-mmt-mmt}.

\begin{deluxetable*}{ccccccccc}
\tablecolumns{9} \tablewidth{0pt} \tablecaption{MMT Fluxes of Narrow Emission Lines for MMT Objects
\label{tbl-mmt-mmt}} \tablehead{ 
\colhead{Object} & 
\colhead{H$\beta$} &
\colhead{[\ion{O}{3}]} & 
\colhead{[\ion{O}{1}]} &
\colhead{H$\alpha$} &
\colhead{[\ion{N}{2}]} & 
\colhead{[\ion{S}{2}]} &
\colhead{[\ion{S}{2}]} \\
\colhead{name} & 
\colhead{(narrow)} &
\colhead{$\lambda$5007} & 
\colhead{$\lambda$6300} &
\colhead{(narrow)} &
\colhead{$\lambda$6583} & 
\colhead{$\lambda$6716} &
\colhead{$\lambda$6731} &
\colhead{$f_{\lambda}$(6563~\AA)}} 
\startdata
NGC   193 &  183  $\pm$ 4 & 340 $\pm$ 4 &148 $\pm$   4 &  626 $\pm$  5 &  979 $\pm$   5 & 286 $\pm$ 5 &  267 $\pm$ 5 & 211 $\pm$ 2\\
NGC   383 &  241  $\pm$ 5 & 426 $\pm$ 5 &254 $\pm$  24 &  594 $\pm$ 70 & 1233 $\pm$  82 & 400 $\pm$ 26 &  382 $\pm$ 24 & 127 $\pm$ 15\\
NGC   541 &  179  $\pm$ 3 & 120 $\pm$ 3 &149 $\pm$   4 &  621 $\pm$  4 &  403 $\pm$   4 &  83 $\pm$ 4 &   70 $\pm$ 4 & 226 $\pm$ 2\\
NGC  1300&  241  $\pm$ 5 & 426 $\pm$ 5 &260 $\pm$  49 & 1326 $\pm$ 124& 1780 $\pm$ 111& 428 $\pm$ 56 &  374 $\pm$ 44 & 173 $\pm$ 16\\
NGC  1497 &  330  $\pm$ 5 & 947 $\pm$ 5 &507 $\pm$  13 & 1002 $\pm$ 57 & 1888 $\pm$  62 & 416 $\pm$ 29 &  203 $\pm$ 21 & 174 $\pm$ 10\\
NGC  2179 &  543  $\pm$ 4 & 964 $\pm$ 4 &199 $\pm$   4 &  727 $\pm$ 45 &  945 $\pm$  42 & 638 $\pm$ 35 &  387 $\pm$ 25 & 317 $\pm$ 20\\
NGC  2329 &  862  $\pm$ 6 &1294 $\pm$ 6 &193 $\pm$   7 &  440 $\pm$ 20 &  945 $\pm$  24 & 196 $\pm$ 28 &  188 $\pm$ 11 & 484 $\pm$ 22\\
NGC  2892 &  148  $\pm$ 6 & 188 $\pm$ 6 & 55 $\pm$  24 &   68 $\pm$ 25 &  230 $\pm$  35 & 188 $\pm$ 23 &  139 $\pm$ 21 &  45 $\pm$ 17\\
NGC  3801 &   21  $\pm$ 4 &  87 $\pm$ 4 & 64 $\pm$   4 &  358 $\pm$  4 &  352 $\pm$   4 & 130 $\pm$ 4 &   78 $\pm$ 4 & 111 $\pm$ 1\\
NGC  3862 &  251  $\pm$ 4 & 212 $\pm$ 4 &208 $\pm$  46 &  704 $\pm$ 40 & 1098 $\pm$  45 & 212 $\pm$ 15 &  137 $\pm$ 46 & 193 $\pm$ 11\\
NGC  4335 &  125  $\pm$ 5 & 333 $\pm$ 5 &221 $\pm$ 111 &  657 $\pm$ 39 & 1144 $\pm$  41 & 132 $\pm$ 5 &  180 $\pm$ 3 & 249 $\pm$ 15\\
NGC  7052 &  236  $\pm$ 4 & 241 $\pm$ 4 &133 $\pm$   4 & 1087 $\pm$  5 & 1264 $\pm$   4 & 267 $\pm$ 4 &  249 $\pm$ 4 & 252 $\pm$ 1\\
UGC  1841 &  618  $\pm$ 6 &1133 $\pm$ 6 &358 $\pm$  21 &  687 $\pm$ 35 & 1476 $\pm$  39 & 339 $\pm$ 21 &  213 $\pm$ 21 & 246 $\pm$ 13\\
UGC  2847 & 6221  $\pm$ 3 &1518 $\pm$ 3 & 980 $\pm$  3 &43225 $\pm$  4 &18570 $\pm$   3&3234 $\pm$ 3& 3336 $\pm$ 3 &1017 $\pm$ 1\\
UGC  7115 &  270  $\pm$ 6 & 284 $\pm$ 6 &377 $\pm$   7 &  455 $\pm$ 38 &  954 $\pm$  43 & 219 $\pm$ 55 &  155 $\pm$ 13 & 182 $\pm$ 15\\
UGC 12064&  152  $\pm$ 5 & 170 $\pm$ 5 &102 $\pm$   5 &  645 $\pm$  5 &  818 $\pm$   5 & 162 $\pm$ 6 &  112 $\pm$ 6 & 145 $\pm$ 1\\
\enddata 
\tablecomments{The emission-line fluxes are in units of $10^{-17}$ erg s$^{-1}$ cm$^{-2}$, and represent the observed  values, not corrected for reddening.}
\end{deluxetable*}

We classified spectroscopically the MMT objects based on their locations in line diagnostic diagrams using the criteria of ~\citet{ho97a}, and the resulting respective classes are recorded in Table~\ref{tbl-1}.  For these 16 objects we also list, for comparison, the spectral classes based on the ~\citet{kewley06} criteria; the consistency is reasonably good.   The uncertain classifications refer to objects for which the sets of conditions involving the low-ionization lines ([\ion{O}{1}], [\ion{N}{2}], [\ion{S}{2}]) do not hold simultaneously for the three ~\citet{vei87} diagnostic diagrams; in these cases, to remain consistent with ~\citet{ho97a}, the [\ion{O}{1}]/H$\alpha$ ratio was given the largest weight, and the corresponding class is listed first, followed by all the classes that may be consistent with the data.
The MMT sources are spectrally classified as mostly L nuclei (of Type 2, with the exception of NGC 383, where a broad H$\alpha$ component is needed for the best spectral fit; see Table~\ref{tbl-broad} for measurements), thus providing a welcome increase in the number statistics for this spectral class;  the MMT sample also adds three T nuclei and one Seyfert galaxy to the Palomar sample.

\subsection{The {\it HST}-STIS Archival Data}\label{hst_data}

The STIS spectra used for this study were  originally acquired mainly 
for the purpose of deriving central black hole masses, and cover the
H$\alpha$ and adjacent strong nebular features.  In all cases, the
grating employed was G750M centered at 6581~\AA\ or 6768~\AA,
which makes available for measurements [\ion{N}{2}] $\lambda\lambda$6548, 
6583, [\ion{S}{2}] $\lambda\lambda$6716, 6731, and at the
6581~\AA\ setting, [\ion{O}{1}] $\lambda$6300.

For ten of the Palomar objects (NGC 3675, 3953, 4258, 4321, 4435, 4486, 4527, 4826, 5879, and 7331), STIS spectra have been obtained via two different programs.   For completeness, Table~\ref{tbl-fl} presents our measurements of the fluxes of the strong emission lines, as described in Section~\ref{lines}, for all of the STIS spectra of these objects, except for the SUNNS spectra (of NGC 4321 and 4435).
We excluded from analysis observations of 18
spectra from program IDs (PID) 8228\footnote{For three of these 10 objects
(two H~{\sc ii}, NGC 2748 and NGC 3949, and one Transition object, NGC
5055), \citet{hug03} show only the 2-D spectra, and characterize them
as displaying weak, patchy extended emission} (five H~{\sc ii}, four T, and one Seyfert),  8607 (five T), and 12187 (two H~{\sc ii} and one T) owing to extremely low S/N.  Two of these sources, NGC 4527 and 7331, are common to two PIDs; for NGC 4527 we could extract a 1-D spectrum for one dataset (PID 8607), while for NGC 7331 none of the programs offered sufficiently good data.  The STIS spectrum of NGC 3516 exhibits an unusual shape that proves unsuitable for spectral modeling using [\ion{S}{2}] as an empirical template for the [\ion{N}{2}] and the narrow H$\alpha$ emission
(see Section~\ref{lines}), and therefore we do not record any measurements for this object.
Thus, STIS data for 17 objects remain unusable for this project.  
The final count of 80 sources with Palomar and STIS spectroscopy, including the SUNNS sources, consists of 20 H~{\sc ii}, 22 T, 14 L2, 13 L1, and 11 S nuclei (both Type 1 and 2).  
With the addition of the MMT objects, the breakdown of the total sample by the ground-based spectral type consists of  20 H~{\sc ii}, 
25 T, 26 L2, 13 L1, and 12 S nuclei (both Type 1 and 2), thus providing quite balanced statistics of bona-fide AGNs (L1 and S), H~{\sc ii},  and ambiguous sources of both L2 and T flavors.

We retrieved from the archive all on-the-fly calibrated frames with
the intention to use the fully rectified and flux- and
wavelength-calibrated 2-D images to further extract the 1-D spectra.
For the majority of objects the S/N is low, and/or
CRs are abundant.  We obtained the best results by performing our own CR correction to the archival
flat-fielded images before feeding these frames back into the \emph{HST}
pipeline software (the {\tt x2d} task under the {\tt stsdas} package,
IRAF)  to generate the calibrated images.

Incident CRs and bad pixels have been removed using the {\tt ccdclip}
algorithm under the {\tt imcombine} task; after the images to
be combined are realigned to account for the offsets parallel to the
slit, pixels with values outside a determined range are assigned the
median value.  The range of acceptable pixel values, and thus the
parameters involved in this procedure (e.g., hsigma, mclip, nkeep) are
adjusted for each individual case after close comparisons of the
central 10 rows (of interest for the 1-D extraction) in both the
initial and the corrected image spectra, to ensure that valid emission
features were not damaged.  For the observations that consist of only 
one nuclear image spectrum, and which have a high CR density, we
interpolated over affected pixels using an automated detection
algorithm for the strongest events, and did manual intervention for weaker
events comparable in strength to real emission features.

The 1-D object spectra were extracted using the {\tt apall} procedure.
As indicated in Table~\ref{tbl-1}, the majority of objects were
observed through a slit with a $0\farcs2$ width.  Part of these
observations were acquired with a plate scale of $0\farcs05$ pixel$^{-1}$, 
while for others the measurements have been binned to
$0\farcs1$ pixel$^{-1}$ at readout.  For the
unbinned spectra we extracted the central 5 rows (pixels), while for
the cases where data have been rebinned we extracted only the central
2.5 rows, to preserve the angular scale. However, since the galaxies are
at different distances/redshifts, the metric scale mapped out by these
observations is not conserved.  For simplicity,
the object spectra acquired with the $0\farcs1$ slit width (which were
all unbinned spatially, so the plate scale is $0\farcs05$ pixel$^{-1}$)
were also generated using 5-pixel-wide extractions.

\begin{deluxetable*}{lcccccc}
\tablecolumns{7} \tablewidth{0pt} \tablecaption{{\it HST} Fluxes of Narrow Emission Lines for the Sample of Nuclei
\label{tbl-fl1}} \tablehead{ 
\colhead{Object} & 
\colhead{[\ion{O}{1}]} & 
\colhead{H$\alpha$} &
\colhead{[\ion{N}{2}]} & 
\colhead{[\ion{S}{2}]} &
\colhead{[\ion{S}{2}]} \\
\colhead{name} & 
\colhead{$\lambda$6300} & 
\colhead{(narrow)} &
\colhead{$\lambda$6583} & 
\colhead{$\lambda$6716} &
\colhead{$\lambda$6731} &
\colhead{$f_{\lambda}$(6563~\AA)}} 
\startdata
  IC 356       &\nodata     &\nodata          &  \nodata         &\nodata          & \nodata      &\nodata      \\  
 NGC 193   & \nodata     & 59$\pm$53 & 365$\pm$107 & 210$\pm$50 & 310$\pm$50&5$\pm$4\\
 NGC 315   & \nodata    & 207$\pm$2  &  520$\pm$7  &  109$\pm$8 & 122$\pm$7    &   1.3$\pm$0.1	\\
 NGC 383   & \nodata     & 1570$\pm$720 & 3290$\pm$1800 & 1210$\pm$370 & 1210$\pm$1090&0.07$\pm$0.01\\
 NGC 541   & \nodata     & 92$\pm$11 & 225$\pm$12 & 37$\pm$9 & 32$\pm$9& 9$\pm$2\\
 NGC 1052 & 3505$\pm$27  & 4801$\pm$26 & 4088$\pm$75 & 1930$\pm$20 & 2727$\pm$21  &    26$\pm$1	\\
  \enddata 
\tablecomments{The full version of this table is presented as Table~\ref{tbl-fl1whole} at the end of the manuscript; {\it this is because emulateapj does not properly set multi-page tables inside the main text}. \\
The emission-line fluxes are in units of 
$10^{-17}$ erg s$^{-1}$ cm$^{-2}$, and represent the observed 
values, not corrected for reddening. The upper limits recorded here for NGC 4150, NGC 4636, and NGC 6503 are listed 
as $(2\pi)^{1/2} \sigma_{\lambda} (3 \sigma_{c})$, where 
$\sigma_{\lambda}$ is the width of the [\ion{N}{2}] line, and 
$\sigma_{c}$ is the root-mean-squared (RMS) uncertainty per pixel in the local 
continuum.  The last column lists the continuum flux density 
$f_{\lambda}$ at 6563~\AA, in units of $10^{-17}$ erg s$^{-1}$ 
cm$^{-2}$ \AA$^{-1}$.    Objects are in the same order as in 
Table~\ref{tbl-1}.}
\end{deluxetable*}

\begin{figure*}
\epsscale{1.165}
\plottwo{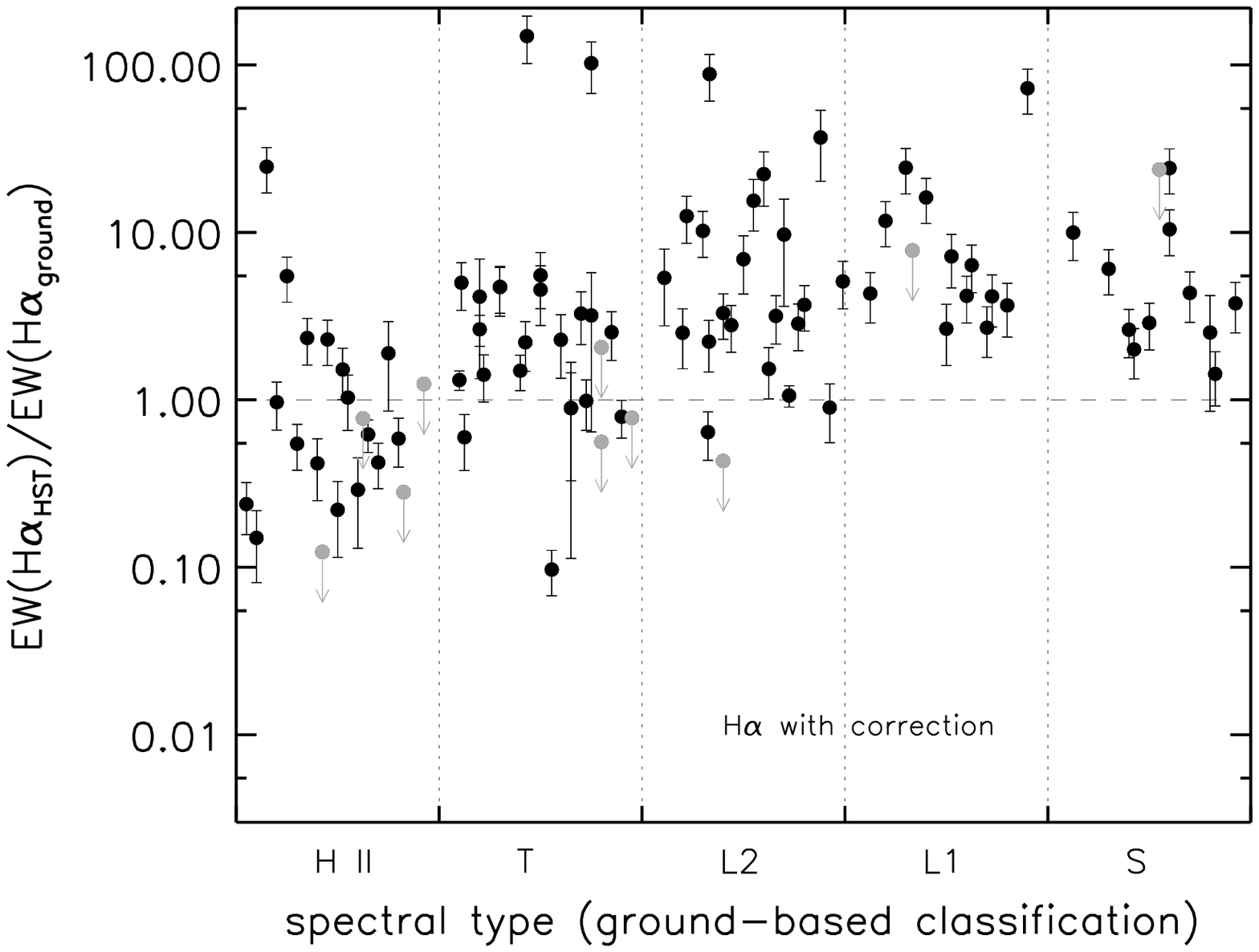}{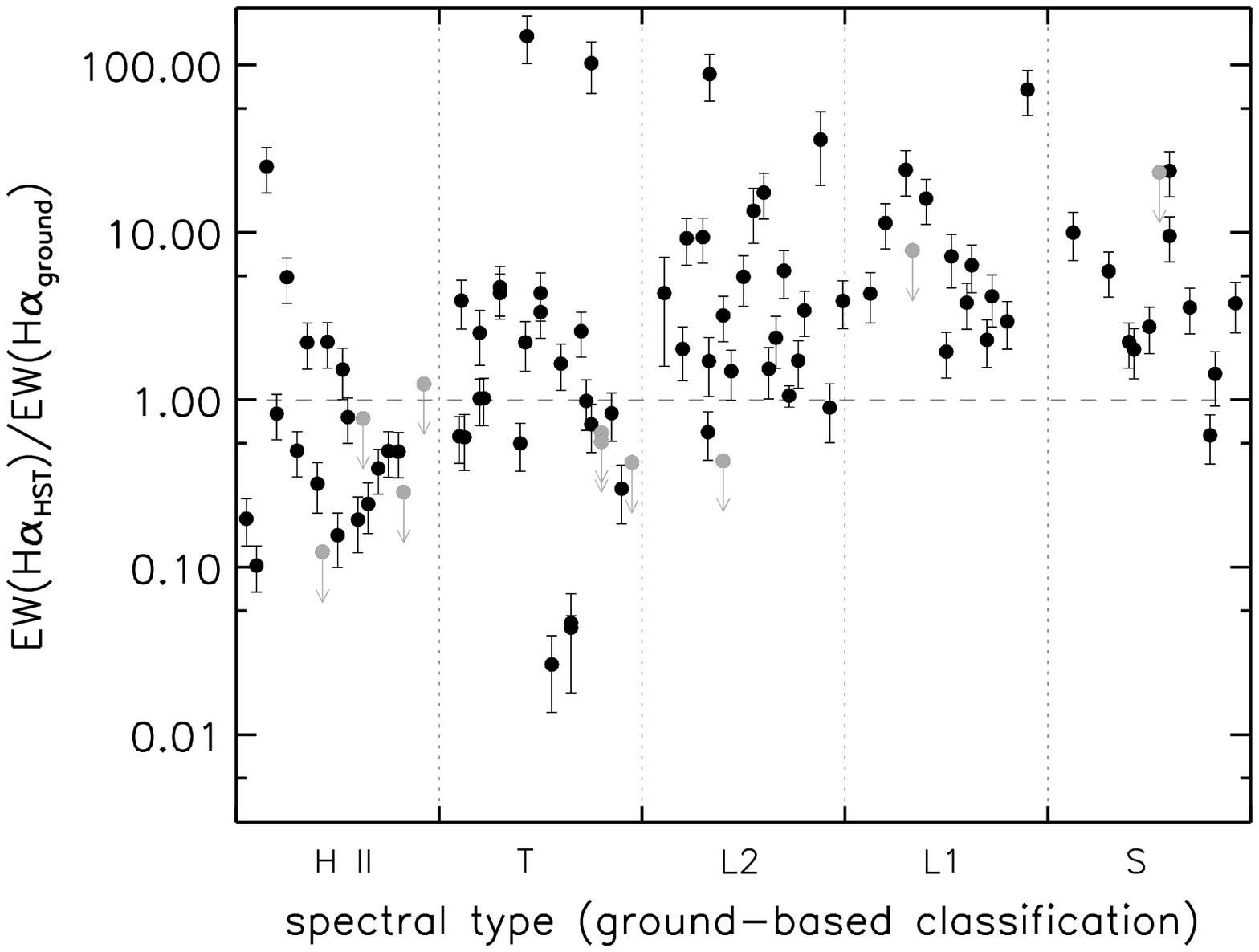}
\caption{Ratios of EWs of the narrow H$\alpha$ line emission
in {\it HST} and ground-based (Palomar and MMT) apertures, on a logarithmic scale, as a function of
ground-based nuclear spectral class, as defined by \citet{ho97a}.  The {\it left} and {\it right} panels reflect measurements of the H$\alpha$ fluxes with and without correction for stellar absorption, respectively (see text).  Upper limits are plotted in gray.
The horizontal spread of points within a given spectral class is for clarity only, with no physical significance.
\label{ew}}
\end{figure*}

\subsection{Measurements of Emission Lines in the \emph{HST}-STIS Spectra} \label{lines}

The STIS aperture is much smaller (two orders of magnitude in
area) than that typically used in ground-based measurements of galaxy
nuclei.   Hence, the corresponding contamination by the host-galaxy
starlight is significantly reduced.  However, the stellar component is
not negligible and it is important to correct for it, as it can 
introduce additional structure in the continuum, and thus might
influence the appearance of the emission features.  Fitting a template
galaxy stellar contribution that would eventually be
subtracted from the nuclear spectrum (as we did for the MMT ground-based data)  is, however, very difficult for most of these {\it HST} observations because of the modest S/N and the relatively
small wavelength coverage.  The strongest effect of starlight
contamination of the nuclear emission is expected to be in the
H$\alpha$ emission, owing to the presence of the Balmer absorption line.
A potentially useful alternative to stellar background light subtraction is to adopt reasonable values for
the strength of the H$\alpha$ absorption, using information offered by
stellar population synthesis models, and correct for it.  

A typical value of the equivalent width (EW) of the H$\alpha$
absorption line can be estimated from measuring all the templates
contained in the \citet{bru03} library of synthetic galaxy spectra.
For three different models of star-formation history (instantaneous
starburst, constant star formation, and exponentially decaying 
star-forming history), each equally represented for three different
metallicity values (subsolar, supersolar, and solar, with $Z = 0.004$,
0.05, and 0.02, respectively, where $Z$ is the total mass fraction of
metals), and ages that range from 0.005 Gyr to 12 Gyr, the EW of 
the H$\alpha$ absorption line varies between 2~\AA\ and 6~\AA,
with most of the cases concentrating around 4~\AA.  
Moreover, studies of the properties of the nuclear stellar populations
\citep[e.g.,][]{sar05} show that, in a large majority of nearby nuclei ($\sim
80$\%), the age is at least 5~Gyr and the chemical enhancement of the
emitting gas is at least solar, narrowing down the possible range in
H$\alpha$ absorption EW to 2.5--4.5~\AA.  Therefore, for the
corrections used in the narrow H$\alpha$ line flux and the line flux
ratio comparison involving this feature, we used for the EW of the
H$\alpha$ absorption line the value of 3.5~\AA\ $\pm$ 1~\AA.  The
adopted error translates into a typical additional uncertainty of $\sim 10$\% in
the resulting line flux ratios, but with considerably larger error
bars for the weak-lined objects.  

To understand the consequences of
our adopted absorption correction, we repeated the correction with
different values of the absorption EW in the range
2.5--4.5~\AA.   We found that the distributions of line ratios overall
and within object subclasses did not change significantly, and the
resulting conclusions were similarly robust.  Nevertheless, because this method of correcting for stellar absorption does not account for the inherent variety in the true strength of the stellar absorption in H$\alpha$, we present the results of this study for narrow H$\alpha$ fluxes with and without the correction described above.

The line fluxes  of the emission features observed in the STIS spectra were measured via spectral modeling following the same technique employed for the stellar continuum subtracted MMT spectra, as discussed above in Section ~\ref{mmt_data}.   To recap, the spectra were fit with multiple Gaussians to match both broad and narrow components of all observed lines, while describing the continuum with a first-order polynomial.  The [\ion{S}{2}] emission was used as an empirical template for the profiles of the [\ion{N}{2}] doublet and the narrow component of H$\alpha$.  To account for the contribution of blueshifted or redshifted wings, in an attempt to assess the presence or absence of a  broad H$\alpha$ component, we employed multiple Gaussians for fitting the emission profile of the NLR.
The resulting measurements of the \emph{HST} fluxes of the strong emission lines, without correction for stellar absorption, are recorded in Table \ref{tbl-fl1} for all objects included in this study.

It is important to note that the procedure described above does not account for the possible
effects of stellar continuum structure on other emission features
considered in this study (e.g., the [\ion{S}{2}] doublet), or for its
effect on the shape of the H$\alpha$ emission line.  In these
particular features, however, these effects are in general expected to be
much smaller than in typical ground-based observations of these
nuclei.  Figure~\ref{ew} provides an illustration of the degree to
which the contrast between the nuclear emission and the continuum is
increased in the \emph{HST} observations; we show this comparison for EWs measured both with and without correction for the H$\alpha$ stellar absorption, as described above.  The plots show the ratio of the EWs of the
narrow H$\alpha$ line emission in the two apertures separately
for all individual spectral types; for the objects observed at two epochs with \emph{HST} (see Table~\ref{tbl-1} for
the Prop IDs), we include both measurements, which are identified by the same value on the abscissa.   Their corresponding H$\alpha$ fluxes differ by $\sim
0.2$ dex when different aperture sizes are employed (0\farcs2 versus
0\farcs1), and are consistent within the errors when the same aperture size is used.  
It is readily apparent that the H$\alpha$ line is stronger in the \emph{HST} spectra of most Seyferts and
LINERs, and of many Transition objects as well.  For the rest of the T 
and H~{\sc ii} nuclei, the uncertainties are large and are
dominated by the limited data quality and structure in the continuum;
it appears thus that our correction for starlight in these sources does not affect the final conclusions of this study.

\subsection{Broad Emission in the {\it HST}-STIS Spectra} \label{broad}

One of the clearest indications of accretion-powered activity is the
presence of broad permitted lines [full width at half-maximum intensity (FWHM)
of a few thousand km s$^{-1}$].  The strongest broad component at
optical wavelengths is expected to be H$\alpha$, and thus detection of
a broad H$\alpha$ component is a promising method of constraining the
origin of the nuclear activity.  The results of the Palomar survey
suggest that broad H$\alpha$ is quite common in local Seyferts 
and a significant fraction of 
LINERs; it is found in at least 10\%\ of all nearby luminous galaxies
\citep{ho97b}.  The true incidence of broad-lined sources must, however,
be higher; especially in the weakly active nuclei, ground-based
observations are unable to distinguish a possibly diluted (and weak) BLR 
from the structure present in the continuum produced by starlight.
\emph{HST} spectroscopy that employs small-aperture observations has the
power to eliminate a significant fraction of contaminating starlight,
and hence to increase the sensitivity to broad wings of line emission.
Thus, an important question to address is whether, at \emph{HST} resolution,
the fraction of nuclei classified as broad lined is larger than that
emerging from ground-based data.

\citet{ho97b} describe the set of objects that show definite or
probable evidence of broad H$\alpha$ emission in their survey.  Our
sample contains 15 of the nuclei with unambiguous broad H$\alpha$ in
the Palomar spectra, and the STIS observations confirm the
presence of the broad Balmer feature in these sources (Table~\ref{tbl-1}).   
In Table~\ref{tbl-broad} we summarize for each of these
nuclei several parameters concerning the broad H$\alpha$ component.
The fractional contribution of the broad component of H$\alpha$ to the
entire H$\alpha +$ [\ion{N}{2}] complex, $f_{\rm blend}$, is given in
column (2), while column (3) indicates its fractional contribution to the
whole H$\alpha$ (narrow + broad) emission, $f_{H\alpha}$.  The
FWHM of broad H$\alpha$ and its velocity shift relative to the narrow
component are recorded in columns (4) and (5), respectively.  Columns (6) and (7)
give the observed (not corrected for reddening) flux and luminosity of
broad H$\alpha$ in units of erg s$^{-1}$ cm$^{-2}$ and erg s$^{-1}$,
respectively, where we used the distances given by \citet{ho97a} to ease the comparison.  
We show in parenthesis the corresponding \citet{ho97b} parameters, when available.

While most of the time the detection and measurement  of a broad H$\alpha$ feature is fairly obvious, there are cases where this process remains ambiguous and therefore debatable.   
For example, NGC 4036 is one of the nuclei with broad H$\alpha$ detections in the Palomar survey, for which 
our spectral fits of the 1-D extraction of the central five rows of the STIS spectrum reveal a broad component, but for which \citet{walsh08} do not find the need for it in any of the individual central rows of the 2-D spectrum derived from the same observational dataset. The disagreement with our findings may be the result of the increased signal, and thus an increased contrast with the continuum, that we obtain by adding the flux in the central few rows.  
Nevertheless, our use of a wider extraction would also decrease the contrast between the broad line and the continuum since the wider aperture adds more continuum.
We will revisit this issue later in the section, where we discuss the new detections of broad H$\alpha$ in the \emph{HST} data.

\begin{figure*}
\epsscale{1.165}
\plottwo{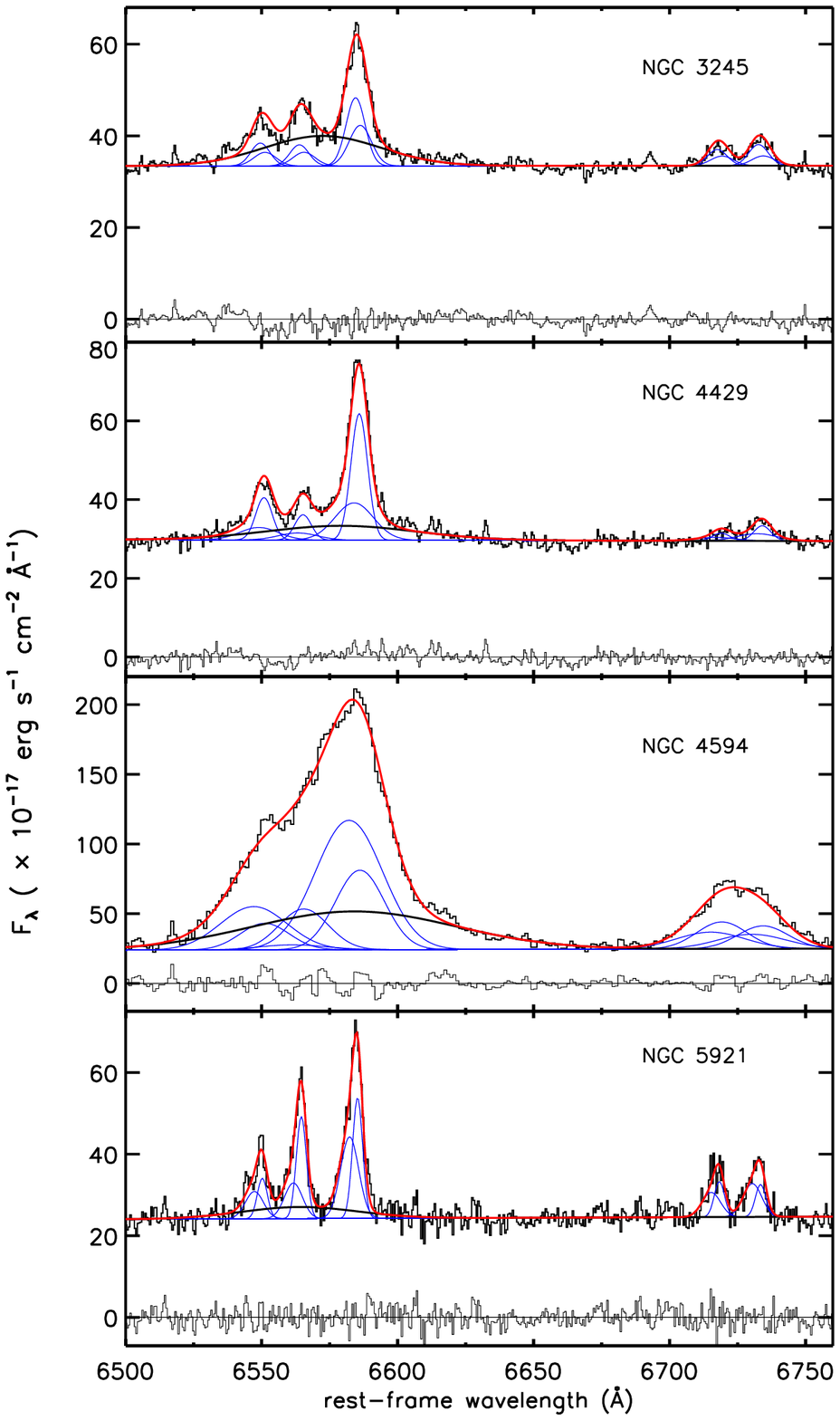}{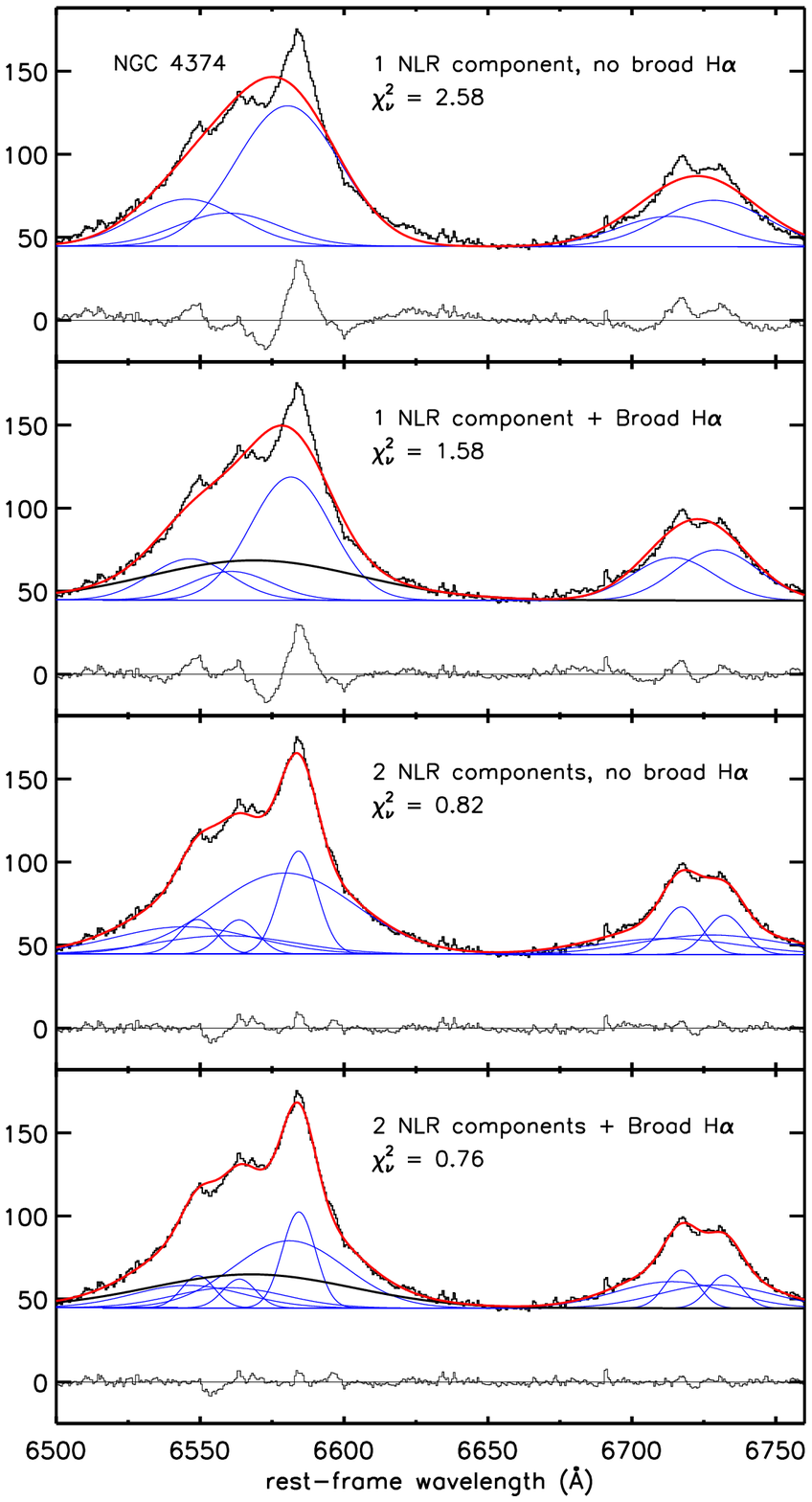}
\caption{STIS spectra and fits of the H$\alpha$ + [N~{\sc ii}] to [\ion{S}{2}] region in all Palomar objects with new detections of the broad H$\alpha$ component: NGC 3245, 4429, 4594, 5921 ({\it left panels}) and NGC 4374 ({\it right panels}), where we show, for comparison, the best fits without and with a broad H$\alpha$ component, with both a single and two components for the NLR; NGC 4736 is omitted here as it is presented in detail by \citet{con12}.  For all spectra, the blue thin lines show the individual Gaussian components, the black line represents the broad H$\alpha$ feature, and the red line is the final fit to the observed spectrum.  The residuals after the subtraction of the model fit are shown at the bottom of each panel.   
\label{fitbroad}}
\end{figure*}

Of the objects that are ambiguous in terms of the
presence/detection of broad H$\alpha$ in the Palomar survey, three
nuclei are included in this study.   While NGC 4501 (a
Seyfert included in the SUNNS sample) is claimed to lack the
broad emission in the STIS aperture (S07), 
NGC 3245 (viewed from the ground as a narrow-lined T nucleus)
requires for a good fit of the H$\alpha +$ [\ion{N}{2}] doublet the
presence of a broad component, with a contribution to the total blend
of $f_{\rm blend} \approx 47$\%.  The detection significance is tested
by fitting the feature with models based on both one- and two-Gaussian
component templates for the [\ion{S}{2}] lines; $f_{\rm blend}$
remains basically unchanged in both cases.  We show in Figure~\ref{fitbroad}, among other sources for which we find new evidence for broad emission in the STIS spectra,
 the individual Gaussian components, the broad H$\alpha$ feature,
and the final spectral fit for both the H$\alpha$ + [\ion{N}{2}] line
blend and the [\ion{S}{2}] $\lambda\lambda$6716, 6731 doublet for NGC 3245.  The
existence of the broad Balmer feature in this nucleus has also been
suggested by \citet{bar01}, based on this same dataset.  The third object with tentative broad H$\alpha$ emission in the Palomar spectrum, NGC 2985, remains, as in the ground-based observations, rather elusive: a moderate ($f_{\rm blend} \approx 44$\%) broad component may be necessary to account for the entire flux of the blend after the narrow components have been modeled based on [\ion{S}{2}] templates;  however, the data are of low quality across all features, and we cannot rule out the possibility that the excess emission originates from weak, intrinsically broad wings in the narrow-line profile.  We decided to include the detection and measurements of this broad feature in our analysis, for the purpose of comparison with the ground-based data.

\begin{deluxetable*}{lccccccc}
\tablecolumns{8} \tablewidth{0pt} \tablecaption{Galaxies with Broad H$\alpha$ Emission 
\label{tbl-broad}} 
\tablehead{ 
\colhead{Object} & \colhead{$f_{\rm blend}$} &
\colhead{$f_{\rm H\alpha}$} & \colhead{FWHM} & \colhead {$\Delta v$} &
\colhead{log $F(\rm H\alpha)$} & \colhead{log $L(\rm H\alpha)$} 
& \colhead{obs. yr.}\\
\colhead{(1)} & \colhead{(2)} &
\colhead{(3)} & \colhead{(4)} & \colhead {(5)} &
\colhead{(6)} & \colhead{(7)} & \colhead{(8)}} 
\startdata
\cutinhead{Palomar Sample}
NGC 315                 & 0.25 & 0.59 & 2870 &1190 & -14.52 & 39.19 & 06/2000 \\ 
                         &(0.29)&(0.64)&(2000)&     &(-13.86)&(39.85)&  (10/1985)\\
NGC 1052                 & 0.34 & 0.54 & 2800 & ~10 & -13.25 & 39.33 & 01/1999\\ 
                         &(0.32)&(0.55)&(1950)&     & (...)  &  (...)&  (02/1984)\\
NGC  2985      & 0.44 & 0.81 & 2250 &210 & -13.78 & 38.96 & 07/2011 \\ 
                       &(0.22)&(0.37)&(2050)&         &(...)        &(...)&  (02/1984)\\
NGC 3031\tablenotemark{a}& 0.72 & 0.90 & 3680 & 360 & -12.15 & 38.22 & 07/1999\\ 
                         &(0.57)&(0.86)&(2650)&     &(-12.02)&(39.17)&  (06/1986)\\
NGC 3227                 & 0.20 & 0.44 & 2680 & ~~0 & -11.78 & 40.93 & 02/1999\\ 
                         &(0.71)&(0.86)&(2950)&     &   (...)& (...) &  (02/1986)\\
NGC 3245                 & 0.47 & 0.79 & 4300 & 420 & -14.39 & 38.38 & 02/1999\\  
                         &(...) &(...) & (...)&     &   (...)& (...) &  (02/1986)\\
NGC 3642\tablenotemark{a}& 0.79 & 0.85 & 1650 & 450 & -13.81 & 39.15 & 10/2000\\ 
                         &(0.35)&(0.51)&(1250)&     &   (...)& (...) &  (02/1984)\\
NGC 3998\tablenotemark{a}& 0.59 & 0.77 & 6320 & 320 & -12.65 & 40.09 & 04/2002\\ 
                         &(0.37)&(0.59)&(2150)&     &(-12.58)&(40.16)&  (02/1984)\\
NGC 4036                 & 0.24 & 0.61 & 2440 & 730 & -14.59 & 38.27 &  04/1999\\ 
                         &(0.14)&(0.39)&(1850)&     &(-13.69)&(39.17)&  (02/1984)\\
NGC 4051\tablenotemark{a}& 0.89 & 0.96 & 6500 & ~~0 & -12.40 & 40.14 & 03/2000\\ 
                         &(0.74)&(0.84)&(1000)&     &   (...)& (...) &  (02/1984)\\
NGC 4258  		 & 0.50 & 0.70 & 1470 &  90 & -13.45 & 38.29 & 08/2000\\ 
NGC 4258   		 & 0.25 & 0.41 & 1310 & -40 & -13.86 & 37.88 & 03/2001\\ 
                         &(0.49)&(0.67)&(1700)&     &(-13.09)&(38.65)&  (02/1984)  \\
NGC 4278  		 & 0.41 & 0.67 & 2880 & ~50 & -14.03 & 38.02 &  05/2000\\ 
                         &(0.19)&(0.38)&(1950)&     &(-13.07)&(38.98)&  (02/1984)\\
NGC 4429  		 & 0.28 & 0.75 & 2650 & 820 & -14.60 & 37.93 &  05/2001\\
                         &(...) &(...) & (...)&     &   (...)& (...) &  (01/1985)\\
NGC 4579\tablenotemark{a}& 0.70 & 0.92 & 6590 &1280 & -12.92 & 39.61 &  04/1999\\
                         &(0.21)&(0.48)&(2300)&     &(-13.08)&(39.45)&  (02/1984)\\
NGC 4594   	         & 0.29 & 0.77 & 4000 & 910 & -13.59 & 39.09 &  01/1999\\
                         &(...) &(...) & (...)&     & (...)  & (...) &  (02/1984)\\

NGC  4636      & 0.76 & 0.98 & 2330 &260 & -14.38 & 38.12 & 04/2001 \\ 
                       &(0.27)&(0.56)&(2450)&         &(-14.18)        &(38.36)&  (02/1985)\\
NGC 4736    		 & 0.80 & 0.93 & 1570 & 140 & -13.38 & 37.96 &  06/2002\\
                         &(...) &(...) & (...)&     & (...)  & (...) &  (02/1984)\\

NGC 5005     & 0.17 & 0.58 & 2610 & ~~5 & -14.41 & 38.32 &  12/2000\\
                         &(0.33)&(0.79)&(1650)&     &(-12.69)&(40.05)&  (02/1984)\\
NGC 5077   		 & 0.34 & 0.58 & 2570 & 550 & -14.24 & 39.05 &  03/1998\\
                         &(0.12)&(0.26)&(2300)&     &(-14.06)&(39.23)&  (02/1984)\\
NGC 5921                 & 0.20 & 0.45 & 2280 & ~40 & -14.82 & 38.06 &  05/2000\\  
                         &(...) &(...) & (...)&     &  (...) &  (...)&  (10/1999)\\
\cutinhead{MMT sample}
NGC 193	 & 0.66 & 0.95& 2460 & 550  & -13.98 & 39.68 & 10/1999\\     	
NGC 383	 & 0.63 & 0.86& 2710 & 480  & -12.99 & 40.79 & 10/2000\\     	
                    &(0.56)&(0.83)&(2780)&(411)&(-13.57)&(40.21)&(11/2008)\\
NGC 4335 & 0.39 & 0.79& 3020 & 370  & -14.16 & 39.55 & 09/1999\\     	
UGC 1841 & 0.43 & 0.81& 3040 & 410  & -13.71 & 40.27 & 11/1999\\     	
\enddata 

\tablecomments{Col. (1): Galaxy name.  When the objects are observed in
different programs, the rows showing the results are listed in
increasing order of the PID number, as in Table ~\ref{tbl-1}.  Col. (2): Fractional contribution
of the broad component of H$\alpha$ to the H$\alpha$ + [\ion{N}{2}]
blend.  Col. (3): Fractional contribution of the broad H$\alpha$ to the
total (narrow + broad) H$\alpha$ emission.  Col. (4): FWHM of broad
H$\alpha$ in units of km~s$^{-1}$.  Col. (5): Velocity shift of the
broad H$\alpha$ component relative to the narrow H$\alpha$ line; the
units are km s$^{-1}$.  Col. (6): Observed (not corrected for
reddening) flux of broad H$\alpha$, in units of erg s$^{-1}$
cm$^{-2}$; Palomar fluxes are available only for objects observed under photometric conditions.  
Col. (7): Observed (not corrected for reddening) luminosity
of broad H$\alpha$, in units of erg s$^{-1}$, assuming distances from
\citet{ho97a}. Col. (8): the year of observation with STIS.
Cols. (2), (3), (4), (6), (7), and (8) show in parentheses the
corresponding parameters measured from the ground with the Palomar \citep{ho97b} and MMT.}

\tablenotetext{a}{Non-Gaussian profile of the
broad component; FWHM and velocity shift are flux-weighted averages of
the respective values in all Gaussians used to fit the broad
H$\alpha$.}
\end{deluxetable*}

In the same \citet{ho97b} study that presents the search for ``dwarf''
nuclei with broad emission, the authors also list the sources where
the total H$\alpha +$ [\ion{N}{2}] blend hints at a probable presence
of weak broad wings of H$\alpha$, but for which a careful profile
fitting analysis indicates null detections.  Among those objects, seven sources 
are included in our study.  For six of them [NGC 4314, 4321, and 4698
(included in the SUNNS sample), and NGC 1961, 2911, and 6500], the profile
fits of the \emph{HST} spectra do not require any broad component.  The
seventh object is NGC 4594, which has also been previously observed with the 
Faint Object Spectrograph (FOS) aboard \emph{HST} using a 0\farcs86-diameter 
circular aperture.  In these
data, the H$\alpha +$ [\ion{N}{2}] complex shows evidence for broad
wings; attributing them to a broad Balmer component, however, remains a
matter of debate \citep{kor96, nic98}; this nucleus exhibits
severely blended velocity structure in the STIS slit (0\farcs1),
which is almost one order of magnitude narrower than that used in the
FOS observation.  While the optical H$\alpha$ + [\ion{N}{2}] emission remains
difficult to interpret, we find that a good fit can be
obtained only by adding a moderately strong broad component ($f_{\rm
blend} \approx 29$\%, Figure~\ref{fitbroad}, Table~\ref{tbl-broad}).

The STIS observations of our MMT sample have been previously presented and discussed by \citet{noel03} for all but NGC 1497, which is discussed by  \citet{walsh08}.    Both of these studies emphasized measurements of the gas kinematics in the vicinity of the supermassive black holes in the centers of these galaxies, and performed spectral fits and measured the nebular emission in individually extracted rows.    As described above, we have performed our own data reduction and extraction of the 1-D spectra for all 16 MMT objects.   Thus, the line fluxes and other parameters (e.g., evidence for H$\alpha$ broad emission) differ somewhat from those listed in \citet{noel03} and \citet{walsh08}.   For example, \citet{noel03} suggest the possible presence of a broad H$\alpha$ component, as measured in the single central row, in six of our MMT sources and two Palomar objects (NGC 4261 and NGC 7626);  
we find that including a broad feature in the spectral fit does not improve the resulting $\chi^2$ per degree of freedom ($\chi_{\nu}^2$) significantly enough to warrant its physical existence in NGC 315, 4261, 7626, and UGC 12064, while we confirm its presence in NGC 193, 383, 4335, and UGC 1841.   
As for NGC 1497, our measurements agree with the findings of \citet{walsh08} that a broad H$\alpha$ line does not  improve the final spectral fit; however, this dataset has rather poor S/N, and the decision for the best spectral fit is basically the most conservative choice that allows for the smallest number of free parameters. 
For some of the galaxy nuclei for which we decided against the presence of the broad feature, apparent broad wings in the H$\alpha$ + [\ion{N}{2}] blend can be fit with two Gaussians for the [\ion{S}{2}] lines and for all the other narrow features (see Table~\ref{tbl-1}).   We adopted the same method when fitting the host-galaxy-corrected (pure emission-line) spectra acquired from the ground with the MMT, and we found that in only one of these systems (NGC 383) a broad H$\alpha$ component is needed for the best spectral fit (Table~\ref{tbl-broad}).

Arguments against detecting broad H$\alpha$ emission, or for mismatches in the measurements of this component, could also stem from employing different spectral fitting methods.   For example, in a recent discussion of the relative efficiency of  \emph{HST} versus ground-based spectra for detecting BLRs in low-luminosity AGNs,  \citet{balmaverde14} suggested that one should use the [\ion{O}{1}] profile (which is usually of larger width than that of the [\ion{S}{2}] lines that we use as templates) to model the [\ion{N}{2}] and the narrow H$\alpha$ contributions to the blend, 
because the stratification in density and ionization within the NLR causes differences in the location of the emitting region of the various lines, and thus in their line profiles; when the [\ion{O}{1}] profile is used, these authors find that there is no more need for a broad component in nine of the objects we discuss here (NGC 315, 1052, 2985, 3642, 4036, 4258, 4278, 5005, and 5077).   Nevertheless, it is important to note that over the spatial scales of the {\it HST} spectra, the line width vs. $n_{\rm crit}$ relationship is spatially resolved, and thus the NLR stratification should no longer be an issue; the use of the [\ion{O}{1}]  emission as a template is not justified, as its critical density is much higher than that of the [\ion{N}{2}] and [\ion{S}{2}] lines.

\subsubsection{New Evidence for Broad H$\alpha$ in the STIS Spectra} \label{newbroad}

As only briefly mentioned above, in addition to the nuclei whose ground-based spectra hinted
at a possible presence of the broad Balmer emission feature, the STIS
data reveal {\it eight objects} in which broad H$\alpha$ emission was
not identified in the ground-based data; five of these cases were
previously reported \citep[NGC 193, 3245, 4335, 4594 and UGC 1841;][]{noel03, walsh08}  and three new cases are given
in this work.  The ones we newly discovered via this analysis are in the centers of one LINER 2 galaxy (NGC 4736) and two Transition objects (NGC 4429, 5921).    

NGC 4736 displays beyond doubt a strong broad H$\alpha$ line.   The STIS spectrum, the associated measurements, and a discussion of this system's rather peculiar nature, in conjunction with  the high-resolution X-ray, radio, and variable UV sources detected at the optical nucleus, are presented by \citet{con12}.    The fractional contribution of the broad H$\alpha$ component to the total flux of the blend is 80\%.   Interestingly, this LINER has possibly the least luminous BLR known to date ($L_{H \alpha} = 2.2 \times 10^{37}$~erg s$^{-1}$).   

The detections of the broad Balmer lines in NGC 4429 and NGC 5921 are, on the other hand, marginal, with a contribution of 28\% and 20\%, respectively; however, these fractions are not significantly sensitive to the fitting model employed.   For these two sources, as well as for NGC 3245 and NGC 4594 \citep[discussed by][]{walsh08}, the two-Gaussian models for the narrow emission were not sufficient for an adequate fit to the broad wings of the H$\alpha$ + [\ion{N}{2}] blend, thus supporting the existence of a weak BLR.    Figure~\ref{fitbroad} illustrates the profile decompositions for these systems. 

While we have not been able to properly account for the underlying stellar light, we argue that our claims for new H$\alpha$ broad components are secure despite the relatively low fluxes of these lines.   
As emphasized by Filippenko \& Sargent (1985;
e.g., their Figure 12) and \citet{ho97a}, the underlying stellar
continuum near H$\alpha$ could mimic, if not correctly subtracted, a weak
broad component caused by a broad dip around 6700~\AA\ produced by TiO
in late-K and early-M stars.   An analysis of the synthetic galaxy
spectra of \citet{bru03} indicates that, for the spectral range
covered by the STIS observations, the presence of a strong bump near
H$\alpha$ blueward of the 6700~\AA\ TiO absorption feature would
translate into a pronounced slope in the continuum.  None of the
objects shown in Figure~\ref{fitbroad} display such a feature,
suggesting that the effects of starlight contamination do not contribute
significantly in creating an apparent broad H$\alpha$ profile.

Finally, there are a few nuclei in our sample for which the presence of broad emission in their STIS spectra has been previously discussed elsewhere, and for which conclusions regarding the detection and measurement of the broad H$\alpha$ emission appear discordant.   We discuss here possible explanations for the different interpretations.
For NGC 4261, NGC 7626, and UGC 12064, we do not find evidence for a broad H$\alpha$ component, while \citet{noel03} requires it for fits of the spectra extracted from the central row.    For NGC 7626 we acknowledge the presence of broad wings in the H$\alpha$ + [\ion{N}{2}] blend  (see Table~\ref{tbl-1}); however, fits involving two Gaussians for the narrow forbidden lines are associated with a $\chi_{\nu}^2$ value that is not improved by adding a broad line.   The discrepancy between our conclusions and those of \citet{noel03} could be explained by the narrow-line gas kinematics in these nuclei, for which they find clear evidence. A rapidly rotating NLR that is spatially resolved to some extent can reveal blueshifted emission on one side of the nucleus and redshifted emission on the other;  in this case, a wide (i.e., several pixels) extraction blends these features together and could result in a BLR impostor, which might be what we are observing in our $0\farcs2$ extraction of the 1-D spectrum.  
Another apparently discrepant case is NGC 4036, for which \citet{walsh08} did not require broad H$\alpha$ in the fit to individual row spectra in any of the slit positions, and for which we measure a contribution by such a component of 24\% of the blend in the $0\farcs2$ aperture that combines the central five rows; as indicated in their Figure 2g, it is possible that, in this case, our detection is not real, as it could arise instead from the unresolved high-velocity material in the inner NLR, similar to the case presented above.  The same explanation can apply to the presence of the broad Balmer line in the ground-based spectrum \citep{ho97b}.

NGC 4374 and NGC 4486 are other prominent examples of complex NLR kinematics, which is most likely due to the fact that their BHs are very massive.  
NGC 4374's nuclear emission measured with STIS has been previously interpreted by \citet{bower98} as arising from two separate gas components, fit by two Gaussian components, while \citet{maciejewski01} and \citet{barth01a} explain the line profiles by a rotating disk model rather than requiring two dynamically distinct components.  
The more recent row-by-row analysis of \citet{walsh10} does not rule out a broad H$\alpha$ component in the three innermost CCD rows; however, they show that the NLR disk in this system is very rapidly rotating, and much of the velocity structure in the broad blended line profile is spatially resolved with highly non-Gaussian line profiles at any one location, a large gradient in mean velocity, and strong changes in asymmetry across the nucleus.  

We have explored various spectral fits of the five-pixel 1-D extraction of this spectrum and find that while a single NLR component is improved by adding a broad H$\alpha$ component, the overall fit remains rather poor; also, there is only a marginal 8\% increase in $\chi_{\nu}^2$ in the model involving two NLR components with the addition of a broad H$\alpha$ line that would contribute $39$\% of the H$\alpha$ + [\ion{N}{2}] emission blend (see Fig.~\ref{fitbroad}).  Interestingly, for both fits using two NLR components with and without broad H$\alpha$, the  [\ion{N}{2}]/H$\alpha$ and [\ion{S}{2}]/H$\alpha$ line flux ratios remain consistent within the uncertainties.
We thus conclude that this object does not show evidence for a broad H$\alpha$ emission feature.   When the same sequence of fitting procedure is applied to NGC 4486, for observations from both {\it HST} programs, we find no significant improvement in the fit by adding a broad H$\alpha$ line, consistent with the conclusions of \citet{walsh13} who investigate the NLR kinematics on a row-by-row basis.

It is interesting to note that, with very few exceptions, there is a significant
redward velocity offset of the broad H$\alpha$ component with respect
to the narrow component in the STIS spectra, regardless of whether this is 
fitted with single or multiple Gaussians.  This trend is present in a minority of the
Palomar detections --- only 5 out of 34 confident cases.
\citet{ho97b} suggested that these redshifts may be caused by
systematic errors in determining the continuum level or from
assumptions about the profiles of the narrow lines.  Another plausible
explanation for these redshifts, also invoked by \citet{ho97b} by
analogy with the results of spectropolarimetry of NGC 1068 by
\citet{ant85} and \citet{mil91}, is scattering of the broad-line
photons by outflowing electrons in the NLR.   On the other hand, the possibility that our broad H$\alpha$ detections arise from unresolved high-velocity narrow-line emission, as discussed above, should not result in a general redward offset.    

\subsubsection{Decade-Scale Variability in the BLR?} \label{blrvar}

The overall characteristics of the broad  H$\alpha$ emission measured in the STIS aperture are somewhat surprising, especially when compared to those measured from the ground.  
The fractions $f_{\rm blend}$ and $f_{H\alpha}$, and the FWHM(broad
H$\alpha$) are usually larger in the \emph{HST}  spectra, as expected if
the narrow emission is spatially resolved and a significant amount of
its light is not included in the small aperture.  

Nevertheless, with very few exceptions, the STIS broad H$\alpha$ fluxes are smaller than those measured in the ground-based aperture; in median, the reduction in the broad H$\alpha$ flux exceeds a factor of eight.
A similar trend was noticed by S07 for the subsample of
six SUNNS broad-lined objects.    Since the broad feature is expected to arise from
subparsec scales, it should not suffer from aperture losses; thus,
S07 interpreted this result as
possibly arising from variability in the broad emission, on timescales
that are shorter than a decade (i.e., the interval between the
Palomar and \emph{HST} observations). 

The reason why variability can explain these differences comes from the different detection thresholds for a BLR component in the ground-based and space-based data. Since the larger aperture employed in the ground-based observations includes a higher fraction of the stellar light surrounding the nucleus, there is a weaker contrast between the BLR and the associated continuum, and thus a lower likelihood to detect the broad Balmer line as a separate component in the H$\alpha$ + [\ion{N}{2}] emission blend. Thus, while the chance for the luminosity to increase or decrease is the same in a (simple) variability scenario, AGNs in which the BLRs are low in the ground-based data and high in the space-based data would simply be less likely to provide us with a broad H$\alpha$ detection and hence with a flux measurement in the low state (i.e., from the ground). Similarly, AGNs in which the BLRs are high in the ground-based data and low in the space-based data would exhibit a lower flux in the broad component measured by \emph{HST}, and hence a negative flux offset similar to what we see in the comparison of the Palomar and the \emph{HST} nebular emission.



\begin{figure}
\epsscale{1.25}
\plotone{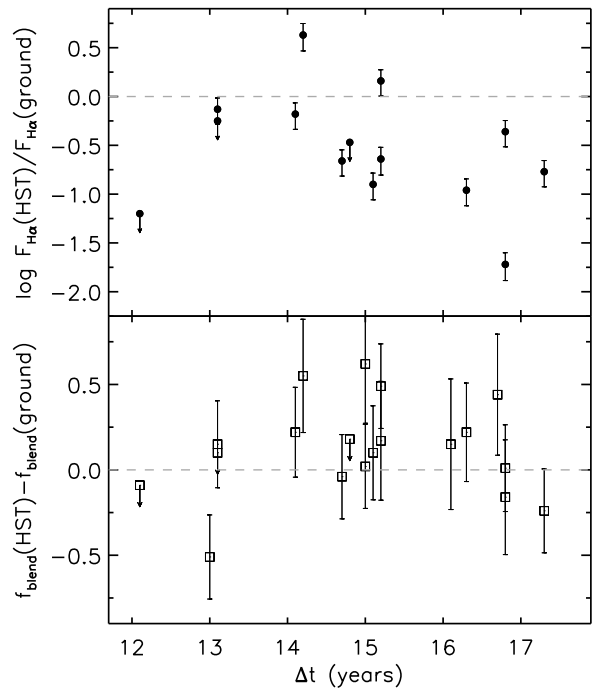}
\caption{ The differences in log $F_{\rm H\alpha}$, the flux of the broad
  H$\alpha$ component [{\it upper panel}], and in $f_{\rm blend}$, the fraction with which
  this feature contributes to the total H$\alpha$ + [\ion{N}{2}] emission [{\it lower panel}],
  measured in the STIS and the ground-based spectra.  The broad H$\alpha$ flux is lower in the \emph{HST} apertures, measured at later times, with a general statistical trend of increasing the negative difference with increasing time interval between observations.  However, $f_{\rm blend}$ remains generally higher when measured in the small \emph{HST} apertures, regardless of the time lapse between observations.  
\label{variability}}
\end{figure}

With the increased statistics offered by the present sample we can, probably for the first time,
analyze broad H$\alpha$ variation in more objects and over longer intervals.  In addition to simply
comparing the flux values from the ground-based and \emph{HST} data, it is of
interest to examine their ratio as a function of time between these
observations, to get some sense of the timescales of variability.  If
the scatter in the observed ratio increases significantly with
increasing time interval, we can conclude that the variability
spectrum has significant power on timescales extending beyond a decade.

\begin{figure*}
\epsscale{1.15}
\plotone{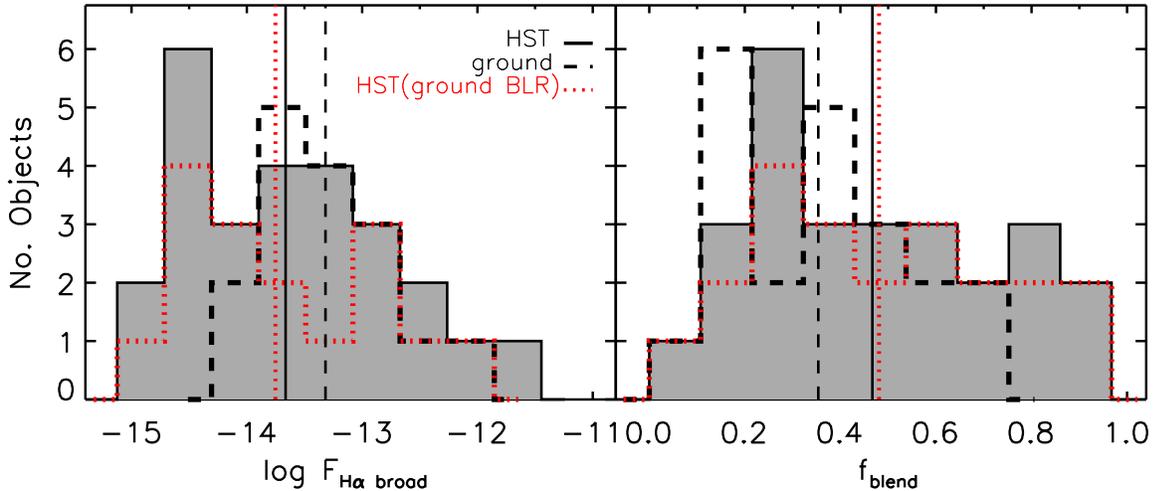}
\caption{A comparison of the ground-based and \emph{HST}-STIS overall distributions of the broad H$\alpha$ flux {\it (left)} and $f_{\rm blend}$, the fraction with which this feature contributes to the total H$\alpha$ + [\ion{N}{2}] emission {\it (right)}.  The whole sample of objects for which broad H$\alpha$ is detected in the STIS apertures regardless of whether they appear as Type 1 from the ground is shown as a black continuous line, while the subsample previously known to exhibit broad emission from the ground is shown as dotted red.    We find that on average, the \emph{HST}  fluxes are lower than those measured from the ground, regardless of whether we consider all of the \emph{HST} detections or just those previously known to be of Type 1.  
\label{habroad_hist}}
\end{figure*}

Figure~\ref{variability} presents the change in  log $F_{H\alpha}$ and in $f_{\rm blend}$ as
a function of time between the Palomar and the STIS observations, for
all galaxies with broad-line emission detected from the ground.  There is evidence of a weak trend between the amplitude of variation and time: an increasingly negative difference with increasing time interval between observations.
There is no apparent trend in  $f_{\rm blend}$; as expected based on the increased contrast with the continuum,  $f_{\rm blend}$ remains generally higher when measured in the small \emph{HST} apertures, regardless of the time interval between observations.  The MMT object for which we have the broad H$\alpha$ comparison data is too isolated from the rest of the sample in its  time interval ($\Delta t \approx -8$ years), and is not plotted here.   A more definitive quantitative analysis of the variability statistics in such objects will require larger samples or additional epochs of observation.

Of course, the variability scenario also predicts new detections of broad-line
sources in the STIS spectra of objects that in the Palomar survey
would have been in a low-flux state, and therefore appeared as Type 2
nuclei owing to sensitivity limits.  The SUNNS sample does not fulfill
this variability prediction; however, in our much larger dataset there
are potentially eight new broad H$\alpha$ detections (\S \ref{newbroad}), thereby bringing 
additional supporting evidence for the validity of this result.   

We also note that our working sample is not subject to biases with respect to the broad-line detection. The sample selection is based purely on the availability of high-quality G750M STIS spectra, with no a priori knowledge of the existence or detectability of a broad H$\alpha$ emission feature. Nevertheless, one might worry that because the ratios presented in Figure ~\ref{variability} can be computed only for those cases where we detect broad H$\alpha$ in both apertures, the results would be biased against those objects in which the broad H$\alpha$ significantly increased over the years (from nondetection in Palomar sample to detection in \emph{HST} data). To address this, we compare in Figure ~\ref{habroad_hist} the overall distributions of the broad H$\alpha$ flux and $f_{\rm blend}$ in both apertures, where we show separately the whole sample of objects with \emph{HST} BLR detection as well as the subsample of objects previously known to exhibit broad emission from the ground. We find that, on average, the \emph{HST} fluxes are lower than those measured from the ground, regardless of whether we consider all of the \emph{HST} detections or just those previously known to be of Type 1. 
By the same token, the BLR component accounts for a higher fraction of the total H$\alpha$ emission in the higher spatial resolution (small aperture) data, which argues again for a clear improvement in the contrast with the continuum as the \emph{HST} observations resolve the nuclear emission. One might also think that the diminution of the broad H$\alpha$ in the small apertures could be caused by cutting out some BLR light from larger scales, and thus that the BLR is not spatially constrained into subarcsecond regions, as has been previously assumed. The increase in $f_{\rm blend}$, however, argues against this possibility.

\section{General Nuclear Properties of the Sample} \label{gen}

We examine in this section several properties of the nuclear nebular
line-emission behavior in conjunction with some of the host-galaxy
characteristics.   Although the sources used in this study consist of
a large number of local galaxy nuclei, spanning quite homogeneously the
entire range of spectral classifications for emission-line galactic
nuclei, they do not form a complete sample in any statistical sense.
We thus investigate whether obvious biases are present and may be
important for understanding the results of the emission-line comparison.  We first look into aperture effects and then the degree to which the optically detected AGN phenomenon extends to luminosities much lower than those usually observed in bona-fide accretion sources.

\subsection{Selection Effects} \label{selection}

\begin{figure*}
\epsscale{1.275}
\plotone{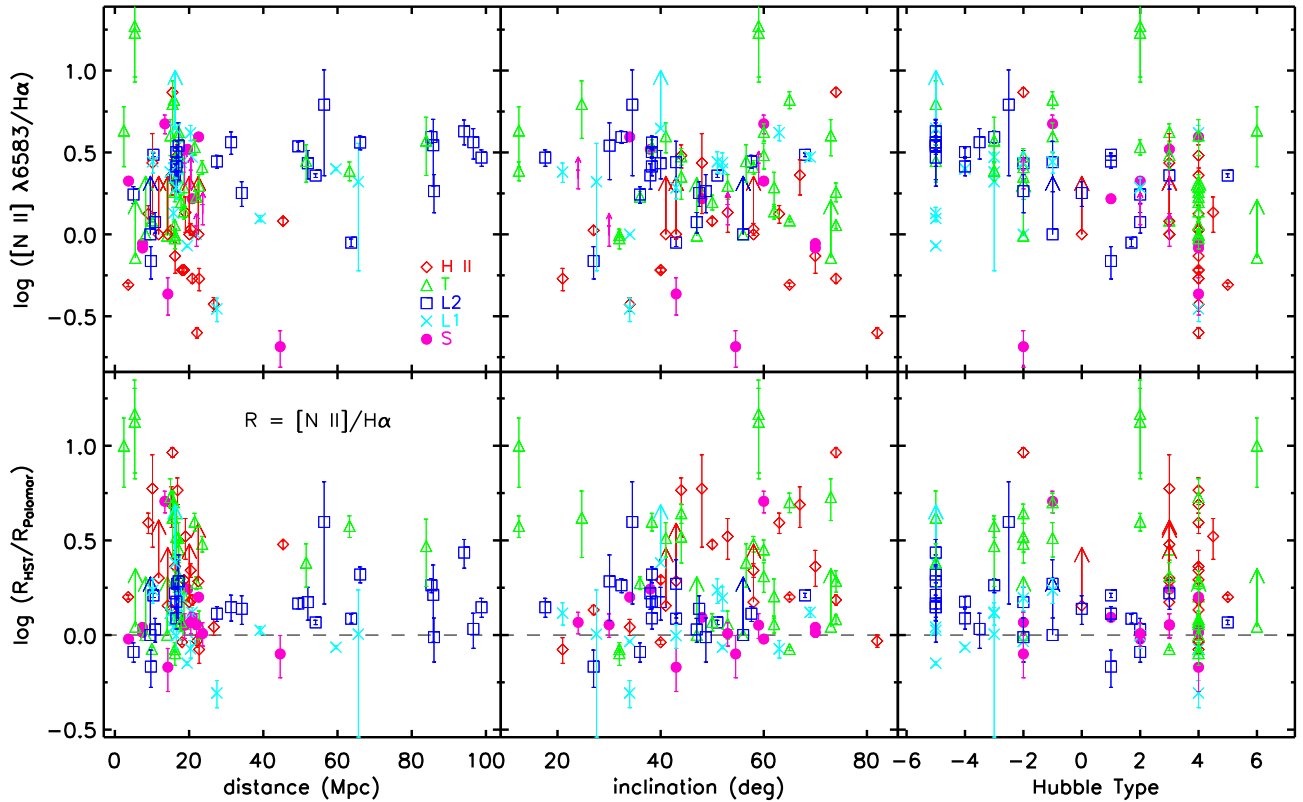}
\caption{The STIS  [N~{\sc ii}] $\lambda$6583/H$\alpha$ line flux ratio, and its value relative to that observed from the ground ($R_{\rm HST}/R_{\rm ground}$),  plotted against the galaxy distance, the host-galaxy inclination, and the morphological (Hubble) type.  The data points are symbol (and color) coded according to the ground-based classification of their nuclear emission as follows: H~{\sc ii} systems (red diamonds), transition nuclei (T, green triangles), narrow-line LINERs (L2, blue squares), broad-line LINERs (L1, cyan crosses), and Seyferts (S, magenta filled dots).   There is no apparent trend in any of these parameters, both for the entire sample of galaxy nuclei and for subsets of different spectral types. 
\label{hadistance}}
\end{figure*}

Because of the fixed slit size, the metric aperture depends on the
distance to the source, and thus the measured flux ratios reflect
nebular properties of different spatial extents within the galaxy
centers.  Such an effect may influence the comparisons of flux ratios;
for example, \citet{sto91} reported an inverse correlation between
[\ion{N}{2}]/H$\alpha$ and distance for a sample of (distant and relatively bright) Seyfert
2s and LINERs that can be attributed to inclusion of increasing
amounts of circumnuclear H~{\sc ii} region emission in more distant
sources.  We investigate the potential influence of this effect in our
sample of nearby nuclei in Figure~\ref{hadistance}, where both this line
flux ratio as measured with \emph{HST}, and the [\ion{N}{2}]/H$\alpha$ ratio measured in the small and large (ground-based) apertures, are shown as a function of galaxy
distance.  The different spectral types, as defined by ground-based
observations, are illustrated by different symbol types (Seyferts
include both Type 1 and Type 2).  The H$\alpha$ flux is not corrected for
the absorption from starlight in the host galaxies (see Section~\ref{lines}).   
The distribution in distances, and consequently in
the metric scales that the \emph{HST} nuclear observations encompass, is
heavily concentrated around $\sim 16$ Mpc\footnote{This is most likely because the galaxies are  members of the Virgo cluster; it translates into an equivalent aperture radius of $\sim 9$ pc, for an area of $0\farcs2
\times 0\farcs25$}, with outliers extending to 96 Mpc (aperture radius of $\sim 55$ pc).  The LINER 2 nuclei
are the only class that shows a large dispersion in distance (6--96
Mpc).  No obvious trend in line ratio with source distance is present
in either the full sample or the subgroups of specific 
spectral types, suggesting that distance-dependent aperture effects
are not a major concern.

For assessing orientation-biased selection effects, we also illustrate in 
Figure~\ref{hadistance} the dependence on the
inclination angle ($i$) of the host galaxies of both the \emph{HST} [\ion{N}{2}]/H$\alpha$
flux ratio and the ratio of this line ratio measured in the small and large apertures.   
We have calculated $i$ based on the isophotal axial ratio $R_{\rm 25} = {\rm log}(a/b)$, where $a$ and $b$ are the major and minor axis diameters measured at a $B$ surface brightness level of 25 mag arcsec$^{-2}$, as in \citet{ho97a}.    Visual inspections and two-sample Kolmogorov-Smirnov (KS) tests indicate that, with the
exception of T and L2 nuclei, all other types of objects present uniform
distributions in their host inclinations.  As found in other studies (e.g., Ho et
al. 1997), galaxies with T nuclei might be marginally more inclined than those
with L2 nuclei.  Similarly
to the above discussion regarding the possible trends of the
[\ion{N}{2}]/H$\alpha$ flux ratio with distance, the more edge-on systems would
be expected to include more circumnuclear H~{\sc ii} emission in the
aperture as well, and consequently to reveal smaller [\ion{N}{2}]/H$\alpha$ ratios.
However, within the whole sample and the individual subsets of different
spectral types, there is no obvious inverse correlation
between the [\ion{N}{2}]/H$\alpha$ ratio and the galaxy inclination.
There is also no statistically supported evidence that
[\ion{N}{2}]/H$\alpha$ ratios are smaller in T than in L2 nuclei.  This
analysis suggests that inclination biases are not significant in
our sample. 

Because for elliptical galaxies the axis ratio does not indicate an inclination angle, we also show in Figure~\ref{hadistance} the dependence on the host morphological type of the \emph{HST} [\ion{N}{2}]/H$\alpha$ flux ratio as well as the ratio of this line ratio measured in the two apertures.   We use here the morphological classification given by the mean numerical Hubble Type index as given in the Third Reference Catalogue of Bright Galaxies (RC3; de Vaucouleurs et al. 1991).    It is apparent that the sample contains virtually every morphological type, and hence provides a good representation of the general galaxy population. 
Ellipticals compose 16\% of the sample, while lenticulars and spiral systems make up 25\% and 60\%, respectively.  Barred galaxies (AB and B) contribute 24\% to the disk systems (S0--Sm) and 34\% to the spiral galaxies.
Once again, we find no apparent influence on the distributions of the line ratios as a function of the host morphology, both for the whole sample of galaxy nuclei and for the subsets of different spectral types.

\subsection{$L$(H$\alpha$) Distribution} \label{halumin}

\begin{figure}
\epsscale{1.2}
\plotone{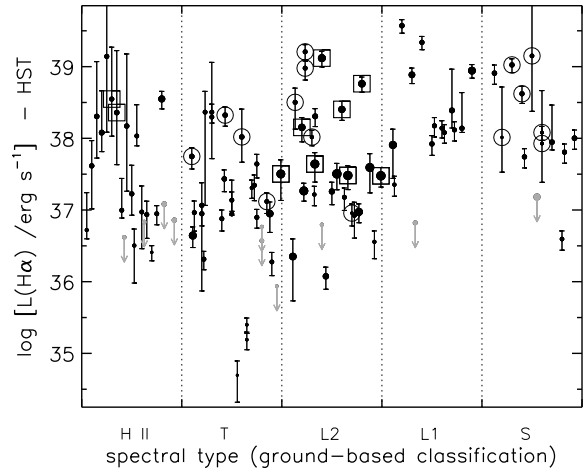}
\caption{The luminosity of the narrow H$\alpha$ line emission, measured in the \emph{HST}  aperture, as a function of ground-based nuclear spectral class.  
The objects for which STIS observations suggest the presence of broad H$\alpha$ features are indicated by large circles, while the large squares indicate the T, L2, and H~{\sc ii} nuclei with narrow lines that exhibit complex velocity structures and need multiple Gaussian profile fitting to assess the presence of a broad component.  The size of each data point is scaled by the distance to the host galaxy. 
Sources maintain the same horizontal locations as those in Figure~\ref{ew}.   
\label{lumin}}
\end{figure}

Of great interest is the degree to which the AGN phenomenon extends to
luminosities fainter than those of the well-established accretion-like
sources (Seyferts and LINER 1s).  Figure~\ref{lumin} illustrates
the range in narrow H$\alpha$ luminosity spanned by these nearby
nuclei in the \emph{HST} aperture, separately for each spectral category.
On average, L1 galaxies are the most luminous among all objects; the more luminous L2 nuclei share a similar distance range with that of the L1s, and are not among the most distant galaxies in our sample;  the ambiguous T objects are generally less
luminous than the S and L1 nuclei, and cover the widest range in the observed power;  
and the H~{\sc ii} galaxies span a similar range in $L$(H$\alpha$) as the Seyferts.

The broad-lined objects are of special interest, particularly in terms
of potential statistical trends toward finding or missing them in regard to their total emitting power.  We thus indicate the nuclei with broad H$\alpha$ components in the STIS spectra (as described in Section~\ref{broad}) by large open circles; we also mark with large open squares
the sources for which the narrow emission lines seem to have contributions
from multiple kinematically distinct regions (i.e., multiple Gaussians
are needed to fit their line profiles) that are interpreted in some
instances by ~\citet{ver97} and ~\citet{gon99} as evidence for
composite S/L + H~{\sc ii} emission.   We indicate these types of objects among the Seyferts as well.  
With only two exceptions, the nuclei with {\bf new} evidence of broad H$\alpha$ are 
among the most luminous ones; the same is true for the objects marked
by squares.  This pattern may reflect selection effects related to
sensitivity; in the more luminous sources, the contrast between the
nuclear and the surrounding emission is higher, thus implying a higher
likelihood to detect a broad component.

\subsection{Is the Nuclear Emission Resolved? } \label{ha}

An important issue in this comparative study is the degree to which
the high spatial resolution \emph{HST} observations are able to resolve the
nuclear emission in these nuclei.  Spectra obtained through a smaller
aperture should be characterized by a reduction in the nebular flux.
In particular, for extended sources of uniform surface brightness, the
decrease in flux is expected to match the simple ratio of the aperture
areas ($\sim -2.2$ dex); smaller changes in the emission-line fluxes would
indicate some degree of central concentration.

\begin{figure}
\epsscale{1.2}
\plotone{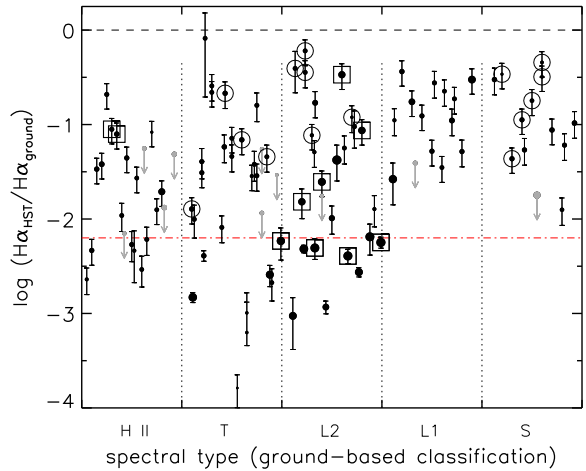}
\caption{Ratios of the narrow H$\alpha$ line emission in \emph{HST} and ground-based (Palomar and MMT) apertures, as a function of ground-based nuclear spectral class.  The size of each data point is scaled by the distance to the host galaxy.  The dashed line corresponds to no change in the H$\alpha$ flux among the two apertures, while the (red) dot-dashed line at $-2.2$ dex corresponds to what the ratio should be for a uniform nebular surface brightness.  
Sources maintain the same horizontal locations as those in Figure~\ref{ew}.   
\label{halpha}}
\end{figure}

Figure~\ref{halpha} presents the ratios of H$\alpha$ flux, in the
narrow component only, measured in the \emph{HST} spectra relative to that
given by the Palomar and MMT observations.  The results are again sorted according to the
ground-based nuclear spectral classification as defined by \citet{ho97a}.  For
almost all sample objects, the H$\alpha$ flux is smaller by at least
0.5 dex in the high-resolution \emph{HST} aperture, which means that, even
for the most distant objects in the sample, the nuclear emission is
resolved.  The symbol size scales again with the galaxy distance, and it can be seen that the ratios do not show any trend with the sampled metric aperture.    

It is readily apparent that the smallest change in the flux ratio is
found in the nuclei that are unambiguously powered by accretion, the
Seyfert and the LINER 1 objects, indicating as expected the highest
degree of central concentration of emission.  The H~{\sc ii},
T, and L2 nuclei show lower values of the flux ratio,
suggesting a shallower gradient in the surface brightness
distribution, and thus a more distributed ionization source.  The behavior of the
T and L2 sources is very similar, arguing that the T nuclei 
are probably {\it not} more prone to contamination by emission from extranuclear
star-forming regions than L2 nuclei, contrary to what has been
suggested by \citet{ho03} based on indications that T nuclei might reside in
more highly inclined host galaxies than those hosting LINERs.  As noted in
Section~\ref{selection}, our sample properties are not strongly influenced
by projection effects, if present; hence, such biases should not
play a major role in distinguishing between these two types of
emission-line nuclei.

Some trends are particularly noteworthy. For 5 out of 21 H~{\sc ii} nuclei, the
\emph{HST} aperture spectra show H$\alpha$ emission at levels that are even
lower than those predicted by scaling down the ground-based value by the
aperture ratio.  These cases show that ``nuclear''  star formation can
in reality span a significant range of radial scales, and may not
continue into the actual center of many galaxies; this is consistent with previous findings of,
for example, \citet{cidfernandes04} and \citet{gon04}, who find that the star formation powering the centers of H~{\sc ii} galaxies resides on scales larger than tens of parsecs.   The same trend is apparent among the most distant L2 and some T nuclei (not only the most distant), thus suggesting that the average surface brightness actually increases with radius in these objects.  This could be caused by, for example, obscuration of a source in the center, or extended nebulosity like diffuse ionized gas (DIG) or warm ionized medium (WIM) that is actually brighter at larger radii.
This behavior can have interesting consequences on the results of the nebular aperture comparison we conduct here, and we address them in the next section.

\section{The nebular excitation and the central engines} \label{nebular}

\begin{figure*}
\epsscale{1.15}
\plottwo{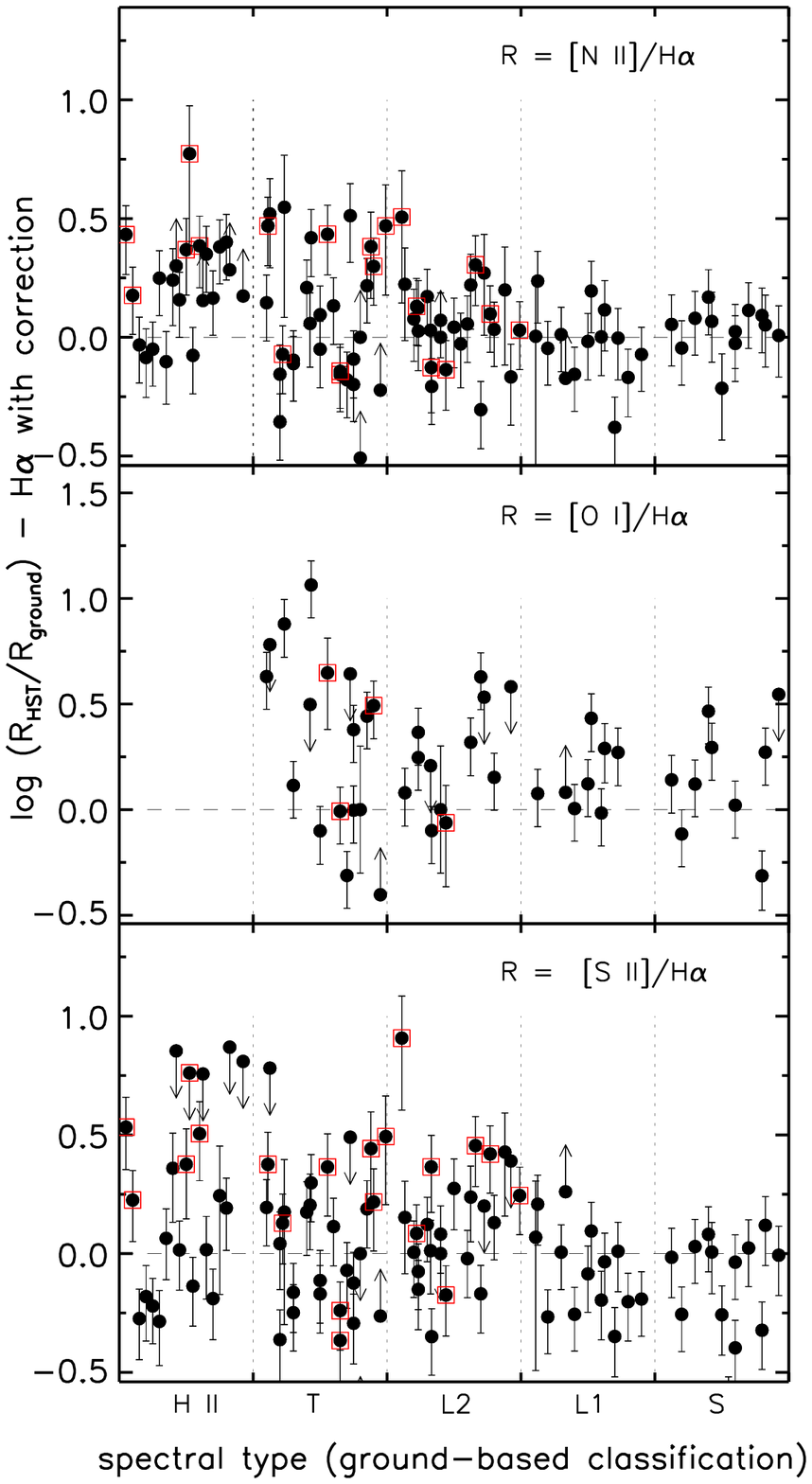}{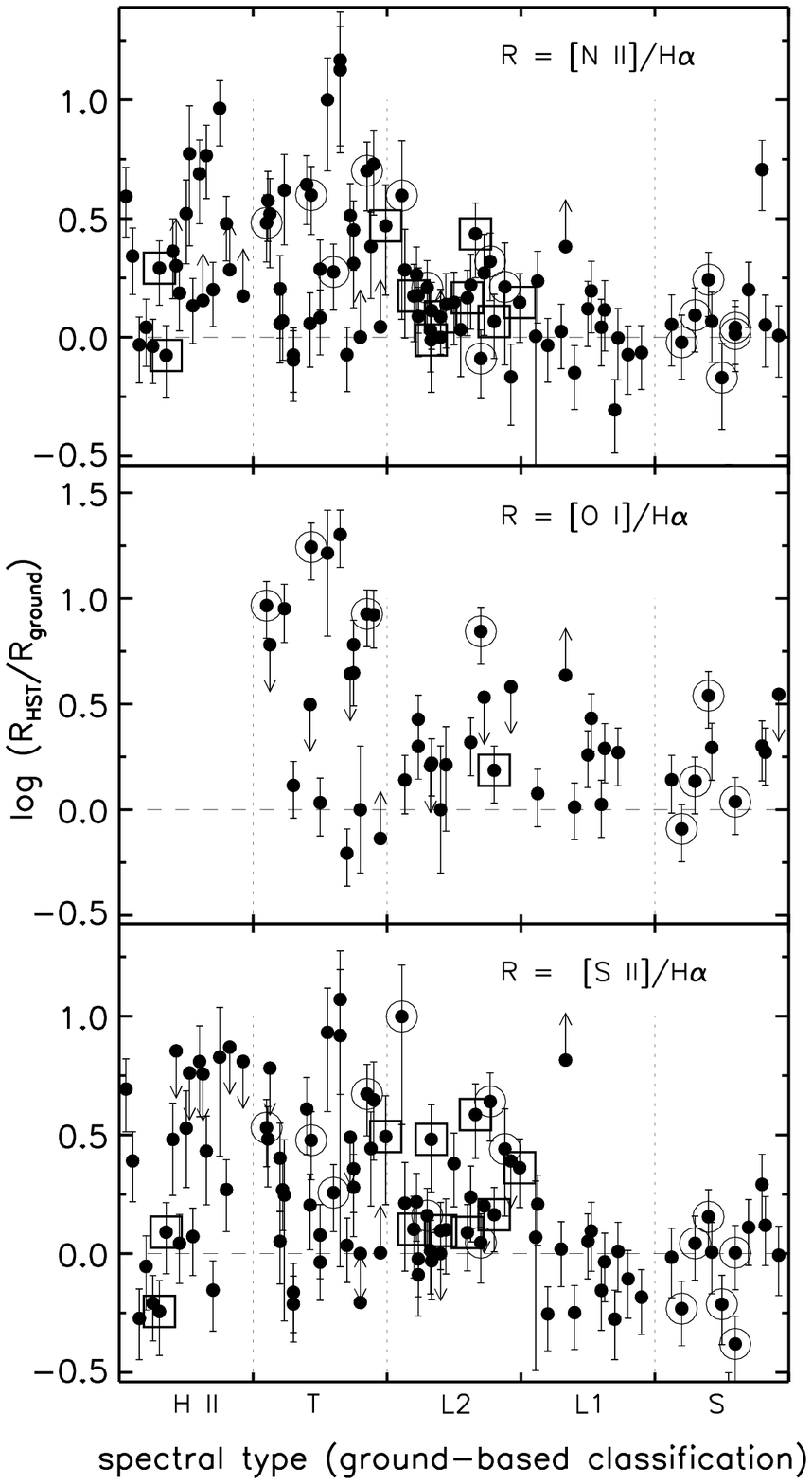}
\caption{Ratios of \emph{HST} to ground-based measured line diagnostic flux ratios, $R_{\rm HST}/R_{\rm ground}$, as a function of (ground-based) spectral class.    The {\it left} and {\it right} panels reflect measurements of the H$\alpha$ fluxes with and without correction for stellar absorption, respectively (see Section~\ref{lines}).   As with Figure~\ref{lumin}, we indicate in the right panels the systems with STIS detections of a broad H$\alpha$ line (large open circles), as well as those whose spectra are fitted with multiple Gaussian profiles to account for broad wings in the emission lines (large open squares).
The red open squares in the left panels mark the sources with H$\alpha$ flux ratio $< -2.2$ dex, with the line corresponding to the simple ratio of the aperture areas (Figure~\ref{halpha}).
Ratios of ratios for a given source maintain their horizontal location in all three panels, and are the same as those in Figure~\ref{ew}.
\label{rofratios}}
\end{figure*}

The central emission structure and its energetics can be further
tested by simple comparisons of the nebular-line flux ratios between
the \emph{HST} and ground-based spectra.   
The present data cannot directly probe gradients in the ionization process via the ionization parameter, which affects mostly the unavailable bluer regions of the optical spectrum, including the [\ion{O}{3}] emission. Nevertheless, various diagnostics are possible, as follows.

(1) Ratios of collisionally excited forbidden lines relative to a Balmer recombination line (e.g., [\ion{N}{2}]/H$\alpha$) play a key role in the standard spectroscopic classification diagrams; higher values would change the spectral classification of a T nucleus into either a Seyfert or a LINER.   Moreover, an increase in these ratios toward the more central regions could reflect one or a combination of the following causes: a hardening of the ionizing spectrum, an increase in the density (if still below the critical value), an increase in the metallicity, and a decrease of contamination by star formation \citep[e.g., based on models calculated with MAPPINGS III codes for AGN ionization;][]{gro04a, yan12}. 

(2) The [\ion{S}{2}] $\lambda$6716/$\lambda$6731 ratio provides information on the range of the particle density ($n_e$) for the emitting clouds; smaller values of the [\ion{S}{2}] $\lambda$6716/$\lambda$6731 ratio toward the nucleus indicate an increase in $n_e$.  

(3)  Particularly strong tests of nonstellar ionization are supplied by [\ion{O}{1}] $\lambda\lambda$6300, 6364, as these features are very weak in normal H~{\sc ii} regions. An increase in this ratio toward the nucleus would translate into a higher likelihood that the excitation is provided by hard photons such as those of an AGN power-law continuum that penetrate much deeper into an optically thick cloud, creating an extensive partially ionized zone and hence strong low-ionization forbidden lines.

\begin{figure*}
\epsscale{1.04}
\plotone{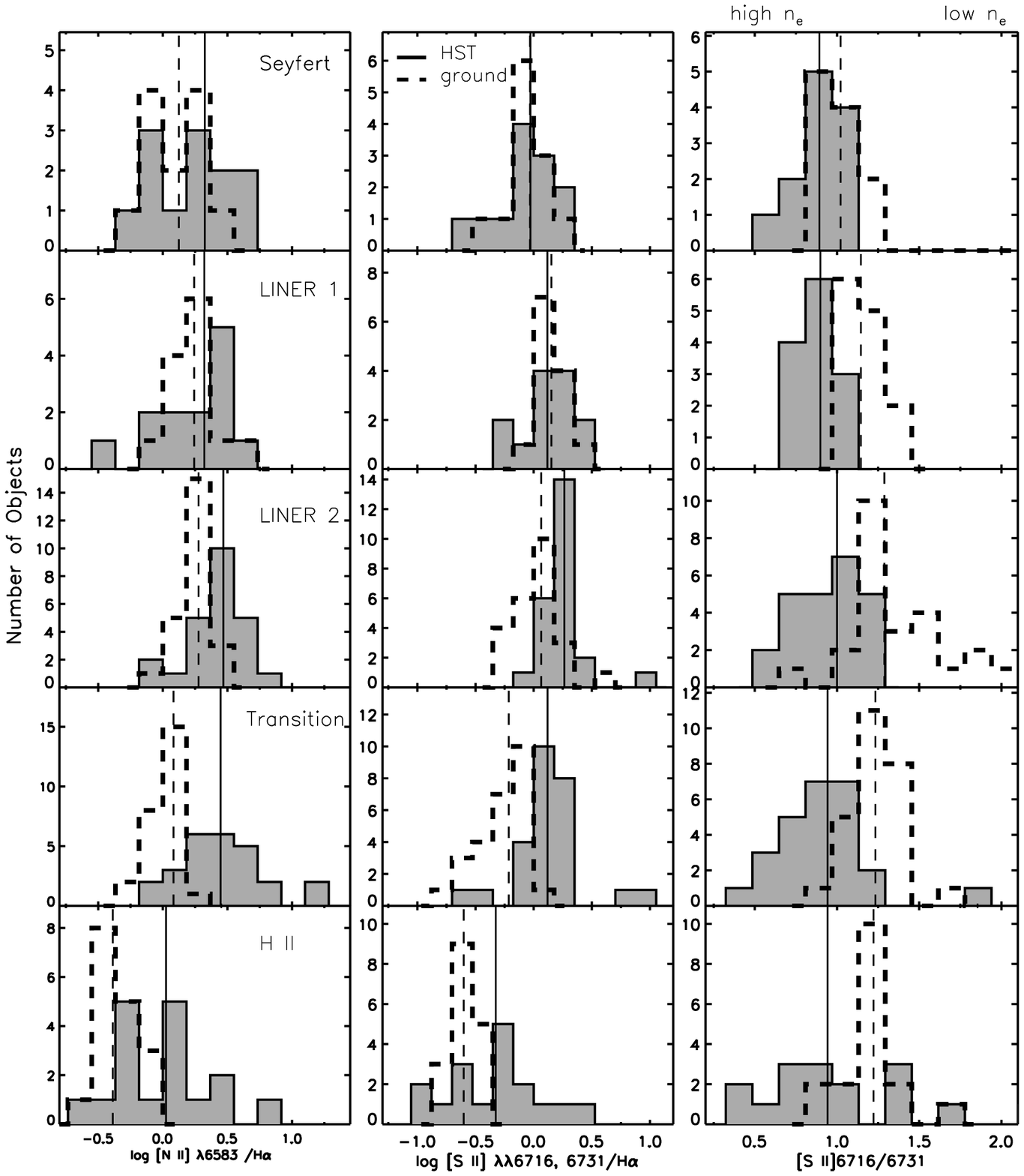}
\caption{ Distributions in [\ion{N}{2}]  $\lambda$6583/H$\alpha$ ({\it left}) and
  [\ion{S}{2}] $\lambda \lambda$6716, 6731/H$\alpha$ ({\it right}) as
  measured with \emph{HST} (continuous lines) and from the ground (dashed lines).
  Median values are indicated by vertical continuous and dashed lines,
  respectively.   The small-aperture measurements show a definite increase in the
  average values for L2, T, and H~{\sc ii} nuclei, and the shapes of the distributions are
  changed; the \emph{HST} measurements exhibit generally wider histograms,
  with more objects toward the high values of these line flux ratios.   When pairs of T, L1, L2, and S nuclei are compared, the KS probabilities are significantly higher for line ratios and fluxes measured with {\it HST} than from the ground, suggesting an increased difficulty in ruling out the ``null hypothesis" of a common parent population.
\label{histograms}}
\end{figure*}

\subsection{Line-Ratio Gradients} \label{linegradients}

Figure~\ref{rofratios} presents the ``ratios of ratios'' that compare $R =$ [\ion{N}{2}]/H$\alpha$,
[\ion{S}{2}] $\lambda\lambda$6716, 6731/H$\alpha$ and [\ion{O}{1}]
$\lambda 6300$/H$\alpha$ measured from the \emph{HST} data to
those obtained from the ground-based spectra.   For the sake of completeness, we show these ratios of ratios calculated with and without correction for stellar absorption (see Section~\ref{lines}) in the left and the right panels, respectively.   We indicate in the right panel the systems with \emph{HST} detections of a broad H$\alpha$ line (large open circles), as well as those whose spectra are fitted with multiple Gaussian profiles to account for broad wings in the emission lines (large open squares), as in Figure~\ref{lumin}.

There is a general pattern that can be seen in all of these plots: for
the majority of the L1 and S nuclei, these ratios show little scatter
around unity, while H~{\sc ii}, T, and L2 nuclei reveal a wide range of
values for these ratios.  The L1 and S nuclei exhibit no clear average change in [\ion{N}{2}]/H$\alpha$ between the two apertures, a systematic decrease in [\ion{S}{2}]/H$\alpha$ in the small aperture, and a systematic increase in the [\ion{O}{1}]/H$\alpha$ line ratio in the more nuclear spectra.  For the H~{\sc ii}, T, and L2 nuclei, the majority of sources have larger [\ion{N}{2}]/H$\alpha$, [\ion{S}{2}]/H$\alpha$, and possibly
[\ion{O}{1}]/H$\alpha$, in the small-aperture observations.  
As revealed in Figure~\ref{hadistance}, there is no obvious overall correlation with distance to the host galaxy, host inclination, or host morphology in either the [\ion{N}{2}]/H$\alpha$ or $R_{\rm HST}/R_{\rm ground}$ values that could skew this comparison, and we tested that this remains true for the other ratios.

Figure~\ref{histograms} illustrates the distributions in the [\ion{N}{2}]/H$\alpha$, 
[\ion{S}{2}]/H$\alpha$, and [\ion{S}{2}] $\lambda$6716/$\lambda$6731 line ratios measured in the small and large
apertures (where we excluded the upper and lower limits);  there is a clear increase in the median values of the forbidden to H$\alpha$ line ratios for H~{\sc ii}, T, and
L2 nuclei in the more central regions, with not much change between the small- and large-aperture 
measurements of S and L1 nuclei.    

The aperture effects on the ratio [\ion{S}{2}] $\lambda$6716/$\lambda$6731 that serves as a density indicator \citep{ost89} are
also  illustrated in Figure ~\ref{histograms}, separately for each nuclear spectral type.   
It is quite apparent that in the more central nuclear regions 
mapped by the STIS observations the gas densities are larger (i.e.,
smaller [\ion{S}{2}] $\lambda$6716/$\lambda$6731 flux ratios), {\it in all
types of objects}.   A few exceptions are among the H~{\sc ii} nuclei,
but there are generally large error bars associated with these measurements, so they are not necessarily
excluded from the general trend.  The T nuclei show a
particularly pronounced difference in the gas density measured in the
two apertures, with the largest fraction ($\sim 90\%$) of objects exhibiting greater than 50\% increase in their $n_e$.

It is noteworthy that the differences between the ground-based and \emph{HST}
ratios of forbidden to H$\alpha$ lines are generally small.   Although
as described in Section~\ref{ha}, the flux sampled in the \emph{HST}
apertures is typically a small fraction of that contributing to the
Palomar and MMT spectra, and the spatial scale of the observed regions differs
by at least an order of magnitude, there are very few sources that
might indicate a change in their classification from ambiguous
(narrow-lined) nuclei to a more secure AGN-like behavior -- that is, a rightward shift in the line-diagnostic diagrams.  The
relatively small fraction of objects with newly detected broad
Balmer lines is consistent with this result.  

To aid in discussing and understanding the possible effects of extended DIG/WIM nebulosity at larger radii, we also mark with red open squares (only in the left panels of Figure~\ref{rofratios}) the peculiar sources for which the H$\alpha$ flux ratio falls below the $-2.2$ dex line corresponding to the simple ratio of the aperture areas (Figure~\ref{halpha}). 
As expected, these particular sources are generally among those exhibiting an increase in the forbidden to H$\alpha$ ratios, owing to the drastic decrease in the Balmer line intensity; when the H$\alpha$ emission is not corrected for stellar absorption (right panels), all of these objects show an increase in the flux ratios.  Thus, especially for the more distant objects, the small-aperture data support the geometric stratification of the extended low-ionization emission regions, which was recognized as the dominant power source in a great majority of (early-type) galaxies spectrally classified as Type 2 LINERs \citep[i.e., the LIERs, with LINER-like emission which is {\it not} nuclear; e.g.,][]{sarzi11, yan12}
Nevertheless, these particular sources are not the only ones showing changes in the flux ratios, suggesting that  the changes in the flux ratios are not entirely caused by this effect.

The objects with newly detected broad H$\alpha$ emission exhibit an increase in the narrow-line ratios in spectra acquired at smaller scales.  Thus, at least for these objects, the scenario in which the small-aperture emission may be dominated to a greater degree by a central accretion source seems applicable.   
Nuclei whose ground-based spectra indicate a potential mix of velocity fields (i.e., distinct multiple Gaussian profiles required to fit the narrow emission) behave similarly in both apertures, as their spectra remain consistent with very little or no change in the nebular line ratios; they are also among the most distant targets, which might mean that, if the ``composite'' picture holds, their \emph{HST} nuclear observations could still include light from circumnuclear star-forming regions.

With the expectation of revealing an increasingly similar behavior of the T and L2 nuclei to that of bona-fide AGNs (L1 and S nuclei) in their more nuclear spectra, we statistically compared the distributions  in the line ratios as well as in the H$\alpha$ fluxes  and luminosities, for all possible galaxy-type--pair combinations.   For the ground-based measurements, the KS probabilities are generally low, revealing potentially common parent populations only for L1--L2 systems ($p_{\rm KS} \la 0.2$),  with basically no chance for commonality with T and S systems or between the latter two types ($p_{\rm KS} << 10^{-3}$ for all these pairs of spectral types).   
The KS probabilities are, however, boosted to $p_{\rm KS} \approx 0.3$--0.8 for the \emph{HST} line-ratio distributions for all possible combination pairs of Type 1 and 2 LINERs, T nuclei, and S nuclei, with higher values for the [\ion{S}{2}]/H$\alpha$ and the [\ion{S}{2}] $\lambda$6716/$\lambda$6731 ratios, especially for the S--T and L2--T pairs, suggesting an increased difficulty in ruling out the ``null hypothesis" of a common parent sample.

The trends we measure in the ratios of collisionally excited forbidden lines to H$\alpha$ are not particularly well mapped by current theoretical models for AGN ionization  \citep[e.g., by classical dust-free AGN photoionization codes, MAPPINGS III;][]{dop96, gro04a, allen08}. For example, they do not reproduce the high [\ion{S}{2}]/H$\alpha$ ratios we measure in the \emph{HST} observations, and thus remain powerless in addressing electron densities or gradients in $n_e$ as high as we find here. 
The models predict a tight clustering of the line-ratio values for relatively broad ranges of variation in the parameters that can cause gradients like those we measure in our sample: for example, an increase by 0.5 dex in the average [\ion{S}{2}]/H$\alpha$ or [\ion{N}{2}]/H$\alpha$ measured for T nuclei is consistent with either (i) an increase in $n_e$ by at least an order of magnitude ($10^3$ to $10^4$), (ii) a factor of $\sim 3$ increase in metallicity, (iii) a hardening of the continuum corresponding to a change of the spectral index $\alpha$ from $2$ to $-1.4$, or (iv) a combination of these changes.  While the degeneracy in identifying the principal cause for such gradients cannot be broken, our data  probe a clear increase in $n_e$, which proves important for understanding of the nuclear emission in these nuclei.  
We explore in more detail the potential insights from these results in the following subsections.

\subsection{The Effects of Gradients in Electron Density} \label{ne}

That the emission in the small aperture is dominated to a greater degree by a central
accretion-powered source is not, however, the only reason for the somewhat stronger forbidden lines relative to H$\alpha$ in the more nuclear spectra.  The enhancement in these ratios can
also be simply a direct consequence of the trend
toward larger electron densities at small radii.   As revealed by the patterns in the [\ion{S}{2}] doublet ratio, this is happening for virtually all nuclear types.  As long as the density remains below the critical value, $n_{\rm crit}$, the line emissivity increases $\propto n^2$.  Because large densities may lead to suppression of the infrared fine-structure
line cooling, a boost may result in the optical forbidden lines like
[\ion{N}{2}], [\ion{S}{2}], and [\ion{O}{1}].   Interestingly, the
[\ion{S}{2}]/H$\alpha$ ratios generally {\it decrease} in the small
aperture for L1 and S nuclei, suggesting that the densities in these
sources extend to values beyond the [\ion{S}{2}] critical densities.

Looking closer, it becomes clear that there is varied behavior for the different line ratios studied here, which can be understood as a consequence of the different critical densities of the forbidden lines.   For comparison, $n_{\rm crit}$ for  [\ion{S}{2}], [\ion{N}{2}], and [\ion{O}{1}]  are  (1.5--3.9) $\times 10^3$ cm$^{-3}$, $8.7 \times 10^4$ cm$^{-3}$, and $1.8 \times 10^6$ cm$^{-3}$, respectively, thus spanning several orders of magnitude.    
While $n_e$ generally increases in the {\it HST} aperture, it is possible that a significant fraction of objects reach $n_e$ ranges similar to or greater than the (relatively small) $n_{\rm crit}$ of [\ion{S}{2}] $\lambda\lambda$6716, 6731; this could act to collisionally suppress the [\ion{S}{2}]/H$\alpha$ ratio in these small apertures, thereby leading to the measured decrease in this ratio relative to the one in ground-based spectra.

In this scenario, only a few objects would reach $n_e$ values higher than $n_{\rm crit}$ of the [\ion{N}{2}] $\lambda\lambda$6548, 6583, with even fewer reaching the $\sim 10^6$ cm$^{-3}$ values of $n_{\rm crit}$ for [\ion{O}{1}] $\lambda\lambda$6300, 6364, leading to increasingly less suppression in these features; thus, there is an increasing fraction of objects with ratio of ratios $R_{\rm HST}/R_{\rm ground} > 1$ for [\ion{N}{2}] and [\ion{O}{1}], respectively, which is what we find in our comparison.   This effect has been seen directly in previous work illustrating measurements of the line flux ratios as a function of position along the slit; for example, \citet{bar01} and ~\citet{walsh08} revealed for NGC 4579 and additional objects a strong trend where the [\ion{S}{2}] ratio rises toward the high-density limit closer to the nucleus, while the [\ion{O}{1}]/H$\alpha$ and [\ion{O}{1}]/[\ion{S}{2}] ratios spike upward in that innermost region.

There is yet another physical explanation for the observed small change in the line ratios with distance from the center: the line ratios would be left unchanged if both the density of ionizing photons and of electrons diminish with radius.   A physical basis for such a balance has been described by \citet{gro04a, gro04b}, who argue that the response of gas pressure to radiation pressure acting on dust in the emitting clouds will produce a local excitation source that becomes independent of the external flux, and hence distance from a central radiation source.  The results of this study for L1 nuclei and Seyferts are at least qualitatively in accord with their model.

Within this picture, it is then interesting to investigate to what
degree this effect might contribute to the lack of change in the T
and L2 nuclei as well, given that these sources' emission is, by definition, dominated by low-ionization regions.  
The T and L2 systems with new detections of broad components are not among those with the largest gradient in the gas densities.  Thus, for these objects in particular, the potential counteracting changes in $n_e$ are
probably not very strong, and a significant aperture dependence is
seen.  On the other hand, the rest of the T nuclei, with the greatest
changes in $n_e$ among the entire sample, would be expected to show
weak gradients, and this is in accord with the observations.  This pattern does
not seem to apply for L2 nuclei as well;  the gradients in $n_e$ are large among L2 galaxies, but the values of $n_e$ in the \emph{HST} apertures remain systematically lower than in the T, S, and L1 nuclei.  Thus, gradients in $n_e$
could be, at least in part, the cause for the similar line ratios in the small and large apertures.

We note that because the measured density is weighted by luminosity, the difference in density gradients between the different types of objects may not reflect a difference in the physical density gradients, but could be due to the difference in their luminosity profiles.  As evidenced by the smaller difference between the two apertures, the Seyferts have a sharper luminosity profile; thus, the densities measured in both large and small apertures are more heavily weighted toward the central density. The T nuclei have a shallower luminosity profile, which might translate into a larger density difference.   Nevertheless, this point does not invalidate any of the results presented here.

\subsection{L2 vs. L1 Nuclei} \label{l1l2}

That Type 1 and 2 L nuclei do not show significantly greater similarity in their nuclear nebular emission brings
back the question of whether these two object types are fundamentally
the same and just appear different from various angles owing
to obscuration, or have real intrinsic differences --- for example, the BLR could be completely missing in the Type 2 systems, perhaps as a consequence of the evolution of the AGN environment as a function of the accretion rate.  

To summarize our findings, the L1 and L2 nuclei span different ranges in narrow H$\alpha$ luminosities and
fluxes as measured with STIS (see Section ~\ref{gen}).  
The forbidden to Balmer line-ratio distributions of these nuclei do not exhibit an increase in the KS likelihood of being drawn from the same parent populations when measured in the more nuclear STIS spectra.  
The values of their ratio of ratios in [\ion{N}{2}]/H$\alpha$ and [\ion{S}{2}]/H$\alpha$ seem to follow different trends: they are almost scattered in opposite directions around the unit value, with L1 nuclei showing lower [\ion{S}{2}]/H$\alpha$ values in the \emph{HST} spectra relative to those of Palomar and MMT (Figure~\ref{rofratios}). 
Also, the average $n_e$ values and general distributions remain dissimilar.   
While the largest gradients in $n_e$ among all LINERs are registered in L2 nuclei, L1s exhibit higher gas densities than L2s in both ground-based and space-based observations: median $n_e$(L1) $\approx
8.5 \times 10^3$ cm$^{-3}$, $n_e$(L2) $\approx 5 \times 10^3$
cm$^{-3}$ in the \emph{HST} spectra, and $n_e$(L1) $\approx 1.5 \times 10^3$
cm$^{-3}$, $n_e$(L2) $\approx 0.6 \times 10^3$ cm$^{-3}$ in the ground-based
data.   

A possible impediment to understanding these differences can be that the L2 nuclei employed in this study cover
a wider distribution in galaxy distances than the L1, T, or S nuclei, which are more concentrated near 20 Mpc.  
In an attempt to match the two samples of LINERs in both mean/median and distribution of distances, we ran the KS comparison tests on the line ratios using L2 subsamples that exclude some or all of the five most distant sources (see Figure~\ref{hadistance}).  
In all cases, the subsamples of L2 nuclei manifest the same behavior as the full L2 sample, which means that differences in metric aperture do not contribute significantly to the differences in the nebular emission behavior between L1 and L2 nuclei.

\begin{figure}
\plotone{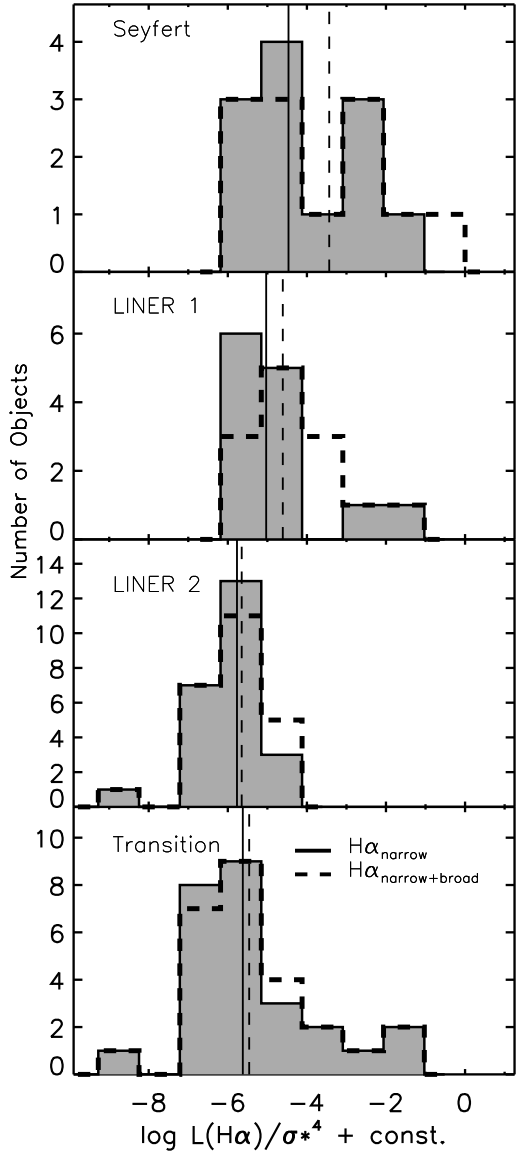}
\caption{Distributions and average values of $L_{{\rm H}\alpha}/\sigma^4$ for S, L1, L2, and T nuclei.    $L_{{\rm H}\alpha}$ is the H$\alpha$ luminosity measured in the  \emph{HST} apertures, and is calculated both with and without contributions from the broad component; the corresponding distributions are shown as dashed and continuous lines, respectively.    The distribution means are drawn as vertical lines.
\label{hist_ledd}}
\end{figure}

The differences we found between the L1 and L2 sources appear to support both an obscured BLR and the missing BLR interpretation for the L2 sources, with both of these phenomena being linked to the low luminosities and accretion levels in the latter systems.  

On the one hand, our findings of lower average $n_e$ in the more nuclear regions of L2s than in L1s, that could explain their positive gradients in [\ion{S}{2}]/H$\alpha$ ratios which are in contrast to the negative gradients in L1s, offer support to the unified scheme for orientation-dependent obscuration in LINERs: an obscuring medium blocks the innermost (and densest) part of the narrow-line region in addition to the BLR \citep[e.g.,][]{bar99}.
Greater obscuration is expected for lower-luminosity accretion sources \citep[i.e., the receding torus model;][]{lawrence91, simpson03, con04} which, again, is consistent with the lower H$\alpha$ luminosities for the L2 nuclei than for the Type 1 LINERs. 
Nevertheless, as discussed in detail by \citet{ho08}, there remains very little observational evidence for any obscuration or absorption in these particular systems; thus, the idea that L2 nuclei are more obscured that the L1s remains speculative. 

On the other hand, if differences between the two types of LINERs are linked to the disappearance of the BLR in the Type 2 systems, then we should also see differences in their nuclear activity; for example, this phenomenon is expected at very low luminosity levels \citep{nic00, nic03, lao03}, and extremely low accretion rates or efficiency in accretion  \citep{elitzur09, elitzur14}.    To test this idea, we have compared $L_{{\rm H}\alpha}/\sigma^4$ for objects of different spectral types. We employ $L_{{\rm H}\alpha}/\sigma^4$ as a proxy for the Eddington ratio $L/L_{\rm Edd}$, with $L_{{\rm H}\alpha}$ being the H$\alpha$ line luminosity measured in the \emph{HST} spectra, which we use as a gauge of the accretion luminosity, and $\sigma^4$ as a measure of the black hole mass, via the black hole mass vs. stellar velocity dispersion ($M_{\rm bh} - \sigma$) relation that has been well measured for both inactive galaxies \citep[e.g.,][]{gebhardt00, ferrarese00} and active systems \citep[e.g.][]{bennert14}. We use the $\sigma$ values from \citet{ho09b} for the Palomar objects and from \citet{beifiori12} for the MMT galaxies.    
Figure ~\ref{hist_ledd} shows the distributions and average values of $L_{{\rm H}\alpha}/\sigma^4$ for S, L1, L2, and T nuclei.   
The distributions, average, or median values of the L1 and L2 nuclei are indeed different, with low KS probabilities ($p < 0.1$) for being drawn from the same population;  the L1 systems are at least $0.7$ dex stronger in the median $L_{{\rm H}\alpha}/\sigma^4$ values, regardless of whether $L_{{\rm H}\alpha}$ values include or exclude the broad H$\alpha$ emission.   
This comparison argues, then, for ``BLR nakedness'' in the L2 systems.  For these galaxies, the lack of detectable dense gas at small radii could simply extend beyond the BLR to encompass the inner NLR.

One could note that the flatter gradients in the line ratios of the L2 nuclei have another possible explanation that may have nothing to do with accretion power: if they are also dominated by some kind of stellar photoionization on the scales probed by \emph{HST}, they will display flat gradients in line ratios.   Nevertheless, the line gradients for this class of objects are only weak in  [\ion{N}{2}]/H$\alpha$, while they are quite significant in [\ion{S}{2}]/H$\alpha$ and especially in [\ion{O}{1}]/H$\alpha$,  arguing against dominant stellar photoionization at the parsec scales probed by the STIS observations, for at least half of the sample. 

Thus, the parallel between L and S nuclei of Types 1 and 2 extends nicely to the smallest physical scales probed to date, from the morphology of the narrow-line nebular emission of both L2 and Seyfert 2 nuclei \citep{pog89} appearing less concentrated than their Type 1 counterparts (Figure~\ref{halpha}), to a potentially strong relation between the strength of accretion and the existence (or disappearance) of a broad-line component.  Although we do not argue for any particular physical model here, it is possible that the trends we see in the line-ratio gradients, the gradients in $n_e$, and the differences in $L_{{\rm H}\alpha}/\sigma^4$ for the L1 and L2 nuclei mirror the differences between the Type 1 and 2 Seyferts or quasars, which are likely caused by a systematic change in the shape of the accretion-disk continuum and its interplay with the ambient line-emitting regions \citep[e.g.,][]{shen14}.

\subsection{T Nuclei and Traditional AGNs}

The prevailing paradigm of what T nuclei are involves a weak central AGN with compact circumnuclear star formation that all fall in the aperture of a typical ground-based measurement (e.g., Kennicutt et al. 1989).  The composite model \citep[e.g.,][]{ho93} predicts that the small aperture should help isolate the accretion-powered component such that the small-aperture spectrum is more AGN-like.
The results of the aperture comparisons presented here show unprecedented evidence for the fact that T objects behave more like the accretion-type sources in their more nuclear regions.     We have found for T nuclei (i) a three-fold increase in the incidence of broad-lined nuclei, (ii) the largest of the predicted gradients in the line ratios (Figure~\ref{rofratios}), and (iii) a shift in their overall line-ratio distributions that supports a closer similarity of their nebular properties with those of the S, L1, and L2 nuclei in the small aperture. The KS tests suggest a less likely probability of these systems not sharing a common parent population.    All of these trends are consistent with the composite picture for T nuclei.

\begin{figure}
\plotone{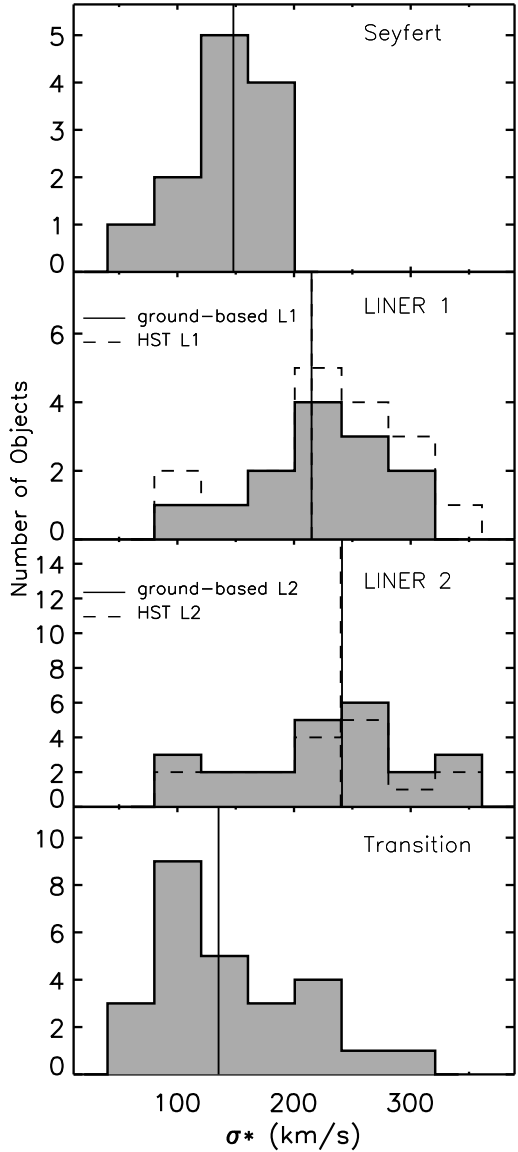}
\caption{Distributions and average values of $\sigma*$ for S, L1, L2, and T nuclei.    For L1 and L2 systems we consider both the ground-based and the  \emph{HST} spectral classification (i.e., the new detection of the broad H$\alpha$ feature, and thus the reclassification as Type 1); the corresponding distributions are shown in continuous and dashed lines, respectively.    The distribution means are drawn as vertical lines.
\label{hist_sigma}}
\end{figure}

The central AGN-like source in the proposed models has been generally considered to be a LINER; however, previous studies did not have the sensitivity or the statistical power to investigate in detail the nature of the central object in the composite model.   This work offers new quantitative insights into its nature.
We found an increase in similarity between the properties of T and L2 nuclei in the $\sim 10$ pc scale regions; there are comparable distributions of fluxes, line ratios and their gradients, and possibly accretion rates as well, as probed by the $L_{{\rm H}\alpha}/\sigma^4$ parameter (Figure ~\ref{hist_ledd}). 
Nevertheless, the transition to a more AGN-like behavior in the more nuclear regions is different in the T and L2 galaxies: the T nuclei 
show the strongest line-ratio gradients, thus exemplifying most positively the expected results of the aperture comparison test proposed here.   
Interestingly, when we compared the stellar velocity dispersions ($\sigma*$), as a measure of $M_{\rm bh}$, for our samples of S, L1, L2, and T galaxies, we found that  T and S nuclei show consistent (low) average values for $\sigma*$, and thus for $M_{\rm bh}$, while the L1 and L2 galaxies exhibit increasingly larger ones (Figure ~\ref{hist_sigma}).   These characteristics suggest greater likeness in the T nuclear environments with Seyferts than with L2 systems.  

A possible account for these trends is provided by the idea that T nuclei are the intermediate phase in the evolution from a star-formation-dominated nucleus (i.e., an H~{\sc ii} galaxy) toward an unveiled compact accretion source (i.e., a Seyfert), where the accretion rate rises toward the values seen in S galaxies, while the L2 systems are the manifestation of the post-Seyfert  \citep[and possibly post-L1 phase; e.g.,][]{elitzur14}, where the accretion wanes down to eventual quiescence \citep{con08, con09, sch07, sch10}.   This scenario supports (i) the closeness of the T systems with the H~{\sc ii} systems in the ground-based observations, as traces of surrounding stellar activity should be more pronounced than in L2s, while (ii) exhibiting increasing AGN-like nature in their most nuclear regions, as evidenced via both new detections of broad H$\alpha$ lines and strong gradients in the line ratios.

If T nuclei are an intermediate phase in the evolution from a star-formation-dominated nucleus toward a dominance of Seyfert-like excitation, it is thus possible that our observations catch these systems in various evolutionary stages.   This would naturally explain the wide range in their nuclear properties, and in particular, the span of the line ratios as well as the small geometrical segregation between H~{\sc ii} and AGN-like emission when more than 90\% of the nebulosity is resolved as the aperture is reduced by a factor of $\sim 100$ in area and the measured flux drops by an order of magnitude or more (i.e., Figures~\ref{ew},~\ref{halpha}, and~\ref{rofratios}).  

One could argue that if a galaxy has a cold gas supply, its nucleus could go through H~{\sc ii}, LINER, and Seyfert phases episodically, in whatever order.   Correlations between these spectral types and the stellar population of the host galaxy could then simply be a reflection of differences in gas supply and thus different AGN duty cycles in various galaxy types.   Nevertheless, there is a systematic trend in the masses of the black holes in these galaxies, suggesting that the black hole grows along a T $\rightarrow$ S $\rightarrow$ L sequence, which is readily apparent even within our small, inhomogeneous, and statistically incomplete sample 
(Figure ~\ref{hist_sigma})\footnote{Note also that this trend remains consistent with the proposed Type 1 $\rightarrow$ Type 1.2/1.9 $\rightarrow$ Type 2 evolutionary sequence advanced by \citet{elitzur14}.}.   Markedly, the time scales necessary to transform from one galaxy type to the next are not unreasonable: considering a radiative efficiency value $\epsilon \approx 10^{-6}$ \citep[e.g.,][]{ree82, nar94, ho08}, the $e$-folding time in the BH mass is $t_e \approx 4.5 \times 10^8 \frac{\epsilon}{\lambda(1 - \epsilon)} \approx 4.5 \times 10^2/\lambda$;  with $\lambda = L/L_{\rm Edd} = 5 \times 10^{-4}$, a growth in mass from $2 \times 10^7~{\rm M}_{\sun}$ (for
H{\sc~ii}s) to $2 \times 10^8~{\rm M}_{\sun}$ (for LINERs) is a few Myr.  Thus, such an evolutionary scenario is physically possible.   

Further constraints on this evolutionary scenario could come from investigations of the amount and properties of stellar light in the very central regions, as probed by the \emph{HST} apertures, in comparison with those measured at $\sim 100$ pc scales.    
Unfortunately, high-quality spectroscopy at high (i.e., \emph{HST}) spatial resolution  for a sample that is as large and homogeneously distributed in spectral types as the one discussed in this study is not yet available for a rigorous investigation of these issues.   The \emph{HST} data presented here have low S/N that does not allow accurate fits of the underlying starlight and thus secure identification of their nuclear stellar properties. 
Also, the lack of [\ion{O}{3}]/H$\beta$ measurements at high spatial resolution precludes a full account for a possible reclassification of the T into S nuclei in the small aperture.

\begin{figure*}
\epsscale{1.15}
\plottwo{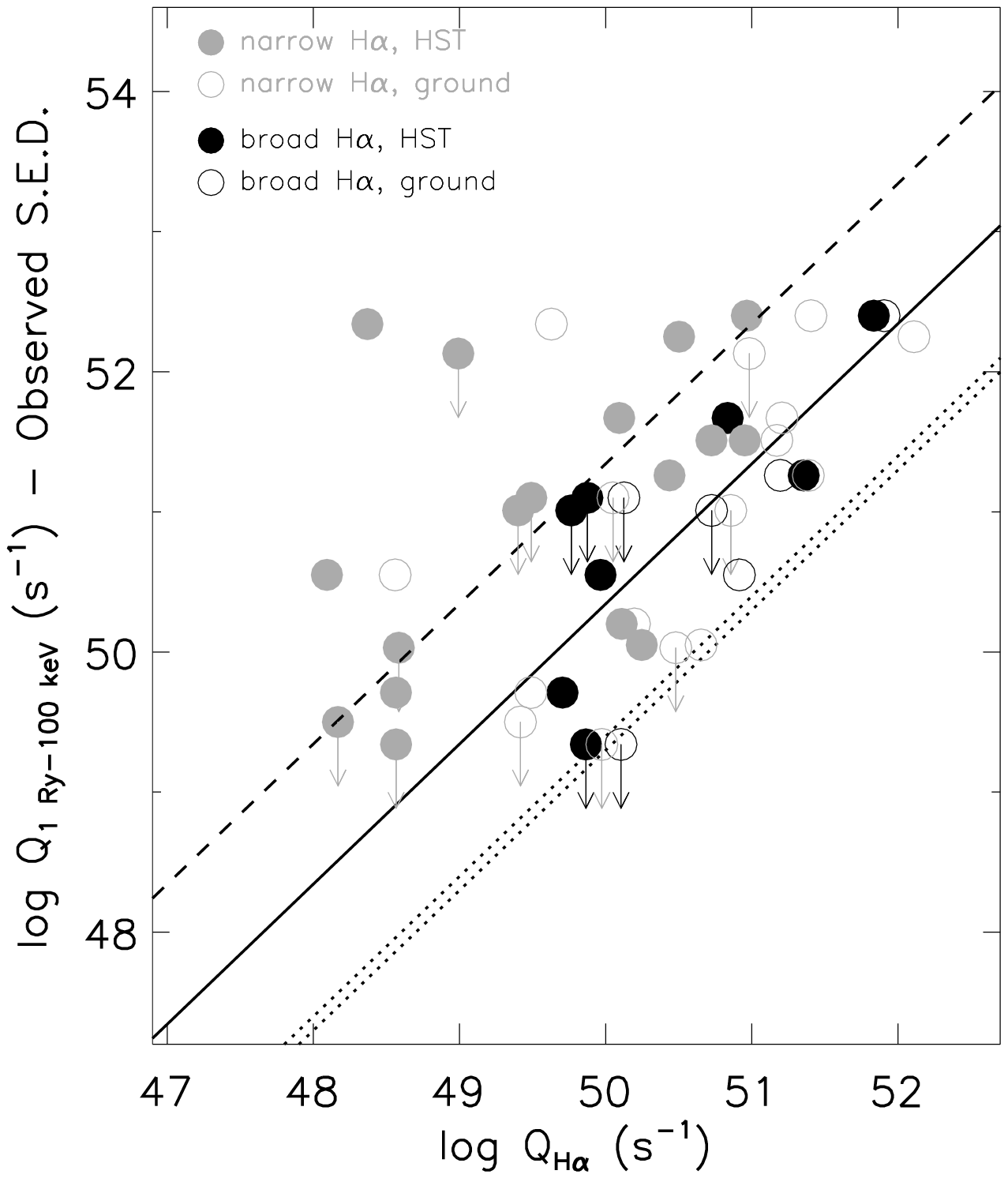}{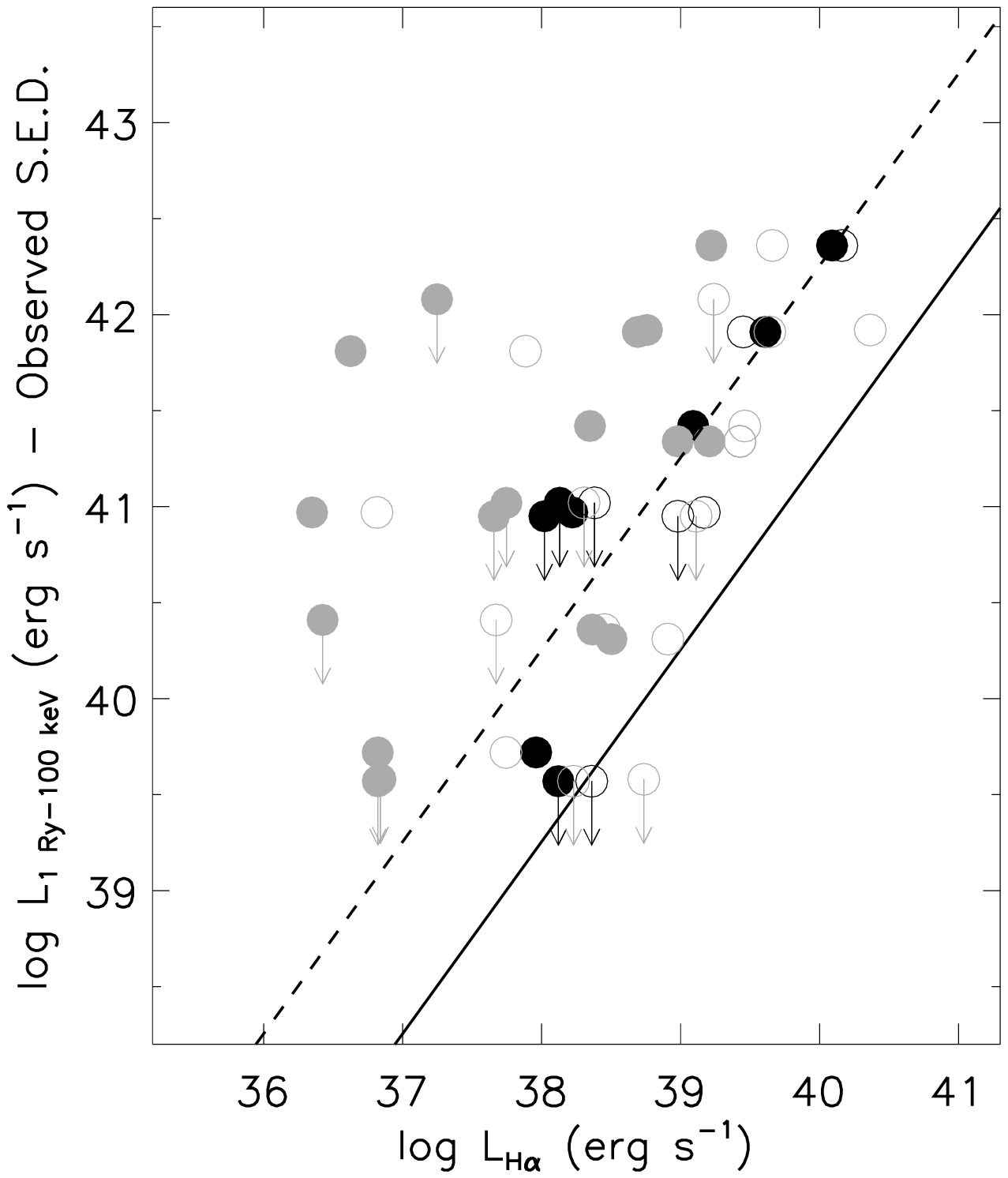}
\caption{The 1~Ry -- 100~keV photon rate ({\it left}) and ionizing luminosity ({\it right}) from EHF, calculated based on the observed SED  as a function of the H$\alpha$ luminosities and photon rates, respectively, for 16 objects from our sample.     We show $Q_{{\rm H}\alpha}$ calculated for both narrow (gray) and broad components (when detected; black), from both ground-based and space-based spectra.   Arrows denote upper limits on the ionizing photon rates.   The solid line illustrates the minimum balance ($Q_{\rm 1~Ry--100~keV} = 22\, Q_{{\rm H}\alpha}$; $L_{\rm 1~Ry--100~keV} = 18\, L_{{\rm H}\alpha}$), considering Case B recombination and covering factor $f_c = 1$; the dashed line illustrates the minimum requirement for a fraction $f_c = 0.1$ of the ionizing photons absorbed by the nebula, and the double-dotted line corresponds to $Q_i = 0.25\, Q_{{\rm H}\alpha}$ (see text for explanations).  
\label{qenerg}}
\end{figure*}

\subsection{Comparison of Ionizing and Emission-Line Power}

Another way of testing the dominant power source in these low-luminosity AGNs can be an assessment of whether AGN photoionization can power the measured emission-line luminosities, for both the broad and the narrow H$\alpha$ components.    
Following EHF, and using the data therein, we run  an energy-budget test via a direct comparison of the H$\alpha$ luminosity and count rate with the ionizing luminosity $L_i = L_{\rm 1~Ry--100~keV}$ and the ionizing photon rate $Q_i = Q_{\rm 1~Ry--100~keV}$, that are obtained by integrating the observed spectral energy distributions (SEDs) or by scaling a template quasar SED.     

Based on photoionization models (Cloudy, v94.0; Ferland et al. 1998) computed by \citet{lewis03} for a wide range of ionization parameters, densities, and metallicities, EHF find that energy balance in a line-emitting nebula requires $L_i > (18 \pm 2) L_{{\rm H}\alpha}/f_c$, where $f_c$ is the covering factor, or the fraction  of the ionizing luminosity of the AGN that is absorbed by the line-emitting gas.    
Thus,  with only  a fraction 10\% of ionizing photons being absorbed by the line-emitting gas, the minimum energy balance condition for AGN ionization is given by $L_i/L_{{\rm H}\alpha} > 180$, for both narrow and broad emission features.     
At the same time, a minimum requirement for photon balance can be quite different for the two emission regions: for the narrow-line component $Q_i > 2.2~ Q_{{\rm H}\alpha}$, corresponding to Case B recombination (i.e., one H$\alpha$ photon is emitted for every 2.2 recombinations);  for the denser and much closer (to the center) BLR, the number of H$\alpha$ photons that can be produced for each ionization can be at least 7--8 times higher than the standard Case B estimate, or  $Q_i \ga 0.25~ Q_{{\rm H}\alpha}$ \citep[i.e., potential contributions to the H$\alpha$ emission via collisional excitation should be minimal for the NLR, but important for the broad component; e.g.,][]{ost89}.  

$Q_{\rm 1~Ry--100~keV}$ and $L_{\rm 1~Ry--100~keV}$  measurements from EHF are available for 16 of our objects, and we present them in Figure~\ref{qenerg} as a function of $Q_{{\rm H}\alpha}$ and $L_{{\rm H}\alpha}$ calculated separately for the narrow (gray) and broad (when detected; black) components, from both ground-based and space-based spectra.   
The sample includes two T, one S, five L1, and eight L2 nuclei, as classified from the ground. 
The ground-based narrow H$\alpha$ measurements reflect merely a reiteration of the results of the EHF analysis, and are generally consistent with a deficit of ionizing photons.   Nevertheless, the small-aperture narrow nebular emission  clearly moves away from the deficit, higher than the lines corresponding to the various energy-budget limits discussed above, showing that the (weak) AGN is definitely capable of powering the line emission within the compact region immediately surrounding it.    As probably expected, the energy balance is satisfied for the broad H$\alpha$ emission in all cases; all of the points are above the line corresponding to the minimum requirement for AGN photoionization.   

These results support with new and homogeneous data similar trends presented previously by EHF.    For ten of the objects included in our sample, EHF also illustrated a comparison of the ionizing photon rate with the rate required to power the H$\alpha$ emission in the compact regions mapped by \emph{HST}, and their conclusion was that the ``little monster" is generally capable of powering the line emission close to it.   It should be noted, however, that their measurements consisted of a mix of narrow-band imaging and spectroscopy, and that the recorded H$\alpha$ emission included both the narrow and the broad components.    The separation of the broad and narrow emission is important, as the energy-balance mechanisms operating in these two regions are expected to be qualitatively different, and we are able to pursue this with our data for the first time.

\section{Summary \& Conclusions}

In this paper we have presented the results of an analysis of the
nebular emission in a sample of $\sim100$ nearby galaxy nuclei, with the
purpose of understanding the often ambiguous underlying energy sources in these 
objects.  Our investigation employs a comparison of the nebular line
emission sampled by large (ground-based) and small (\emph{HST}) spectroscopic
apertures.  With this approach, we have examined the spatial
distribution of emission-line surface brightness and the excitation 
structure, to identify signatures of accretion sources and their
properties.
The conclusions of this analysis are as follows.\\

(1)  The nuclear emission is clearly resolved by the STIS aperture, as
    the narrow H$\alpha$ flux is generally less than that measured in
    the Palomar spectra.  Sometimes the difference in H$\alpha$ flux is
    larger than what the simple ratio of aperture areas would imply
    for a uniform distribution; especially when measured in the H~{\sc ii} nuclei, 
   this might suggest that star formation may not extend to the very central regions.  
    The difference in flux between the two apertures is at least as strong in the T nuclei as in
    the narrow-lined LINERs, suggesting that the geometry of the nebular material
    in these spectral categories does not differ substantially.
    
(2)   In spite of the enhanced sensitivity offered by the \emph{HST} data, there are only three Transition and five LINER 2 nuclei, as defined from the ground, for which broad H$\alpha$ emission is newly detected and interpreted as originating from an AGN-type ionizing source.   
These statistics imply at least a three-fold increase in the incidence of the broad-lined nuclei among  Transition Objects, and an overall increase by at least 30\% in the true incidence of galaxies harboring actively accreting supermassive black holes in their centers, relative to the detections achieved with ground-based spectroscopy.   

(3)  While we confirm that most of the previously reported broad H$\alpha$ detections from the ground are robust, measured broad-line emission fluxes are generally lower in the \emph{HST}
    spectra than those observed in the Palomar survey.  
    This trend probably reflects a ground-based detection bias for high-state objects; thus, the
    data reveal novel evidence for broad-line variability on a
    decade-long timescale, with at least a factor of three in amplitude.

(4)  Although the \emph{HST} observations offer an order of magnitude increase in contrast, there is only a minority of objects that show the expected increasing AGN domination on small scales.   While the radial gradients in the emission-line ratios are modest on average, there is variation among spectral types and a clear dependence on the critical density $n_{\rm crit}$ of each forbidden line considered: the T nuclei are the sources showing the most significant changes in all of the nebular line ratios between the two apertures, the L2 systems exhibit moderate gradients, while the L1s and Seyferts exhibit an increase only in the [\ion{O}{1}]/H$\alpha$ ratio, the transition with the highest $n_{\rm crit}$.

(5)  For all types of nuclei, the gas densities are generally higher in
    the more central regions.  These gradients in the gas densities
    provide a sound explanation for the general lack of change in the nebular line ratios,
    particularly for the traditional accretion sources like L1 and
    S nuclei: the higher densities potentially flatten the ionization
    gradients and suppress forbidden emission from low-$n_{\rm crit}$ transitions in the densest, more
    central regions.   The T nuclei exhibit the strongest density gradients,
    thereby possibly explaining the wide range in the nebular-line flux ratios (i.e., the few negative gradients measured in the lowest $n_{\rm crit}$ transition, [\ion{S}{2}] $\lambda\lambda$6716, 6731) and the lack of change in the  [\ion{N}{2}]/H$\alpha$ flux ratios.

(6) The three-fold increase in the incidence of the BLR, along with the presence of the largest line-ratio gradients in the T nuclei, argue for an AGN-like nature of these sources in their most nuclear regions.   The subtle differences in the line-ratio gradients between T and L2 galaxy nuclei, along with a statistical difference in their $M_{\rm bh}$, support the proposed H~{\sc ii} $\rightarrow$ T $\rightarrow$ S $\rightarrow$ L sequence in which T nuclei are intermediate phases in galaxy evolution from star forming to a brief optically detected AGN state (i.e., the Seyfert nucleus), followed by a transition to quiescence via the L1 and then L2 states.   This picture offers new insight into the composite model for these sources:  the central accretion source that is surrounded by star-forming regions is more likely a Seyfert than a LINER, unlike previously thought.

(7)  The data bring additional support to the idea that L2 nuclei are the lower luminosity, lower accretion efficiency counterparts of the L1 systems;  in this scenario, the weak levels of accretion are responsible for the disappearance of the BLR.   
An energy-balance analysis shows that when one isolates the nuclear emission properly, there is really no energy-budget problem from accretion in the L2 galaxies, consistent with their interpretation as endpoints in the waning stages of accretion as galaxies transition from an AGN to quiescence.

(8) The H~{\sc ii} nuclei remain separated from the other types of
    sources in almost all of their nebular properties.  If these
    objects harbor an AGN in their centers, this must be weak enough
    to be entirely buried into the star-forming emission (and/or dust)
    that remains dominant at $\sim 10$~pc scales.

(9)  Possible selection effects that could skew these results are minimal, as both the line ratios measured in the \emph{HST} apertures and the ratios of ratios measured in the two apertures do not vary as a function of either the physical scale of the employed aperture or the inclination of the host galaxy.  \\

\acknowledgements 
We thank the referee for comments that significantly helped improve the paper.
This work has been supported by NASA through grants GO-12187 and AR-11749
from the Space Telescope Science Institute (STScI), which is operated by the
Association of Universities for Research in Astronomy, Inc., under
NASA contract NAS5-26555.   A.V.F. is also grateful for the financial 
support of NSF grant AST-1108665, the TABASGO Foundation, and the 
Christopher R. Redlich Fund. 
L.C.H. acknowledges additional support from the Kavli Foundation, Peking University, and the Chinese Academy of Science through grant No. XDB09030102 (Emergence of Cosmological Structures) from the Strategic Priority Research Program.  Research by A.J.B. has been supported in part by NSF grant AST-1412693. 
A.C. and C.A.C. acknowledge support by the 4-VA Collaborative at James Madison University.
New observations reported here were obtained at the MMT Observatory, a joint facility of the Smithsonian Institution and the University of Arizona.  This research has made use of the NASA/IPAC Extragalactic Database (NED) which is operated by the Jet Propulsion Laboratory, California Institute of Technology, under contract with NASA.  The data presented herein were obtained from the Multimission Archive at STScI (MAST); support for MAST for non-\emph{HST} data is provided by the NASA Office of Space Science via grant NAG5-7584 and by other grants and contracts.

\clearpage

\newpage

\LongTables 

\begin{deluxetable*}{lrrllcccl}
\tablecolumns{9} \tablewidth{0pt} \tablecaption{The Nuclei Sample, {\it HST}-STIS Observations
\label{tbl-1whole}} \tablehead{ 
\colhead{Object} & 
\colhead{Prop.} & 
\colhead{$D_L$\tablenotemark{a}} &
\colhead{Spectral} & 
\colhead{Aperture} &
\colhead{Central $\lambda$} & 
\colhead{Plate} &
\colhead{Palomar} &
\colhead{Notes}\\
\colhead{Name} & 
\colhead{ID} & 
\colhead{(Mpc)} &
\colhead{Type\tablenotemark{b}} & 
\colhead{Size} &
\colhead{(\AA)} & 
\colhead{Scale\tablenotemark{c}} &
\colhead{/MMT\tablenotemark{d}} &
\colhead{}} 
\startdata 
  IC 356      &12187&18.90&  T2      &0$\farcs$25 $\times$ 0$\farcs$1 & 6581 &0.05  &P& (3)\\  
  NGC 193  & 8236 & 56.40 &L/L   &0$\farcs$25 $\times$ 0$\farcs$2 & 6768 &0.10 &M& (1)\\  
 NGC 315   & 8236 &59.60&  L1.9   &0$\farcs$25 $\times$ 0$\farcs$1 & 6768 &0.05  &P& (1)\\  
 NGC 383   & 8236 & 65.62 &L/L1.9   &0$\farcs$25 $\times$ 0$\farcs$1 & 6768 &0.05 &M& (1)\\
 NGC 541   & 8236 & 63.24 &L:T/T   &0$\farcs$25 $\times$ 0$\farcs$2 &6768  &0.10 &M& \\
 NGC 1052 & 7403 &19.48&  L1.9   &0$\farcs$25 $\times$ 0$\farcs$2 & 6581 &0.05  &P& (1)\\
 NGC 1300 & 8228 & 17.84 &L/S &0$\farcs$25 $\times$ 0$\farcs$2 & 6768 &0.05 &M& \\
 NGC 1497 & 7354 & 86.00 &L/T   &0$\farcs$25 $\times$ 0$\farcs$1 & 6768 &0.05 &M& (2)\\
 NGC 1961 & 9106 &54.10&  L2      &0$\farcs$25 $\times$ 0$\farcs$2 &6581  &0.10  &P & (2)\\
 NGC 2179 & 9068 & 34.20 &L/L   &0$\farcs$25 $\times$ 0$\farcs$2 & 6581 &0.05 &M& \\
 NGC 2329 & 8236 & 85.32 &L/T:   &0$\farcs$25 $\times$ 0$\farcs$2 & 6768 &0.10 &M& \\
 NGC 2685 & 8607 &13.47&  S2/T2&0$\farcs$25 $\times$ 0$\farcs$2 & 6581 &0.10 &P& \\  
 NGC 2748 & 8228 &21.61&  H       &0$\farcs$25 $\times$ 0$\farcs$2 & 6768 &0.05  &P & (3)\\  
 NGC 2892 & 8236 & 96.60 &L/T:   &0$\farcs$25 $\times$ 0$\farcs$2 & 6768 &0.10 &M& \\
 NGC 2903 & 8228 & 9.03 &  H       &0$\farcs$25 $\times$ 0$\farcs$2 & 6768 &0.10   &P & \\  
 NGC 2911 & 7354 &52.00&  L2      &0$\farcs$25 $\times$ 0$\farcs$1 & 6768 &0.05  &P & (2)\\
 NGC 2964 & 8228 &20.43&  H       &0$\farcs$25 $\times$ 0$\farcs$2 & 6768 &0.05   &P& \\  
 NGC 2976 & 8591 & 3.60 &  H       &0$\farcs$25 $\times$ 0$\farcs$1 & 6581 &0.05   &P & \\  
 NGC 2985 &12187&21.50&  T1.9   &0$\farcs$25 $\times$ 0$\farcs$1 & 6581 &0.05 &P& (1)\\  
 NGC 3003 & 8228 &22.12&  H        &0$\farcs$25 $\times$ 0$\farcs$2 & 6768 &0.10 &P  & \\  
 NGC 3031 & 7351 & 3.68 &S1.5/L1&0$\farcs$25 $\times$ 0$\farcs$1 & 6581 &0.05 &P& (1), M81\\
 NGC 3162 & 8228 &26.66&  H        &0$\farcs$25 $\times$ 0$\farcs$2 & 6768 &0.10 &P& \\  
 NGC 3227 & 7403 &20.85&  S1.5    &0$\farcs$25 $\times$ 0$\farcs$2 & 6581 &0.05 &P& (1)\\
 NGC 3245 & 7403 &23.53&  T2:      &0$\farcs$25 $\times$ 0$\farcs$2 & 6581 &0.05 &P& (1)\\
 NGC 3254 & 8228 &32.15&  S2       &0$\farcs$25 $\times$ 0$\farcs$2 & 6768 &0.05 &P& (3)\\  
 NGC 3310 & 8228 &18.10&  H        &0$\farcs$25 $\times$ 0$\farcs$2 & 6768 &0.10 &P& \\  
 NGC 3516 & 8055 &38.90&  S1.2   &0$\farcs$25 $\times$ 0$\farcs$2 & 6581 &0.05 &P& (1),(4)\\
 NGC 3521 & 8228 &12.19& H/L2:: &0$\farcs$25 $\times$ 0$\farcs$2 & 6768 &0.10 &P& (3)\\  
 NGC 3627 & 8607 &  9.96&  T2/S2 &0$\farcs$25 $\times$ 0$\farcs$2 & 6581 &0.10 &P& M66\\  
 NGC 3642 & 8228 &27.50&  L1.9   &0$\farcs$25 $\times$ 0$\farcs$2 & 6768 &0.05 &P& (1)\\
 NGC 3675 & 8607 &17.20&  T2      &0$\farcs$25 $\times$ 0$\farcs$2 & 6581 &0.10 &P& \\  
 NGC 3675 &12187&17.20&  T2      &0$\farcs$25 $\times$ 0$\farcs$1 & 6581 &0.05 &P& \\  
 NGC 3684 & 8228 &23.81&  H        &0$\farcs$25 $\times$ 0$\farcs$2 & 6768 &0.10 &P& (3)\\
 NGC 3686 & 8228 &18.61&  H        &0$\farcs$25 $\times$ 0$\farcs$2 & 6768 &0.05 &P& (2)\\
 NGC 3705 & 8607 &18.43&  T2       &0$\farcs$25 $\times$ 0$\farcs$2 & 6581 &0.10 &P& (3)\\
 NGC 3756 & 8228 &21.43&  H        &0$\farcs$25 $\times$ 0$\farcs$2 & 6768 &0.05 &P & (3)\\
 NGC 3801 & 8236 & 44.55 &S/S   &0$\farcs$25 $\times$ 0$\farcs$2 & 6768 &0.10 &M& \\
 NGC 3862 & 8236 & 94.12 &L/T:   &0$\farcs$25 $\times$ 0$\farcs$2 & 6768 &0.10 &M&  (2)\\
 NGC 3917 & 8607 &17.44&  T2       &0$\farcs$25 $\times$ 0$\farcs$2 & 6581 &0.10 &P& (3)\\
 NGC 3949 & 8228 &18.34&  H        &0$\farcs$25 $\times$ 0$\farcs$2 & 6768 &0.10 &P& (3)\\
 NGC 3953 & 8228 &18.62&  T2       &0$\farcs$25 $\times$ 0$\farcs$2 & 6768 &0.10 &P& \\
 NGC 3953 & 8607 &18.62&  T2       &0$\farcs$25 $\times$ 0$\farcs$2 & 6581 &0.10 &P& \\
 NGC 3998 & 7354 &21.93&  L1.9    & 0$\farcs$25 $\times$ 0$\farcs$1 & 6768 &0.05 &P& (1)\\
 NGC 4036 & 7403 &20.80&  L1.9    &0$\farcs$25 $\times$ 0$\farcs$2 & 6581 &0.05 &P& (1)\\
 NGC 4041 & 8228 &22.70&  H     &0$\farcs$25 $\times$ 0$\farcs$1 & 6768 &0.05 &P& (2)\\
 NGC 4051 & 8228 &14.28&  S1.2 &0$\farcs$25 $\times$ 0$\farcs$2 & 6768 &0.10 &P& (1)\\  
 NGC 4088 & 8228 &16.24&  H    &0$\farcs$25 $\times$ 0$\farcs$2 & 6768 &0.05 &P& \\    
 NGC 4100 & 8228 &20.83&  H    &0$\farcs$25 $\times$ 0$\farcs$2 & 6768 &0.10 &P& \\    
 NGC 4150 & 8607 &14.12&  T2   &0$\farcs$25 $\times$ 0$\farcs$2 & 6581 &0.10 &P& \\  
 NGC 4212 & 8228 &19.01&  H    &0$\farcs$25 $\times$ 0$\farcs$2 & 6768 &0.05 &P& NGC 4208\\    
 NGC 4258 & 8228 &  7.45&  S1.9 &0$\farcs$25 $\times$ 0$\farcs$2 & 6768 &0.05 &P& (1), M106\\
 NGC 4258 & 8591 &  7.45&  S1.9 &0$\farcs$25 $\times$ 0$\farcs$1 & 6581 &0.05 &P& (1), M106\\
 NGC 4261 & 8236 &31.32&  L2   &0$\farcs$25 $\times$ 0$\farcs$1 & 6768 &0.05 &P& 3C 270\\  
 NGC 4278 & 7403 &15.83&  L1.9 &0$\farcs$25 $\times$ 0$\farcs$2 & 6581 &0.05 &P& (1)\\
 NGC 4303 & 8228 &15.16&  H    &0$\farcs$25 $\times$ 0$\farcs$2 & 6768 &0.10 &P& M61\\  
 NGC 4321 & 8228 &16.17&  T2   &0$\farcs$25 $\times$ 0$\farcs$2 & 6768 &0.05 &P& M100\\
 NGC 4335 & 8236 & 66.10 &L/L   &0$\farcs$25 $\times$ 0$\farcs$2 & 6768 &0.10 &M&  (1)\\
 NGC 4374 & 7124 &17.18&  L2    &0$\farcs$25 $\times$ 0$\farcs$2 & 6581&0.10 &P& M84, 3C 274\\  
 NGC 4414 & 8607 &18.31&  T2:  &0$\farcs$25 $\times$ 0$\farcs$2 & 6581 &0.10 &P& (3) \\
 NGC 4429 & 8607 &15.88&  T2   &0$\farcs$25 $\times$ 0$\farcs$2 & 6581 &0.10 &P& (1) \\  
 NGC 4435 & 9068 &16.58&  T2/H &0$\farcs$25 $\times$ 0$\farcs$2 & 6581 &0.05 &P& \\  
 NGC 4486 & 8666 &16.57&  L2     &0$\farcs$25 $\times$ 0$\farcs$2 & 6581&0.05 &P& M87\\  
 NGC 4486 &12162&16.57&  L2     &0$\farcs$25 $\times$ 0$\farcs$1 & 6581&0.05 &P& M87\\  
 NGC 4526 & 9163 &15.52&  H      &0$\farcs$25 $\times$ 0$\farcs$2 & 6768 &0.05 &P& \\  
 NGC 4527 & 8228 &14.99&  T2     &0$\farcs$25 $\times$ 0$\farcs$2 & 6768 &0.05 &P& (3)\\ 
 NGC 4527 & 8607 &14.99&  T2     &0$\farcs$25 $\times$ 0$\farcs$2 & 6581 &0.10 &P& \\  
 NGC 4536 & 8228 &15.29&  H       &0$\farcs$25 $\times$ 0$\farcs$2 & 6768 &0.10 &P& \\  
 NGC 4552 &  8472 &15.56&  T2:   &0$\farcs$25 $\times$ 0$\farcs$1 & 6581&0.10 &P& M89\\  
 NGC 4569 & 8607 &12.35&  T2      &0$\farcs$25 $\times$ 0$\farcs$2 & 6581 &0.10 &P& (3), M90\\  
 NGC 4579 & 7403 &19.58&  S1.9    &0$\farcs$25 $\times$ 0$\farcs$2 & 6581 &0.05 &P& (1), M58 \\
 NGC 4594 & 7354 &10.39&  L2      &0$\farcs$25 $\times$ 0$\farcs$1 & 6768 &0.05 &P& (1), M104 \\
 NGC 4636 & 8472 &16.24&  L1.9  &0$\farcs$25 $\times$ 0$\farcs$2 & 6581 &0.10 &P& (1)\\  
 NGC 4736 & 8591 &  5.02&  L2       &0$\farcs$25 $\times$ 0$\farcs$1 & 6581 &0.05 &P& (1), M94 \\  
 NGC 4800 &12187&22.50&  H         &0$\farcs$25 $\times$ 0$\farcs$1 & 6581 &0.05 &P& (3)\\  
 NGC 4826 & 8591 &  5.34&  T2       &0$\farcs$25 $\times$ 0$\farcs$1 & 6581 &0.05 &P& M64\\  
 NGC 4826 & 8607 &  5.34&  T2     &0$\farcs$25 $\times$ 0$\farcs$2 & 6581 &0.10 &P& \\  
 NGC 5005 & 8228 &20.39&  L1.9  &0$\farcs$25 $\times$ 0$\farcs$2 & 6768 &0.05 &P& (1)\\ 
 NGC 5055 & 8228 &  8.29&  T2     &0$\farcs$25 $\times$ 0$\farcs$2 & 6768 &0.10 &P& (3), M63\\  
 NGC 5077 & 7354 &39.20&  L1.9  &0$\farcs$25 $\times$ 0$\farcs$1 & 6768 &0.05 &P& (1)\\
 NGC 5248 & 8228 &16.88&  H       &0$\farcs$25 $\times$ 0$\farcs$2 & 6768 &0.05 &P& \\  
 NGC 5364 & 8228 &19.51&  H       &0$\farcs$25 $\times$ 0$\farcs$2 & 6768 &0.10 &P& (3)\\  
 NGC 5377 &12187&27.50&  L2      &0$\farcs$25 $\times$ 0$\farcs$1 & 6581 &0.05 &P& \\  
 NGC 5806 &12187&25.54&  H       &0$\farcs$25 $\times$ 0$\farcs$1 & 6581 &0.05 &P& (3)\\  
 NGC 5879 & 8228 &16.12&  T2/L2&0$\farcs$25 $\times$ 0$\farcs$2 & 6768 &0.10 &P& \\
 NGC 5879 & 8607 &16.12&  T2/L2&0$\farcs$25 $\times$ 0$\farcs$2 & 6581 &0.10 &P& \\
 NGC 5905 & 9177 &45.33&  H       &0$\farcs$25 $\times$ 0$\farcs$1 & 6581 &0.05 &P& \\  
 NGC 5921 & 8228 &21.96&  T2      &0$\farcs$25 $\times$ 0$\farcs$2 & 6768 &0.05 &P& (1)\\  
 NGC 6500 & 7354 &63.68&  L2      &0$\farcs$25 $\times$ 0$\farcs$1 & 6768 &0.05 &P& (2)\\
 NGC 6503 & 8607 &  5.52&  T2/S2&0$\farcs$25 $\times$ 0$\farcs$2 & 6581 &0.10 &P& \\  
 NGC 6951 & 8228 &22.57&  S2      &0$\farcs$25 $\times$ 0$\farcs$2 & 6768 &0.05 &P& \\  
 NGC 7052 & 8236 & 51.55 &T/L   &0$\farcs$25 $\times$ 0$\farcs$1 & 6768 &0.05 &M& \\
 NGC 7331 & 8228 &14.01&  T2      &0$\farcs$25 $\times$ 0$\farcs$2 & 6768 &0.05 &P& (3)\\  
 NGC 7331 & 8607 &14.01&  T2      &0$\farcs$25 $\times$ 0$\farcs$2 & 6581 &0.10 &P& (3)\\  
 NGC 7626 & 8236 &49.42&  L2::    &0$\farcs$25 $\times$ 0$\farcs$2 & 6768 &0.10 &P& (2)\\                                     
UGC 1841 & 8236 & 85.80 &L/L:   &0$\farcs$25 $\times$ 0$\farcs$2 & 6768 &0.10 &M&  (1)\\
UGC 2847 & 8591 &  2.43 &T:/H:T&0$\farcs$25 $\times$ 0$\farcs$1 & 6581 &0.05 &M& IC 342\\
UGC 7115 & 8236 & 98.80 &L/T:   &0$\farcs$25 $\times$ 0$\farcs$2 & 6768 &0.10 &M&  (2)\\
UGC 12064& 8236 & 83.90 &T/L   &0$\farcs$25 $\times$ 0$\farcs$1 & 6768 &0.05 &M& \\
\enddata 
\tablenotetext{a}{The mean distance of all redshift-independent values, from NASA/IPAC Extragalactic Database (NED).}
\tablenotetext{b}{Spectral class for the Palomar objects is from \citet{ho97a}; spectral class for the MMT objects is based on our measurements of emission-line fluxes in the MMT spectra (see Table~\ref{tbl-mmt-mmt}) and spectral classification criteria of  \citet{ho97a}/Kewley et al. (2006). With the exception of NGC 383, all of the MMT objects are of Type 2 (narrow emission lines only) based on their  MMT spectra.  Uncertain classifications are followed by a colon.}  
\tablenotetext{c}{In arcsec pixel$^{-1}$; the spatial pixel size on STIS is intrinsically
0$.\farcs$05, but some spectra were obtained with a
2-pixel binning readout mode along the spatial direction, producing a spatial scale of 0$\farcs$1 in the readout.}
\tablenotetext{d}{The origin of the ground-based data; P for Palomar, M for MMT.}
\tablenotetext{e}{On the \emph{HST} spectral measurements: (1) Broad H$\alpha$ is required to fit the H$\alpha$ 
+ [\ion{N}{2}] emission complex.  (2) Broad wings are apparent in the
H$\alpha$ + [\ion{N}{2}] blend; however, models
of two Gaussians for the [\ion{S}{2}] lines do not require a broad H$\alpha$.  (3) Poor-quality {\it HST}-STIS
spectra, no measurements.  (4) Not suitable for modeling or deblending the emission lines using [\ion{S}{2}] as an empirical template (see Sections ~\ref{mmt_data} and ~\ref{lines}).}
\end{deluxetable*}

\clearpage

\begin{deluxetable*}{lcccccc}
\tablecolumns{7} \tablewidth{0pt} \tablecaption{{\it HST} Fluxes of Narrow Emission Lines for the Nuclei Sample
\label{tbl-fl1whole}} \tablehead{ 
\colhead{Object} & 
\colhead{[\ion{O}{1}]} & 
\colhead{H$\alpha$} &
\colhead{[\ion{N}{2}]} & 
\colhead{[\ion{S}{2}]} &
\colhead{[\ion{S}{2}]} \\
\colhead{name} & 
\colhead{$\lambda$6300} & 
\colhead{(narrow)} &
\colhead{$\lambda$6583} & 
\colhead{$\lambda$6716} &
\colhead{$\lambda$6731} &
\colhead{$f_{\lambda}$(6563~\AA)}} 
\startdata
  IC 356       &\nodata     &\nodata          &  \nodata         &\nodata          & \nodata      &\nodata      \\  
 NGC 193   & \nodata     & 59$\pm$53 & 365$\pm$107 & 210$\pm$50 & 310$\pm$50&5$\pm$4\\
 NGC 315   & \nodata    & 207$\pm$2  &  520$\pm$7  &  109$\pm$8 & 122$\pm$7    &   1.3$\pm$0.1	\\
 NGC 383   & \nodata     & 1570$\pm$720 & 3290$\pm$1800 & 1210$\pm$370 & 1210$\pm$1090&0.07$\pm$0.01\\
 NGC 541   & \nodata     & 92$\pm$11 & 225$\pm$12 & 37$\pm$9 & 32$\pm$9&9$\pm$2\\
 NGC 1052 &3505$\pm$27  &4801$\pm$26 & 4088$\pm$75 &1930$\pm$20 & 2727$\pm$21  &    26$\pm$1	\\
 NGC 1300 & \nodata     & 541$\pm$47 & 851$\pm$49 & 295$\pm$40 & 313$\pm$41&69$\pm$20\\
 NGC 1497 & \nodata     & 495$\pm$119 & 909$\pm$94 & 411$\pm$68 & 516$\pm$74&51$\pm$16\\
 NGC 1961 &53$\pm$2     & 165$\pm$3  &  377$\pm$7  &  117$\pm$4 & 141$\pm$6    &   4.4$\pm$0.2	\\
 NGC 2179 &38$\pm$15& 85$\pm$12 & 152$\pm$15 & 85$\pm$12 & 66$\pm$11&25$\pm$14\\
 NGC 2329 & \nodata     & 212$\pm$17 & 833$\pm$26 & 164$\pm$15 & 141$\pm$14&25$\pm$2\\
 NGC 2685 & 11$\pm$3    &  18$\pm$2  &   86$\pm$3  &    6$\pm$2 &  24$\pm$3    &    19$\pm$1	\\
 NGC 2748 & \nodata     & \nodata    & \nodata     & \nodata    &  \nodata     &   \nodata	\\
 NGC 2892 & \nodata     & 286$\pm$57 & 1044$\pm$70 & 114$\pm$31 & 121$\pm$31&14$\pm$2\\
 NGC 2903 & \nodata     &  54$\pm$5  &   72$\pm$6  &   32$\pm$5 &   18$\pm$4   &   4.0$\pm$1.0	\\
 NGC 2911 &  \nodata    &  44$\pm$8  &  119$\pm$8  &  39$\pm$9  &   50$\pm$8   &   3.6$\pm$0.3	\\
 NGC 2964 & \nodata     &  83$\pm$5  &   89$\pm$5  &   25$\pm$4 &   26$\pm$4   &    13.1$\pm$0.6\\
 NGC 2976 & \nodata     & 166$\pm$2  &   82$\pm$2  &   28$\pm$4 &   9$\pm$3    &   4.7$\pm$0.7	\\
 NGC 2985 &801$\pm$34 & 381$\pm$38 & 1301$\pm$51 & 344$\pm$22 & 354$\pm$21 & 43$\pm$1\\  
 NGC 3003 & \nodata     & 345$\pm$11 &   87$\pm$6  &   25$\pm$4 &   19$\pm$4   & 0.093$\pm$0.06	\\
 NGC 3031 &6440$\pm$170 &6360$\pm$190&13440$\pm$160&1980$\pm$70 & 3130$\pm$70  &   1.2$\pm$0.5	\\
 NGC 3162 & \nodata     & 141$\pm$7  &   53$\pm$5  &   18$\pm$4 &   13$\pm$4   &     8$\pm$1	\\
 NGC 3227 & 637$\pm$19  &2032$\pm$24 & 3351$\pm$13 & 749$\pm$18 &  777$\pm$17  &    21$\pm$1	\\
 NGC 3245 &  62$\pm$7   &  84$\pm$7  &  217$\pm$11 &  57$\pm$4  &   66$\pm$4   &    33$\pm$2	\\
 NGC 3254 & \nodata     &  \nodata   &  \nodata    &  \nodata   &  \nodata     &   \nodata	\\
 NGC 3310 & \nodata     &3539$\pm$60 & 2139$\pm$43 &  273$\pm$23&  296$\pm$23  &    16$\pm$1	\\
 NGC 3516 &\nodata     &\nodata     &  \nodata     &\nodata      & \nodata      &\nodata     \\  
 NGC 3521 & \nodata     &  \nodata   &  \nodata    &  \nodata   &  \nodata     &   \nodata	\\
 NGC 3627 & 14$\pm$3    & 172$\pm$3  &  209$\pm$3  &   67$\pm$3 &  71$\pm$3    &    16$\pm$1	\\
 NGC 3642 & \nodata     & 271$\pm$27 &   95$\pm$12 &   77$\pm$8 &   73$\pm$8   &   16$\pm$1	\\
 NGC 3675 & 12$\pm$5    &  22$\pm$3  &   94$\pm$3  &   11$\pm$3 &  23$\pm$3    &   26$\pm$2	\\
 NGC 3675 &90$\pm$37 & 124$\pm$20 &  378$\pm$29 & 82$\pm$19 & 76$\pm$18& 64$\pm$2 \\  
 NGC 3684 & \nodata     &  \nodata   &  \nodata    &  \nodata   &  \nodata     &   \nodata	\\
 NGC 3686 & \nodata     & 850$\pm$20 &  516$\pm$16 &   31$\pm$9 &  47$\pm$10   &   29$\pm$1	\\
 NGC 3705 & \nodata     & \nodata    & \nodata     & \nodata    & \nodata      &   \nodata	\\
 NGC 3756 & \nodata     &  \nodata   &  \nodata    &    \nodata &  \nodata     &   \nodata	\\
 NGC 3801 & \nodata     & 79$\pm$14 & 132$\pm$26 & 10$\pm$9 & 9$\pm$8&9$\pm$2\\
 NGC 3862 & \nodata     & 286$\pm$47 & 1217$\pm$65 & 227$\pm$21 & 318$\pm$24&33$\pm$4\\
 NGC 3917 & \nodata     & \nodata    & \nodata     & \nodata    & \nodata      &   \nodata	\\
 NGC 3949 & \nodata     &  \nodata   &  \nodata    &    \nodata &  \nodata     &   \nodata	\\
 NGC 3953 & \nodata     &  28$\pm$6  &   54$\pm$7  &   18$\pm$6 &  30$\pm$7    &   6.3$\pm$0.7	\\
 NGC 3953 & \nodata     &  21$\pm$2  &   30$\pm$2  &    6$\pm$2 &  10$\pm$2    &   11.1$\pm$0.5	\\
 NGC 3998 &  \nodata    &6475$\pm$2  & 6466$\pm$5  &1585$\pm$8  & 1984$\pm$6   &   62$\pm$6	\\
 NGC 4036 & 135$\pm$6   & 161$\pm$9  &  479$\pm$12 & 158$\pm$4  &  171$\pm$4   &   20$\pm$1	\\
 NGC 4041 & \nodata     & 371$\pm$26 &  199$\pm$28 &   66$\pm$11&  70$\pm$12   &   8.2$\pm$0.8	\\
 NGC 4051 & \nodata     &5790$\pm$410& 2510$\pm$610&  614$\pm$68& 628$\pm$94   &   104$\pm$7	\\
 NGC 4088 & \nodata     &  31$\pm$5  &   23$\pm$4  &    6$\pm$3 &  11$\pm$3    &   3$\pm$0.7	\\
 NGC 4100 & \nodata     & 286$\pm$10 &  144$\pm$7  &   37$\pm$6 &  26$\pm$5    &   3$\pm$0.5	\\
 NGC 4150 & \nodata     &   $<$15.6 &   15$\pm$1  &    6$\pm$1 &   5$\pm$1    &   12$\pm$1	\\
 NGC 4212 & \nodata     &  39$\pm$8  &   53$\pm$8  &   27$\pm$7 &  11$\pm$6    &   5.1$\pm$0.4	\\
 NGC 4258 & \nodata     &1274$\pm$19 & 1052$\pm$61 &  544$\pm$17& 651$\pm$18   &   41$\pm$2	\\
 NGC 4258 &750$\pm$9    &1811$\pm$13 & 1591$\pm$4  &  325$\pm$30& 376$\pm$29   &   24$\pm$1	\\
 NGC 4261 & \nodata     &  15$\pm$2  &   56$\pm$3  &   26$\pm$3 &  22$\pm$3    &   1.1$\pm$0.1	\\
 NGC 4278 & 201$\pm$14  & 462$\pm$29 &  622$\pm$37 & 205$\pm$29 & 273$\pm$34   &   15$\pm$1	\\
 NGC 4303 & \nodata     & 392$\pm$16 &  415$\pm$16 &   81$\pm$12& 106$\pm$13   &   40$\pm$2	\\
 NGC 4321 & \nodata     & 632$\pm$17 &  628$\pm$15 &   83$\pm$9 & 103$\pm$10   &   19$\pm$2	\\
 NGC 4335 & \nodata     & 180$\pm$17 & 654$\pm$12 & 183$\pm$14 & 191$\pm$15&39$\pm$4\\
 NGC 4374 &410$\pm$20 & 900$\pm$340 & 3130$\pm$75 & 850$\pm$40 &1280$\pm$35 & 44$\pm$12 \\  
 NGC 4414 & \nodata     & \nodata    & \nodata     &   \nodata  & \nodata      &   \nodata	\\
 NGC 4429 &36$\pm$6     &  44$\pm$5  &  289$\pm$8  &   31$\pm$5 &  45$\pm$5    &   30$\pm$2	\\
 NGC 4435 & \nodata     &  23$\pm$2  &   69$\pm$3  &    9$\pm$2 &  15$\pm$3    &   13$\pm$0.6	\\
 NGC 4486 &2930$\pm$120 & 4900$\pm$290 &15580$\pm$240 &2000$\pm$130 & 4050$\pm$130 &210$\pm$20 \\  
 NGC 4486 &2330$\pm$110 & 2900$\pm$340 & 7530$\pm$270 &1160$\pm$150 & 1910$\pm$140 &150$\pm$30 \\  
 NGC 4526 &  \nodata    &  31$\pm$2  &  227$\pm$6  &   21$\pm$8 &  45$\pm$35   &   29$\pm$2	\\
 NGC 4527 & \nodata     &    \nodata &  \nodata    &  \nodata   &  \nodata     &   \nodata	\\
 NGC 4527 & 4.1$\pm$3.5     &   7$\pm$2  &   28$\pm$2  &    6$\pm$2 &   9$\pm$3    &   4.2$\pm$1.8	\\
 NGC 4536 & \nodata     &  34$\pm$7  &   77$\pm$9  &   36$\pm$10&  42$\pm$11   &   5.6$\pm$0.5	\\
 NGC 4552 &930$\pm$230 & 1500$\pm$1100 &  5000$\pm$810 &740$\pm$220 & 730$\pm$170  &49$\pm$7 \\  
 NGC 4569 & \nodata     & \nodata    & \nodata     &   \nodata  & \nodata      &   \nodata	\\
 NGC 4579 &1518$\pm$21  & 912$\pm$14 & 3014$\pm$36 & 935$\pm$17 & 1047$\pm$18  &   56$\pm$2	\\
 NGC 4594 &  \nodata    & 806$\pm$19 & 2466$\pm$22 & 886$\pm$28 &  886$\pm$20  &   24$\pm$1	\\
 NGC 4636 &20$\pm$11 & $<$21 &  93$\pm$46 &85$\pm$19 & 114$\pm$19  &19$\pm$2 \\  
 NGC 4736 & 503$\pm$31   &  300$\pm$32 &  525$\pm$25  &   248$\pm$18 &  216$\pm$18    &   64$\pm$7	\\
 NGC 4800 &\nodata     &\nodata     &  \nodata     &\nodata      & \nodata      &\nodata     \\  
 NGC 4826 & \nodata     &   5$\pm$2  &   77$\pm$3  &   19$\pm$3 &  10$\pm$2    &   27$\pm$2	\\
 NGC 4826 &11$\pm$8     & 7.4$\pm$4  &  137$\pm$5  &   30$\pm$5 &  37$\pm$5    &   47$\pm$2	\\
 NGC 5005 & \nodata     & 278$\pm$28 & 1160$\pm$36 &  342$\pm$14& 380$\pm$14   &   22$\pm$1	\\
 NGC 5055 & \nodata     &    \nodata &  \nodata    &  \nodata   &  \nodata     &   \nodata	\\
 NGC 5077 &  \nodata    & 418$\pm$18 &  522$\pm$2  & 247$\pm$7  &  286$\pm$11  &   4.3$\pm$0.2	\\
 NGC 5248 & \nodata     &  25$\pm$4  &   77$\pm$5  &    5$\pm$3 &  12$\pm$2    &   13$\pm$0.7	\\
 NGC 5364 & \nodata     &    \nodata &  \nodata    &  \nodata   &  \nodata     &   \nodata	\\
 NGC 5377 &93$\pm$26 & 224$\pm$18 &  624$\pm$26 &176$\pm$13 & 173$\pm$16  &81$\pm$3 \\  
 NGC 5806 &\nodata     &\nodata     &  \nodata     &\nodata      & \nodata      &\nodata      \\   
 NGC 5879 & \nodata     &  29$\pm$3  &   52$\pm$4  &   17$\pm$3 &  17$\pm$3    &   2.6$\pm$0.6	\\
 NGC 5879 &~~~8$\pm$3 &  44$\pm$2  &   50$\pm$2  &   17$\pm$2 &  22$\pm$1    &   5.3$\pm$0.2	\\
 NGC 5905 & \nodata     &~144$\pm$5~ & ~174$\pm$5~ &   26$\pm$6 &  38$\pm$7    &   9.4$\pm$0.4	\\
 NGC 5921 & \nodata     & 181$\pm$12 &  307$\pm$12 &   95$\pm$8 & 104$\pm$9    &   23$\pm$2 	\\
 NGC 6500 &  \nodata    & 271$\pm$13 &  241$\pm$6  &  171$\pm$15&~154$\pm$24   &   3.1$\pm$0.7	\\
 NGC 6503 &~~~2.1$\pm$1.7~  &  $<$23.8   &   17$\pm$2  &    9$\pm$3 &   9$\pm$3    &   6.7$\pm$1.2	\\
 NGC 6951 & \nodata     & 145$\pm$7  &  572$\pm$11 &   85$\pm$7 &  89$\pm$6    &   11$\pm$1	\\
 NGC 7052 & \nodata     & 280$\pm$71 & 785$\pm$35 & 184$\pm$44 & 184$\pm$44&0.02$\pm$0.01\\
NGC 7331 & \nodata     &   \nodata  &  \nodata    &   \nodata  &  \nodata     &   \nodata	\\
 NGC 7331 & \nodata     &   \nodata  &  \nodata    &   \nodata  &  \nodata     &   \nodata	\\
 NGC 7626 & \nodata     &  87$\pm$4  &  299$\pm$10 &   65$\pm$6 &  71$\pm$6    &   8.4$\pm$0.4  \\
UGC 1841 & \nodata     & 445$\pm$159 & 1558$\pm$394 & 459$\pm$69 & 527$\pm$77&5$\pm$1\\
UGC 2847 &260$\pm$90& 702$\pm$267 & 3024$\pm$357 & 377$\pm$154 & 536$\pm$140&631$\pm$49\\
UGC 7115 & \nodata     & 257$\pm$27 & 756$\pm$35 & 257$\pm$14 & 230$\pm$14&26$\pm$2\\
UGC 12064 & \nodata     & 378$\pm$141 & 1413$\pm$148 & 250$\pm$30 & 250$\pm$20&0.01$\pm$0.006\\
\enddata 
\tablecomments{The emission-line fluxes are in units of 
$10^{-17}$ erg s$^{-1}$ cm$^{-2}$, and represent the observed 
values, not corrected for reddening. The upper limits recorded here for NGC 4150, NGC 4636, and NGC 6503 are listed 
as $(2\pi)^{1/2} \sigma_{\lambda} (3\sigma_{c})$, where 
$\sigma_{\lambda}$ is the width of the [\ion{N}{2}] line, and 
$\sigma_{c}$ is the RMS uncertainty per pixel in the local 
continuum.  The last column lists the continuum flux density 
$f_{\lambda}$ at 6563~\AA, in units of $10^{-17}$ erg s$^{-1}$ 
cm$^{-2}$ \AA$^{-1}$.  Objects are in the same order as in 
Table~\ref{tbl-1whole}.}
\end{deluxetable*}

\end{document}